\documentclass[12pt]{article}
\usepackage{amssymb,amsmath,theorem,latexsym}
\title{Mott law as lower bound \\
for a random walk in a random environment}

\author{A. Faggionato$^{1}$, H. Schulz-Baldes$^{2}$, D. Spehner$^{3}$
\\
\\
$^1$ {\small   Weierstrass Institut f\"ur Angewandte Analysis und
Stochastic, 10117 Berlin, Germany}
\\$^2$
{\small Institut f\"{u}r Mathematik, Technische Universit{\"a}t Berlin,
10623 Berlin, Germany}
\\
$^3$
{\small Fachbereich Physik, Universit{\"a}t Duisburg-Essen, 45117 Essen, Germany}
}

\newtheorem{theo}{Theorem}

\newtheorem{prop}{Proposition}
\newtheorem{lemma}{Lemma}
\newtheorem{coro}{Corollary}
\newtheorem{rem}{Remark}
\newtheorem{ex}{Example}

\newcommand{\NN}{{\mathbb N}}
\newcommand{\RR}{{\mathbb R}}

\newcommand{\ZZ}{{\mathbb Z}}

\newcommand{\Aa}{{\cal A}}

\newcommand{\Pp}{{\cal P}}
\newcommand{\Qq}{{\cal Q}}
\newcommand{\PP}{{\bf P}}
\newcommand{\pp}{{\bf p}}
\newcommand{\QQ}{{\bf Q}}
\newcommand{\EE}{{\bf E}}
\newcommand{\Bb}{{\cal B}}
\newcommand{\Gg}{{\cal G}}
\newcommand{\Hh}{{\cal H}}
\newcommand{\Ee}{{\cal E}}
\newcommand{\Ff}{{\cal F}}
\newcommand{\Ww}{{\cal W}}

\newcommand{\Oo}{{\cal O}}

\newcommand{\Nn}{{\cal N}}
\newcommand{\Cc}{{\cal C}}
\newcommand{\Vv}{{\cal V}}

\newcommand{\Ll}{{\cal L}}
\newcommand{\Tt}{{\cal T}}
\newcommand{\uxi}{{\underline{\xi}}}

\newcommand{\supp}{{\mbox{\rm supp}}}
\newcommand{\pro}{\noindent {\bf Proof. }}

\newcommand{\finpro}{\hfill $\Box$}

\let\a=\alpha    \let\d=\delta  
 \let\g=\gamma     \let\k=\kappa
\let\t=\tau
  
   \let\G=\Gamma
\begin{document}

\maketitle

%\date{}

%%%%%%%%%%%%%%%%%%%%%%%%%%%%%%%%%%%%%%%%%%%%%%%%%%%%
\begin{abstract}
We consider a random walk on the support of an ergodic
 stationary  simple point
process on $\RR^d$, $d\geq 2$, which  satisfies a mixing condition w.r.t. the
translations or has a strictly positive density uniformly  on large enough
cubes. Furthermore the point process is furnished with independent random
bounded  energy marks. The transition rates of the random walk decay
exponentially in the jump distances and depend on the energies through a
factor of the Boltzmann-type. This is an effective model for the
phonon-induced  hopping of electrons in disordered solids within the regime of
strong Anderson localization. We show that the rescaled random walk
converges to a Brownian motion whose diffusion coefficient is bounded below
by Mott's law for the variable range hopping conductivity at zero
frequency. The proof of the  lower bound involves estimates for the
supercritical regime of an associated site percolation problem.
\end{abstract}
%%%%%%%%%%%%%%%%%%%%%%%%%%%%%%%%%%%%%%%%%%%%%%%%%%%%

\vspace{.5cm}

%%%%%%%%%%%%%%%%%%%%%%%%%%%%%%%%%%%%%%%%%%
\section{Introduction}
\label{sec-intro}

%%%%%%%%%%%%%%%%%%%%%%%%%%%%%%%%%%%%%%%%%%
\subsection{Main Result}

Let us directly describe the model and the main results of this work,
defering a discussion of the underlying physics to the next section.  Suppose
given an infinite countable set of random points $\{ x_j \} \subset \RR^d$
distributed according to some ergodic stationary simple point process.
One can identify this set with the simple counting  measure
$\hat{\xi}=\sum _{j} \d_{x_j} $ having $\{ x_j \}$ as its support, and
then write $x\in\hat{\xi}$ if $x\in\{ x_j \}$.
The $\sigma$-algebra $\Bb(\hat{\Nn})$
on the space $\hat{\Nn}$ of counting
 measures on $\RR^d$ is
generated by the family of subsets $\{ \hat{\xi} \in \hat{\Nn} :
\hat{\xi} ( B) = n \}$ where $B \subset \RR^d$ is Borel and $n \in
\NN$. The distribution $\hat{\Pp}$ of the point process is a
probability on the measure space $(\hat{\Nn}, \Bb(\hat{\Nn}))$.
It is stationary and ergodic w.r.t. the translations $x\mapsto
x+y$ of $\RR^d$. In the sequel, we need to impose boundedness of
some $\kappa$th moment defined by

\begin{equation}
\label{gaetano}
\rho_\kappa
\;: =\;
\EE_{\hat{\Pp}}\bigl( \hat{\xi} (C_1)^\kappa\bigr) \;.
\end{equation}

\noindent where $C_1=[-\frac{1}{2},\frac{1}{2}]^{ d}$
and $\EE_{\hat{\Pp}}$ is the expectation w.r.t. $\hat{\Pp}$. Then
$\rho=\rho_1$ is the so--called intensity of the process.

\vspace{.2cm}

To  each $x_j$ is associated a random energy mark $E_{x_j}
\in [-1,1]$. These marks are drawn independently and identically
according to a probability measure $\nu$. Again, $\{ (x_j,E_{x_j})
\}$ is naturally
 identified with an element $\xi$ of the
space $\Nn$ of counting measures on $\RR^d\times[-1,1]$, and the
distribution $\Pp$ of the marked process is a measure on $(\Nn,
\Bb(\Nn))$ (with $\Bb(\Nn)$ defined similarly to
$\Bb(\hat{\Nn})$). The distribution $\Pp$ is said to be the {\it
$\nu$--randomization} of $\hat\Pp$~\cite{Kal}. It is stationary
and ergodic w.r.t. $\RR^d$--translations.
In order to assure that $\{ x_j \}$
contains the origin, we consider the  measurable subset
$\Nn_0=\{\xi\in\Nn\,:\,\xi(\{0\}\times[-1,1])=1\}$ furnished with
the $\sigma$-algebra $\Bb(\Nn_0)=\{A\cap \Nn_0\,:\,A\in
\Bb(\Nn)\}$.  The \emph{random environment} is given by a
configuration  $\xi \in \Nn_0$ randomly chosen along  the Palm
distribution $\Pp_0$ associated to $\Pp$. Roughly, one can think
of $\Pp_0$ as
 the probability on $(\Nn_0,\Bb(\Nn_0))$
obtained by  conditioning $\Pp$ to the event $\Nn_0$ (see
Section~\ref{sec-environment}).  Note that  almost each
environment is a simple counting measure, and therefore it can be
identified with its support as we will do in what follows.

 \vspace{.2cm}

 For a fixed environment $\xi \equiv \{ (x_j, E_{x_j} )\}\in\Nn_0$
  let us consider a continuous-time
random walk over the points $\{ x_j \}$ starting at the origin
$x=0$ with transition rates from $x\in\hat{\xi}$ to
$y\in\hat{\xi}$ given by

\begin{equation}
\label{eq-rates}
c_{x,y}(\xi)
\;:=\;
\exp\big(- |x-y|-\beta(|E_x-E_y|+|E_x|+|E_y|)\big)
\mbox{ , }
x \;\neq\; y
\mbox{ , }
\end{equation}

\noindent where $\beta>0$ is the inverse temperature. More
precisely, let $\Omega_\xi=D([0,\infty),\text{supp}(\hat\xi))$ be
the  space of right-continuous paths on the support of $\hat\xi$
having left limits, endowed with the Skorohod topology~\cite{Bil}.
Let us write $(X_t^\xi)_{t\geq 0}$ for a generic element of
$\Omega_\xi$. If  $\PP_0^\xi$ denotes the distribution on
$(\Omega_\xi, \Bb (\Omega_\xi))$ of the  above random walk
starting at the origin, then the set of stationary transition
probabilities  $p_{t}^\xi (y|x) :=\PP^\xi_0 ( X_{s+t}^\xi=y|
X_{s}^\xi = x )$,  $x,y \in \hat{\xi}$, $t \geq 0, s  > 0$ satisfy
the following conditions for small values of $t$~\cite{Br}:

\vspace{0.2cm}

\noindent (C1) $p_t^\xi (y|x) =  c_{x,y}(\xi) \, t + o(t)$ if $x
\not= y$;

\vspace{0.2cm}

\noindent (C2) $p_t^\xi (x|x) = 1- \lambda_x (\xi)\, t + o(t)$
 with $ \lambda_x(\xi) \;: =\; \sum_{y\in\hat{\xi}} c_{x,y}(\xi)$,
where $c_{x,x}(\xi) := 0$.

\vspace{0.2cm}

\noindent It is verified in Appendix~\ref{app-process} that, provided that $\rho_2<\infty$, no
explosions occur and thus the random walk
  is well-defined for $\Pp_0$--almost all $\xi$.

\vspace{0.2cm}

Our main interest concerns the long time asymptotics of the
random walk and the diffusion matrix $D$ defined by
\begin{equation}
\label{eq-Diffmatrix}
(a\cdot Da)
\;=\;
\lim_{t\to\infty}\;\frac{1}{t}
\;\EE_{\Pp_0}\left(\;\EE_{\PP^\xi_0 } \left((X_t^\xi \cdot a)^2
\right)\right)
\;,
\qquad
a\in\RR^d\;,
\end{equation}

\noindent where $(a \cdot b)$ denotes the scalar product of the
vectors $a$ and $b$ in $\RR^d$.  The main results of the work are
 (i) the existence of the limit (\ref{eq-Diffmatrix}) in any dimension
$d\geq 1$ as well as the convergence of the
(diffusively rescaled) random walk to a Brownian motion with
finite covariance matrix $D \geq 0$; (ii) a quantitative lower bound on $D$ in
dimension $d \geq 2$ under given assumptions on the energy
distribution $\nu$ and either one of the following two technical
hypothesis.   Let $\ell$ denote the Lebesgue measure and
$C_N=[-N/2,N/2]^d$. Given $A\subset \RR^d$, let $\Ff_A$ be the
$\sigma$--subalgebra in $\Bb(\hat{\Nn}) $ generated by the random
variables $\hat{\xi}(B)$ with $B\subset A$ and $B\in\Bb(\RR^d)$.

\vspace{.2cm}

%%%%%%%%%%%%%%%%%%%%%%%%%%
\noindent (H1) {\it $\hat{\Pp}$
admits a lower bound $\rho'>0$ on the point density:

\begin{equation}
\label{eq-uniformbound}
\hat{\xi}(C_N)\;\geq\;\rho'\,\ell(C_N)
\;,
\qquad
\forall\;\;N\geq N_0
\;,
\;\;\;\;\hat{\Pp}\mbox{-a.s.}\;,
\end{equation}

\noindent with $\rho'$ and $N_0$ independent on $\hat{\xi}$.}

\vspace{.2cm}

\noindent (H2) {\it $\hat{\Pp}$ satisfies the following mixing
condition: there exists a function $h:\RR_+\to\RR_+$ with
$h(r)\leq c(1+r^{2d+7+\delta})^{-1}$ for some $c,\delta>0$ such
that for any $r_2\geq r_1>1$,

\begin{equation}
\label{eq-mixing}  \left| \hat{\Pp}(A|\Ff_{\RR^d \setminus
C_{r_2}})\,-\,\hat{\Pp}(A)\right| \;\leq\;
r_1^d\,r_2^{d-1}\,h(r_2-r_1)\;, \qquad \forall\; A\in\Ff_{C_{r_1}}
\;, \;\;\;\;\hat{\Pp}\mbox{-a.s.}\;.
\end{equation}
}
%%%%%%%%%%%%%%%%%%%%%%%%%%

  We feel that hypothesis (H1) and (H2) cover nearly all
interesting examples (see, however, Example 2 below). 
The uniform lower bound (H1) holds in the
case of random and quasiperiodic tilings and, more generally, the
so-called Delone sets \cite{BHZ}. The type of mixing condition
(H2) is inspired by decorrelation estimates holding for Gibbs
measures of spin systems in a high temperature phase \cite{Mar}.
It is satisfied for a stationary Poisson point process as well as
for point processes with finite range correlations. Due to the
stationarity of $\hat{\Pp}$, (H2)  implies that $\hat{\Pp}$ is a
mixing, and, in particular, ergodic point process (see
\cite[Chapter 10]{DVJ}). We can now state more precisely the
above-mentioned results.

%%%%%%%%%%%%%%%%%%%%%%%%%%
\begin{theo}
\label{theo-Mott} Let $\hat\Pp$ be the distribution of an ergodic
stationary simple point process on $\RR^d$, let  $\Pp$ be the
distribution of its $\nu$--randomization with a probability
measure $\nu$ on $[-1,1]$, and let $\Pp_0$ be the Palm distribution
associated to $\Pp$.  Assume that  $\rho_{12}<\infty$ and that
\begin{equation}
\label{melone} \xi= \sum _j \d_{(x_j,E_j)}\qquad \Rightarrow
\qquad  \xi \;\not =\; S_x \xi:= \sum _j \d_{(x_j-x,E_j)} \quad
\forall\; x\in \RR^d \setminus\{0\}\;,\; \qquad \Pp\text{ a.s. }
\end{equation}
Condition {\rm (\ref{melone})} is automatically satisfied if $\nu$
is not a Dirac measure. Then:

\vspace{.1cm}

\noindent {\rm (i)}
 The limit in {\rm (\ref{eq-Diffmatrix})} exists and
the rescaled process  $\underline{Y}^{\xi,\varepsilon} =
(  \varepsilon  X_{t\varepsilon^{-2}}^\xi )_{t\geq 0}$ defined on $(\Omega_\xi,\PP^\xi_0)$
converges  weakly in $\Pp_0$-probability as $\varepsilon \to 0 $ to a Brownian motion $\underline{W}_D$ with
covariance matrix $D$. Namely,
for any bounded continuous function $F$ on the path space $D([0,\infty),
\RR^d )$ endowed with the Skorohod topology,

\begin{equation*}
\EE_{\PP^\xi_0}
 \Bigl(
  F \bigl( \underline{Y}^{\xi, \varepsilon}  \bigr)
 \Bigr)
\;\to\;
   \EE
    \Bigl(
     F \bigl(  \underline{W}_D \bigr)
    \Bigr)
\qquad \text{in $\Pp_0$-probability}\;.
\end{equation*}

\vspace{.1cm}

\noindent {\rm (ii)}
Suppose $d\geq 2$   and let either {\rm (H1)} or {\rm (H2)} be
satisfied. Furthermore, suppose that there are some positive
constants $\a$, $c_0$
such that, for any $0<E \leq 1$,
\begin{equation}
\label{lama}
\nu([-E,E])
\;\geq \;
c_0\; E^{ 1+\a}
\;.
\end{equation}

\noindent Then

\begin{equation}
\label{eq-Mott}
D
\;\geq\;
c_1\,
\beta^{-\frac{d (\alpha+1)}{\alpha+1+d}}
%\beta^{-\frac{d(\alpha+1)}{\alpha+1+d}}
\;
\exp\left(-c_2\,\beta^{\frac{\a+1}{\a+1+d} }\right)\;
{\bf 1}_{d}
\;,
\end{equation}

\noindent where ${\bf 1}_{d}$ is the $d\times d$ identity matrix and
$c_1$ and $c_2$ are some positive $\beta$-independent constants.
\end{theo}
%%%%%%%%%%%%%%%%%%%%%%%%%%

\vspace{.2cm}

The  important factor  in the lower bound (\ref{eq-Mott}) is
the exponential factor and not the power law in front of it (on which
we comment below though).
Based on the following heuristics due to Mott \cite{Mot,SE}, we
expect that the expression in the exponential in (\ref{eq-Mott})
captures the good asymptotic behavior of $\ln D$ in the low
temperature limit $\beta \uparrow \infty$ if $\nu([-E,E])\sim c_0
E^{1+\alpha}$ as $E \downarrow 0$. Indeed, as $\beta$ becomes
larger, the rates (\ref{eq-rates}) fluctuate widely with $(x,y)$
because of the  exponential energy factor. The low temperature
limit effectively selects  only jumps between points with energies
in a small interval  $[-E(\beta),E(\beta)]$  shrinking to zero  as
$\beta \uparrow \infty$.  Assuming that $D$ is determined  by
those jumps with the largest rate, one obtains directly the
characteristic exponential factor on the r.h.s. of (\ref{eq-Mott})
by maximizing these rates for a fixed temperature under the
constraint  that the mean density of points $x_j$ with energies in
$[-E(\beta),E(\beta)]$ is equal to  $\rho\,
\nu([-E(\beta),E(\beta)]) \sim c_0\rho E(\beta)^{1 + \alpha}$. One
speaks of {\it variable range hopping} since the characteristic
mean distance $|x-y|$ between sites with optimal jump rates varies
heavily with the temperature. A crucial (and physically
reasonable, as discussed below) element of this argument is the
independence of the energies $E_{x}$. The selection of the points
$\{ x_j \}$ with energies in the window $[-E(\beta),E(\beta)]$
then corresponds mathematically to a $p$-thinning with
$p=\nu([-E(\beta),E(\beta)])$.   It is a well-known fact (see {\it
e.g.} \cite[Theorem~16.19]{Kal}) that an adequate rescaling of the
$p$-thinning of a stationary point process converges in the limit
$p\downarrow 0$ (corresponding here to $\beta\uparrow\infty$) to a
stationary Poisson point process (PPP). Hence one might call the
stationary PPP the normal form of a model leading Mott's law,
namely the exponential factor on the r.h.s. of (\ref{eq-Mott}),
and we believe that proving the upper bound corresponding to
(\ref{eq-Mott}) should therefore be most simple for the PPP. In
dimension  $d=1$, a different behavior of $D$ is
expected~\cite{LB} and this will not be considered here.   Note
that statement (i) does not  necessarily imply that the  motion of
the particle is diffusive at large time, since it could happen
that $D=0$.

\vspace{.2cm}

The preexponential factor in (\ref{eq-Mott}) can be improved to
$\beta^{\frac{(\alpha+1)(2-d)}{(1+\alpha+d)}}$ by means of formal
scaling arguments on the formulas in Sections~\ref{sec-bounds} to
\ref{sec-randomized}. As we are not sure that this is optimal and we
do not control the constant $c_2$ in (\ref{eq-Mott}) anyway,
we choose not to develop this improvement in detail.

%%%%%%%%%%%%%%%%%%%%%%%%%%%%%%%%%%%%%%%%%%
\subsection{Physical discussion}

Our main motivation for studying the above model comes from its importance
for {\it phonon-assisted hopping conduction} \cite{SE} in disordered solids
in which the Fermi level (set equal to $0$ above) lies in a region of {\it
strong Anderson localization}. This means that
the electron Hamiltonian has exponentially localized quantum eigenstates
with localization centers $x_j$ if the corresponding energies $E_{x_j}$ are close
to the Fermi level.  The DC
conductivity of such materials would vanish if it were not for the lattice
vibrations (phonons) at nonzero temperature. They induce transitions between
the localized eigenstates, the rate of which can be calculated from first
principle by means of the Fermi golden rule \cite{MA,SE}.   In the variable
range hopping regime at low temperature, the Markov and  adiabatic
(or rotating wave) approximations can be used to treat quantum mechanically    the
electron-phonon coupling \cite{Spe}.
Coherences between electronic eigenstates with
different energies decay very rapidly under the resulting dissipative electronic
dynamics and one can show that the hopping DC conductivity of the disordered
solid coincides with the conductivity associated with a Markov jump process
on  the set of localization centers $\{x_j\}$, hence justifying the use of
a model of classical
mechanics \cite{BRSW}.  Because Pauli blocking due to Fermi statistics of
the electrons  has to be taken into account, this leads to a rather
complicated exclusion process ({\sl e.g.} \cite{Qua,FM}). If, however, the
blocking is treated in an {\it effective medium}  (or mean field) approximation,
one obtains
a family of independent random walks with rates which
are given by (\ref{eq-rates}) in the limit $\beta \uparrow
\infty$ \cite{MA,AHL}.

\vspace{.2cm}

 Let us discuss the remaining aspects of the model. The
stationarity   of the underlying simple point process $\{x_j\}$
simply reflects that the material is homogeneous, while the independence of
the energy marks is compatible with Poisson level statistics, which is a
general rough indicator for the localization regime and has been proven to
hold for an Anderson model \cite{Min}. The exponent $\alpha$ allows to
model a possible Coulomb pseudogap in the density of states \cite{SE}.

\vspace{.2cm}

Having in mind the Einstein relation between the conductivity and the
diffusion coefficient (which can be stated as a theorem for a number of
models \cite{Spo}), the lower bound (\ref{eq-Mott}) gives  a lower bound on
the hopping DC condutivity.  In the above materials, the DC conductivity
shows experimentally Mott's law, namely
a low-temperature behavior which is
well approximated by the exponential factor in the r.h.s. of (\ref{eq-Mott})
with $\alpha=0$, as predicted by Mott \cite{Mot}  based on the optimization argument
discussed above. In certain materials having a Coulomb pseudogap in the density
of state, Mott's law with $\alpha=d-1$ is observed, as predicted by Efros
and Shklovskii \cite{EF}.
  A first convincing justification of Mott's argument was
given by Ambegoakar, Halperin and Langer \cite{AHL}, who first  reduced the
hopping model to a related  random resistor network, in a manner similar to
the work  of Miller and Abrahams~\cite{MA}, and then pointed out that the
constant $c_2$ in (\ref{eq-Mott})
can be estimated using  percolation theory~\cite{SE}.  Our
proof
of the lower bound (\ref{eq-Mott}) is inspired by this work.  Let
us also mention that the low frequency AC conductivity (response to an
oscillating electric field) in disordered solids  has recently been studied
within a quantum-mechanical one-body approximation in~\cite{KLP}.  Here the
energy necessary for a jump between localized states comes from a resonance
at the frequency of the external electric field rather than a phonon.  It
leads to another well-known formula for the conductivity which is also due
to Mott.

%%%%%%%%%%%%%%%%%%%%%%%%%%%%%%%%%%%%%%%%%%
\subsection{Overview}

Let us develop the main ideas of the proof of Theorem~\ref{theo-Mott}, leaving
precise statements and their proofs to  the following sections.
The model described above is a {\it random walk in a random
environment}.  A main tool used in this work is the contribution of De
Masi, Ferrari, Goldstein and Wick \cite{DFGW} which is based on prior work
by Kipnis and Varadhan \cite{KV}. They construct a new Markov process, called the
{\it environment viewed from the particle}, which allows to translate the
homogeneity of the medium  into properties of the random
walk. In Section~\ref{sec-viewed}, we argue that
$X_t^\xi$ has finite moments w.r.t. $\int {\Pp_0}(d \xi) {\PP}_0^\xi$
(Proposition~\ref{giulietta})
 and study the
generator of the process environment viewed from the particle when the
initial environment is chosen according to the Palm distribution $\Pp_0$
(Propositions~\ref{prop-ergodicity}
and~\ref{prop-gen}),
thus allowing to apply the general Theorem 2.2 of~\cite{DFGW} to deduce
the existence of the limit (\ref{eq-Diffmatrix}).
The  convergence to a Brownian motion stated in
Theorem~\ref{theo-Mott} also follows, but this could have been
obtained (avoiding an analysis of the infinitesimal
generator)  by applying Theorem 17.1 of \cite{Bil}
and Theorem 2.1 of \cite{DFGW}. The results of~\cite{DFGW} also lead
a variational formula for
the diffusion matrix $D$ (Theorem~\ref{theo-DFGW} below).
The  main virtue of this formula is that it allows to bound
$D$ from below through bounds on the transition rates.

\vspace{2mm}

The first step in proving
Theorem~\ref{theo-Mott}(ii) is to define
a new random walk with transition rates
bounded above by the rates (\ref{eq-rates}). This is done
in Section~\ref{sec-bounds}  in the following way.
For a fixed configuration $\xi \in \Nn_0$ of the environment,
consider the set $\{ x_j^c \}=  \{ x_j  \, :\,  |E_{x_j} | \leq E_c  \}$
of all random points
having energies inside a given energy window
$[-E_c,E_c]$ with $0<E_c\leq 1$.
The distribution $\hat{\Pp}^c$ of these points is obtained from $\Pp$ by
a $\delta_c$-thinning with $\delta_c=\nu([-E_c,E_c])$. Given
a cut-off distance $r_c>0$, consider the
random walk on $\supp (\hat{\xi})$ with the
transition rates
$\hat{c}_{x,y}(\xi)= \chi( |x-y| \leq r_c) \chi ( |E_{x}| \leq E_c) \chi ( |E_{y}| \leq E_c)$,
where $\chi$ is the characteristic function.
Since we want this random walk to have a strictly positive diffusion coefficient
in the limit $E_c\to 0$,
one must choose $r_c$ such  that the mean number of points
$x_j$ with energies in $[-E_c,E_c]$
inside a ball of radius $r_c$ is larger than an $E_c$-independent constant $c_3>0$.
This mean number is equal  to $c_4 \delta_c r_c^d$
and is larger than $c_5 E_c^{1+\alpha} r_c^d$ by
assumption~(\ref{lama}), where $c_4$ and $c_5$ are constants depending on $\rho$
and $d$ only.
Hence $r_c = c_6\,E_c^{-(1+\alpha)/d}$.
It is shown in Proposition~\ref{marcopantani}
that the diffusion matrix of the new  random walk is equal to
$\delta_c \,D(r_c,E_c)$, where $D(r_c,E_c)$ is the
diffusion matrix of a random walk on $\{ x_j^c \}$
with energy-independent transition rates $\chi(|x-y| \leq r_c )$.
By  the monotonicity of $D$ in the jump rates and since
$c_{x,y}(\xi) \geq \exp(-r_c-4 \beta E_c) \,\hat{c}_{x,y}(\xi)$, one gets
using also assumption~(\ref{lama}) and the constant $c_0$ therein:
\begin{equation} \label{eq-lower_b_D}
D \;\geq \; c_0\,E_c^{1+\alpha}\, e^{-r_c-4 \beta E_c}\, D(r_c,E_c)\;.
\end{equation}

In Section~\ref{sec-approx}, a lower bound on $D(r_c,E_c)$
is obtained by considering  periodic approximants (in the limit of large periods)
 as in \cite{DFGW}.
The diffusion
coefficient of these approximants can be computed as the resistance of a random
resistor network.
The resistance of the random
resistor network is bounded by invoking estimates
from  percolation theory in Section~\ref{sec-randomized}, hence showing that,
if $r_c$ is large enough,
$D(r_c,E_c) > c_7 \,{\bf 1}_{d}$
where  $c_7>0$ is independent on $E_c,\beta$.
Recalling that $r_c = c_6\,E_c^{-(1+\alpha)/d}$,
an optimization w.r.t. $E_c$ of the right member
of (\ref{eq-lower_b_D}) then yields  $E_c = c_8 \beta^{-\frac{d}{1+\alpha+d}}$
and thus the lower bound (\ref{eq-Mott}).
Let us note that this optimization is the same as in the Mott argument discussed
above and that $E_c\downarrow 0$  and $r_c \uparrow \infty$ as
$\beta \uparrow \infty$.
Moreover, our optimized lower bound results from a
critical resistor network roughly approximating the one appearing in~\cite{AHL}.

\vspace{.2cm}

The paper is organized as follows.  In Section \ref{sec-environment} we
recall some definitions and results about point processes and state some
technical results needed later on.
The statements (i) and (ii) of Theorem~\ref{theo-Mott} are
proven in Section~\ref{sec-viewed} and in Sections~\ref{sec-bounds}
to~\ref{sec-randomized}, respectively.
In
Appendix~\ref{app-process} we show that the continuous-time random
walk in the random environment is well defined by verifying   the absence of
explosion phenomena. Appendix~\ref{app-prooflemma} contains some technical proofs
about  the Palm measure.
Appendix~\ref{app-LPestimate} is devoted to the proof of Proposition~\ref{giulietta}.

\vspace{.2cm}

%%%%%%%%%%%%%%%%%%%%%%%%%%%%%%%%%%%%%%%%%%%
\noindent
{\bf Acknowledgments:}  We would like to thank A. Bovier,
J. \v Cern\'y, B. Derrida, P. A. Ferrari, D. Gabrielli,
A. Ramirez and  R. Siegmund-Schultze  for very useful
comments. The work was supported by the SFB 288, SFB/TR 12 and the
Dutch-German Bilateral Research Group ``Mathematics of random spatial models
from physics and biology".

%%%%%%%%%%%%%%%%%%%%%%%%%%%%%%%%%%%%%%%%%%
\section{The random environment}
\label{sec-environment}

In this section, we recall some properties  of  point processes
(for more details, see~\cite{DVJ,FKAS,MKM,Kal,Tho}).
In the sequel, given a topological space $X$,  $\Bb(X)$ will denote the
$\sigma$-algebra of Borel subsets of $X$.
Given a set $A$,  $|A|$ will denote its
cardinality.  Moreover, given  a probability measure $\mu$, we write $\EE_\mu$
for the corresponding expectation.

%%%%%%%%%%%%%%%%%%%%%%%%%%%%%%%%%%%%%%%%%%
\subsection{Stationary simple marked point processes}

Given a bounded complete separable metric space $K$, consider the
space $\Nn:=\Nn(\RR^d\times K)$ of all counting measures $\xi$ on
$\RR^d\times K$, {\it i.e.} integer-valued measures such that $\xi
( {B} \times K) < \infty$ for any bounded set ${B} \in \Bb(
\RR^d)$.  One can show that $\xi\in\Nn$ if and only if $\xi=\sum_j
\delta_{(x_j,k_j)}$  where $\delta$ is the Dirac measure and $\{
(x_j,k_j)\} $ is a countable  family  of (not necessarily
distinct) points in $\RR^d\times K$ with at most finitely many
points in any bounded set. Then $k_j$ is called the {\it   mark}
at  $x_j$. Given $\xi\in\Nn$, we write $\hat \xi \in \Nn (\RR^d)$
for the counting measure on $\RR^d$ defined by $\hat
\xi({B})=\xi({B}\times K)$ for any ${B} \in\Bb(\RR^d)$. Given
$x\in\RR^d$,  we write $x\in \hat \xi$  whenever $x\in
\text{supp}(\hat \xi)$.  If $\hat{\xi} ( \{ x \}) \leq 1$ for any
$x \in \RR^d$, we say that $\xi\in \Nn$ is {\it simple} and write
$k_{x_j} := k_j$ for any $x_j \in \hat{\xi}$.

\vspace{0.2cm}

A metric on $\Nn$ can be defined in the following way~\cite[Section
1.15]{MKM}.  Fix an element $k^\ast\in K$. Denote by  $B_{r}(x,k)$ and
$B_{r}$ the open balls in $\RR^d \times K$ of radius  $r>0$  centred on
$(x,k)$ and on $(0,k^\ast)$, respectively.  Let  $\xi = \sum_{i\in I} \delta
_{(x_i,k_i)}$ and ${\xi}^\prime =  \sum_{j\in J} \delta _{({x}_j^\prime ,
{k}_j^\prime )}$ be elements of  $\Nn$, where $I$, $J$ are countable
sets. Then $\xi$ and $\xi'$ are close to each other if any point $(x_i,k_i)$
contained in $B_n$ is close to a point $({x}_j^\prime , {k}_j^\prime )$  for
arbitrary large $n$, up to ``boundary effects''.  More precisely, given a
positive integer $n$, let $d_n (\xi, {\xi}^\prime )$ be the infimum over all
$\varepsilon >0$ such that there is a one-to-one map $f$ from a (possibly
empty) subset $D$ of $I$ into a subset of $J$ with the properties:

\vspace{0.2cm}

(i)~$\supp (\xi)   \cap B_{ n-\varepsilon} \subset \{ (x_i, k_i )\,:\,
i \in D \}$;

(ii)~$\supp  ( {\xi}^\prime) \, \cap B_{n-\varepsilon}
  \subset \{ ( {x}_j^\prime , {k}_j^\prime  )\, :\, j \in f(D) \}$;

(iii)~$( {x}_{f(i)}^\prime , k_{f(i)}^\prime )
\in B_{\varepsilon} (x_i,k_i)$ for $i\in D$.

\vspace{0.2cm}

\noindent
One can show that  $d_{\Nn} (\xi, {\xi}^\prime ) = \sum_{n=1}^\infty 2^{-n}
d_n (\xi, {\xi}^\prime  )$ is a bounded metric on $\Nn$ and for this metric
$\Nn$ is complete and  separable.  Moreover, the sets $\{\xi\in \Nn  \,: \,
\xi(B )=n\}$, $B \in \Bb(\RR^d \times K)$,  $n\in\NN$, generate the Borel
$\sigma$-algebra $\Bb (\Nn )$ and $d_\Nn$ generates the coarsest topology
such that  $\xi \in \Nn \mapsto \int \xi ( dx,dk)\,f(x,k)$ is continuous for
any  continuous function $f \geq 0$  on $\RR^d \times K$ with bounded
support.  Finally,  by choosing different reference points $k^\ast$ one
obtains equivalent metrics.

\vspace*{0.1cm}

A {\it marked point process} on $\RR^d$ with marks in $K$ is then a
measurable map $\Phi$
from a probability space into $\Nn$.  We denote by  $\Pp$ its
distribution (a probability measure on  $(\Nn, \Bb ( \Nn))$).
We say that the
process is {\it simple} if  $\Pp$-almost all $\xi \in \Nn$ are simple.
The translations on $\RR^d$ extend naturally to $\RR^d \times K$ by $S_x:
(y,k) \mapsto (x+y,k)$. This induces an action $S$ of the translation group
$\RR^d$ on $\Nn$ by $(S_x\xi)(B )=\xi(S_x B)$, where $B \in \Bb(\RR^d \times
K )$ and $x \in \RR^d$.  For simple counting measures,

$$
S_ x \xi
\;=\;
\sum_{y \in \hat{\xi}} \, \delta_{(y-x,k_y)}
\mbox{ . }
$$

\noindent A  marked point process is said to be {\it stationary} if
$\Pp(A)=\Pp(S_x A)$ for all $x\in\RR^d$, $A \in\Bb (\Nn )$, and
 {\it (space) ergodic}
if the $\sigma$-algebra of translation invariant sets is trivial,
{\it i.e.}, if $A \in\Bb (\Nn )$ satisfies $S_x A = A$
for all $x\in\RR^d$ then  $\Pp(A)\in\{0,1\}$. Due to \cite[Proposition
10.1.IV]{DVJ}, if $\Pp$ is stationary and gives no weight to the trivial
measure without any point (which will be assumed here),  then

\begin{equation}
\label{marchioro}
\Pp\bigl(\xi\in\Nn\,:\, \bigl|\,\text{supp}(\hat{\xi})\,\bigr|=\infty\,
\bigr)\,=\,1\,.
\end{equation}

The marked point processes studied in this work are obtained by the procedure of
randomization, which we recall now together with the related notion of
thinning (see \cite{Kal}).  Let  $\hat{\Phi}$ be  a stationary simple  point
process (SSPP) on $\RR^d$, $\nu$ be a probability measure  on $[-1,1]$ and
$p\in [0,1]$.  The $\nu$--{\it randomization} of  $\hat{\Phi}$ is the
stationary simple marked point process (SSMPP) $\Phi_\nu$ obtained by assigning
to each realization
$\hat{\xi}=\sum_{i\in
I}\d_{x_i}$ of $\hat{\Phi}$  the measure
$\xi=\sum_{i\in I} \d_{(x_i,E_i)}$, where
$\{E_i\}_{i\in I}$ are independent identically distributed
random variables having distribution
$\nu$. Finally, the $p$--{\it thinning} $\hat{\Phi}_p$ of $\hat{\Phi}$ is the
SSPP on $\RR^d$ obtained by assigning   to  each
%realization $\sum_{i\in I}\d_{x_i}$  of $\hat{\Phi}$
  realization $\hat{\xi}$ the measure $\sum_{i\in I} P_i
\,\d_{x_i}$, where $\{P_i\}_{i\in I}$ are independent  Bernoulli
variables with $\text{Prob}(P_i=1)=p$ and
$\text{Prob}(P_i=0)=1-p$. Both the point processes $\Phi_\nu$ and
$\hat{\Phi}_p$ are examples of {\it stationary cluster processes},
also called  {\it  homogeneous cluster fields} (see \cite[Chapter
8]{DVJ} and \cite[Chapter 10]{MKM}). In particular, ergodicity is
conserved by $\nu$--randomization and $p$--thinning
(\cite[Proposition 10.3.IX]{DVJ} and \cite[Proposition
11.1.4]{MKM}). To conclude, let us give a few examples.

%%%%%%%%%%%%%%%%%%%%%%%%%%%%%%%
\begin{ex} \label{ex-PPP}
{\rm A {\it Poisson point process} (PPP) appears, as already
discussed, naturally as limit distribution of thinnings.  Given a
measure $\mu$ on $X$, with $X$ equal to $\RR^d$ or
$\RR^d\times[-1,1]$, the PPP on $X$ with  intensity measure $\mu$
is defined by the two conditions  (i) for any $B\in\Bb(X)$,
$\xi(B)$ is a Poisson random variable with expectation $\mu(B)$;
(ii) for any disjoint sets ${B}_1,\ldots,{B}_n \in\Bb(X )$,
$\xi(B_1),\ldots, \xi(B_n)$ are independent. A PPP on $\RR^d$ is
stationary if and only if its intensity measure $\mu$ is
proportional to the Lebesgue measure, $\mu = \rho\,dx$. In such a
case it is an ergodic process  satisfying the hypothesis (H2) of
Theorem~\ref{theo-Mott} and all moments $\rho_\k$, $\k >0$ in
(\ref{gaetano}) are finite. Its  $p$--thinning is the PPP on
$\RR^d$
 with intensity $p \rho$  while its $\nu$--randomization  is the PPP on $\RR^d\times[-1,1]$
with intensity measure $\rho\,dx\otimes \nu$.
}
\end{ex}
%%%%%%%%%%%%%%%%%%

%%%%%%%%%%%%%%%%%%
\begin{ex}
\label{ex-Anderson}
{\rm
Let us associate to the  uniformly distributed  random variable $y$  in the
unit cube $C_1$   the point measure $\hat{\xi}= \sum_{z \in \ZZ^d}
\delta_{z+y}$.  The  corresponding point process is an ergodic SSPP
satisfying $\rho_\k=1$ for any $\k >0$.
Although this process satisfies (H1), the SSMPP obtained from it 
via $p$--thinning and $\nu$--randomization does not and  does also
not satisfy (H2). However, Theorem 1(ii) is still valid 
for this SSMPP, as can be checked by restricting  the analysis 
of Section 6 to regions
which are unions of boxes of the form $z+[0,1)^d$, $z\in {\mathbb{Z}}^d$
and using the independence of $\hat{\xi}(A)$ and $\hat{\xi}(B)$ 
when $A,B$ are disjoint unions of such boxes.
}
\end{ex}
%%%%%%%%%%%%%%%%%%

Other examples of ergodic SSMPP can be obtained by means of SSPP
with short--range correlations (see \cite[Exercise 10.3.4]{DVJ}).
Of particular relevance for solid state physics are point
processes associated to random or quasiperiodic tilings
\cite{BHZ}, which satisfy the  hypothesis (H1) of
Theorem~\ref{theo-Mott}.

%%%%%%%%%%%%%%%%%%%%%%%%%%%%%%%%%%%%%%%%%%%%%%%%%%%%%%%%%%%%%%%%%%%
\subsection{The Palm distribution}
%%%%%%%%%%%%%%%%%%%%%%%%%%%%%%%%%%%%%%%%%%%%%%%%%%%%%%%%%%%%%%%%%%%

In what follows, it will always be assumed that $\hat \Pp$ and $\Pp$ are
defined as in Theorem \ref{theo-Mott} and that (\ref{melone}) is satisfied if
$\nu$ is a Dirac measure.
In order to shorten notations, we will write $\Nn$ and  $\hat{\Nn}$
for  $\Nn(\RR^d\times[-1,1])$ and $\Nn(\RR^d)$, respectively.
We would like now to ``pick up at random'' a point among $\{ x_j \}$ and
take it as the origin. One thus looks at the following borelian  subset  of $\Nn$:

$$
\Nn_0
\;:=\;
\left\{
\xi\in\Nn \,: \, 0\in \hat \xi
\right\}
\;.
$$

\noindent Since $\Nn_0$ is closed, it defines a  bounded
complete separable metric space. Note that $x \in \hat{\xi}$ if
and only if $S_x \xi \in \Nn_0$.  The Palm distribution $\Pp_0$ on
$\Nn_0$ associated to $\Pp$ is now defined as follows.  Consider
the measurable map $\Gg$ from $ \Nn$ into $ \Nn ( \RR^d \times
\Nn_0)$ given by $ \xi \;\mapsto\; \xi^\ast \;  = \;
\sum_{x\in\hat{\xi}}\, \delta_{(x,S_{x}\xi)} $. Let  ${\Pp}^{\ast}
= \Gg_\ast \Pp$ be the distribution of the marked point process on
$ \RR^d \times \Nn_0$ with  mark space $\Nn_0$, namely $\Pp^*$ is
the image under $\Gg$ of the probability measure $\Pp$  on $\Nn$.
It is easy to show that $\Gg \circ S_x = S_{x}^\ast \circ \Gg$ for
$x\in\RR^d$ where  $S_x^\ast$ is the action on $\RR^d \times
\Nn_0$ of the translations  given by  $( y, \xi) \mapsto (y+x,
\xi)$.
 As a result,  ${\Pp}^{\ast}$ is also stationary.  Then, for any fixed
$A \in \Bb ( \Nn_0)$, the measure $\mu_A({B}) =  \int {\Pp}^{\ast} ( d
\xi^\ast )\,\xi^\ast ( {B} \times A)$ on $\RR^d$
is translation invariant and thus proportional to the Lebesgue measure.
This implies that, for any $N>0$ and any $A \in  \Bb(\Nn_0)$,

$$
\Cc_\Pp ( A)
\;:=\;
\int_{\Nn(\RR^d \times \Nn_0)}
{\Pp}^{\ast} ( d \xi^\ast ) \, \xi^\ast  ( C_1 \times A)
\; =\;
\frac{1}{N^d} \int_{\Nn(\RR^d \times \Nn_0)}
 {\Pp}^{\ast} ( d \xi^\ast ) \, \xi^\ast  ( C_N \times A)
\;.
$$

\noindent
The {\it Palm distribution} associated to $\Pp$ is the probability measure
$\Pp_0$ on $\Nn_0$ obtained from $\Cc_\Pp$ by normalization, namely,  $\Pp_0
= \rho^{-1}\Cc_\Pp$, where $\rho$ is defined in (\ref{gaetano}).  Thus, for any $N>0$,

\begin{equation}
\label{rino}
\Pp_0 ( A)
\;: =\;
\frac{1}{\rho} \frac{1}{N^d} \int_{\Nn}  {\Pp} ( d \xi) \int_{C_N} \hat{\xi}
(d x) \, \chi_A ( S_x \xi)
\;,
\end{equation}

\noindent
where $\chi_A$ is the characteristic function on the Borel set $A \subset
\Nn_0$.  One can show~\cite[Theorem 1.2.8]{FKAS} that for any  nonnegative
measurable function $f$ on $\RR^d \times \Nn_0$

\begin{equation}
\label{aida}
\int_{\RR^d}dx\int_{\Nn_0} \Pp_0(d\xi)\;f(x,\xi)
\;=\;
\frac{1}{\rho}\int_{\Nn}\Pp(d\xi)\int_{\RR^d}
\hat{\xi}(dx)\;f(x,S_x \xi)
\mbox{ , }
\end{equation}

\noindent which is used in~\cite{DVJ} as the definition of
$\Pp_0$. Similarly, there is a Palm distribution $\hat{\Pp}_0$ on
$\hat{\Nn}_0:= \{\hat{\xi} \in \hat{\Nn} : 0 \in \hat{\xi} \}$
associated to the distribution  $\hat{\Pp}$ of a SSPP on $\RR^d$.

\vspace{.2cm}

It is known that the Palm distribution of a stationary  PPP on
$\RR^d$ with distribution $\hat{\Pp}$ (Example~\ref{ex-PPP} above)
is the convolution $\hat{\Pp}_0  = \hat{\Pp} \ast
\delta_{\delta_0}$ of $\hat{\Pp}$ with the Dirac measure at
$\hat{\xi}=\delta_0$ ({\it i.e.} $\hat{\Pp}_0$ is simply obtained
by adding a point at the origin). The Palm distribution of a PPP
on $\RR^d\times [-1,1]$ with intensity measure $\rho\,dx\otimes
\nu$ is  the convolution  $\Pp_0  = \Pp \ast \zeta $ where $\zeta$
is the distribution of a marked point process obtained by
$\nu$--randomization of $ \delta_{\delta_0}$.  The Palm
distribution associated to the SSPP in Example~\ref{ex-Anderson}
is   $\hat{\Pp}_0 = \delta_{\sum_{x\in\ZZ^d}\d_x}$.  Its
$\nu$--randomization is the Palm distribution of the
$\nu$--randomization of Example~\ref{ex-Anderson}.

\vspace{.2cm}

We collect in the lemma below  a number of results on the Palm distribution
which will be needed in the sequel. Their proofs are given in
Appendix~\ref{app-prooflemma}.

%%%%%%%%%%%%%%%%%%%%%%%%%%%%%%%%%%%%%%%%%%%%%%%%
\begin{lemma}
\label{simmetria}
\noindent {\rm (i)}
Let  $k:\Nn_0\times\Nn_0 \to \RR$ be a measurable  function  such that
$\int \hat{\xi} ( dx) \,|k (\xi, S_x \xi )|$ and

$\int \hat{\xi} ( dx) \,|k ( S_x \xi, \xi )|$
are in $L^1(\Nn_0, \Pp_0)$.
Then
\begin{equation*}
\int \Pp_0(d\xi) \int \hat{\xi} ( dx) \;  k(\xi,S_x \xi)
\;=\;
\int \Pp_0(d\xi) \int \hat{\xi} ( dx) \; k(S_x \xi, \xi) \mbox{ . }
\end{equation*}

\vspace{.1cm}

\noindent {\rm (ii)}
Let $\Gamma \in \Bb (\Nn )$ be  such that
$S_x\Gamma=\Gamma$ for all $x\in\RR^d$. Then  $\Pp(\Gamma)=1$ if and only if
$\Pp_0(\Gamma_0)=1$

with $\Gamma_0 = \Gamma\cap\Nn_0$.

\vspace{.1cm}

\noindent {\rm (iii)}
Let  $\Pp$ be ergodic and  $A,B\in\Bb(\Nn_0 )$
be such that  $B\subset A$, $\Pp_0(A\setminus B)=0$ and $S_x \xi \in A$ for

any $\xi \in B$ and any $x \in \hat{\xi}$. Then
$\Pp_0 (A) \in \{ 0, 1 \}$.

\vspace{.1cm}
\noindent {\rm (iv)}
Let $A_j\in \Bb(\RR^d) $ for $j=1,\dots,n$.  Then
\begin{equation}\label{ursula}
\EE_{\Pp_0}
\left(\,\prod _{j=1}^n \hat{\xi}(A_j)\,\right)
\;\leq\;
\frac{c}{\rho}\, \EE_{\Pp}\bigl(\,\hat{\xi}(C_1)^{n+1} \,\bigr)
\;+\;
 \frac{c}{\rho}\,\sum_{j=1}^n \EE_{\Pp} \bigl(\,\hat{ \xi}(\tilde{A}_j)^{n+1}
 \, \bigr)
\;,
\end{equation}

where $\tilde{A}_j:=\cup_{x\in C_1} \bigl(A_j +x\bigr)$ and
$c$ is a  positive constant depending on $n$.

\end{lemma}
%%%%%%%%%%%%%%%%%%%%

%%%%%%%%%%%%%%%%%%%%
\begin{rem}
\label{silenzio}
{\rm  Here we point out a simple geometric property of point measures
$\xi$ within the set
\begin{equation}\label{franci}
\Ww\,:=\,\bigl\{\xi\in\Nn_0\,:\,
S_x\xi\not = \xi\;\forall x\in \RR^d\setminus\{0\}\,\bigr\}\,.
\end{equation}
which will be fundamental in order to apply  the methods
developed in \cite{KV} and \cite{DFGW}.
Let us  consider  a sequence  $\{x_n\}_{n\geq 0}$   of
 elements in  $\text{supp}(\hat{\xi})$ with $x_0=0$ and set
 $\xi_n:=S_{x_n}\xi$.  The $\xi_n$ can be thought of as the
 \emph{environment viewed  from the point $x_n$}.  Due  to the definition of
 $\Ww$, $\{x_n\}_{n\geq 0}$ can be recovered from
 $\{\xi_n\}_{n\in\NN}$ by means of the  identities
$$
x_{n+1}-x_n\;=\; \Delta(\xi_n, \xi_{n+1} ),\qquad n\in \NN\, ,
$$
where the function $\Delta:\Ww\times \Nn_0\rightarrow \RR^d$ is defined as
\begin{equation} \label{eq-delta}
\Delta(\xi',\xi'')
\;:=\;
\begin{cases}
x & \text{ if $\xi'' =S_x\xi'$}\, , \\
0 & \text{ otherwise }\,.
\end{cases}
\end{equation}
 Note that, by Lemma \ref{simmetria}(ii), condition
(\ref{melone}) is equivalent to $\Pp_0(\Ww)=1$. }
\end{rem}
%%%%%%%%%%%%%%%%%%%%

%%%%%%%%%%%%%%%%%%%%%%%%%%%%%%%%%%%%%%%%%%%%%%%%%%%%%%%%%%%%
\section{Variational formula}
\label{sec-viewed}
%%%%%%%%%%%%%%%%%%%%%%%%%%%%%%%%%%%%%%%%%%%%%%%%%%%%%%%%%%%%

The main object of this section is to show the following result,
implying Theorem~\ref{theo-Mott}(i).

%%%%%%%%%%%%%%%%%%%%%%%%%%
\begin{theo}
\label{theo-DFGW}
Let  $\Pp$ satisfy the assumptions of {\rm Theorem \ref{theo-Mott}(i)}.
 Then
the limit {\rm (\ref{eq-Diffmatrix})} exists and $D$ is given by
the variational formula

\begin{equation}
\label{eq-Dvarformula}
(a\cdot D \, a)
\;=\;
\inf_{f\in L^\infty(\Nn_0,\Pp_0)}
\;
\int\Pp_0(d\xi)\,\int \hat{\xi} (dx)\;
c_{0,x}(\xi)\;
\bigl(
a\cdot x\,+\,\nabla_x f(\xi)\,
\bigr)^2
\;, \qquad
a\in\RR^d\;,
\end{equation}

\noindent with

\begin{equation}
\label{eq-nabla}
\nabla_x f(\xi)
\;: = \;
f(S_x\xi)-f(\xi)
\mbox{ . }
\end{equation}

\noindent Moreover, the rescaled process
$\underline{Y}^{\xi,\varepsilon}: = (  \varepsilon
X_{t\varepsilon^{-2}}^\xi )_{t\geq 0}$ defined on $(\Omega_\xi,\PP^\xi_0)$
 converges weakly in $\Pp_0$--probability
as $\varepsilon \to 0 $ to a Brownian motion $\underline{W}_D$ with
covariance matrix $D$.
\end{theo}
%%%%%%%%%%%%%%%%%%%%%%%%%%

The proof is based on the theory of Ref.~\cite{KV} and~\cite{DFGW}
and, in particular, Theorem~2.2 of \cite{DFGW}.  Because of the
geometric disorder and the possibility of jumps between any of the
random points, the application of this general theorem to our
model is technically considerably more involved than in the case of the
lattice model with jumps to nearest neighboors studied
in~\cite[Section 4]{DFGW}. As a preamble, let us state a result
on the process $X_t^\xi$ proven in
Appendix~\ref{app-LPestimate} which will be used several times
below.

%%%%%%%%%%%%%%%%%%%%%%%%%%%%%%%%%%%%
\begin{prop}
\label{giulietta}
Let $\Pp$ satisfy $\rho_\kappa<\infty$ for some integer $\k>3$.
 Then,
given $t>0$ and  $0<\g< \kappa-3$,
$$
\EE_{\Pp_0} \EE_{\PP_0^\xi} \left( |X_t^\xi|^\g \right)
\; <\;
\infty \mbox{ . }
$$
\end{prop}
%%%%%%%%%%%%%%%%%%%%%%%%%%%%%%%%%%%%

%%%%%%%%%%%%%%%%%%%%%%%%%%%%%%%%%%%%
\begin{rem}
\label{virgilio} {\rm   From the variational formula of the
diffusion matrix $D$ given in Theorem \ref{theo-DFGW} one can
easily prove (see {\it e.g.}~\cite{DFGW}) that $D$  is a multiple
of the identity whenever  $\Pp$ is isotropic ({\it i.e.}, it is
invariant under  all rotations by $\pi/2$ in a coordinate plane).
In this case, the arguments leading to a lower bound on $D$ are
slightly simpler (and can be easily adapted to the general case).
Therefore, in order to simplify the discussion and without loss of
generality, in the last Sections \ref{sec-approx} and
\ref{sec-randomized} we will assume  $\Pp$ to be
isotropic.}
\end{rem}
%%%%%%%%%%%%%%%%%%%%%%%%%%%%%%%%%%%%

%%%%%%%%%%%%%%%%%%%%%%%%%%%%%%%%%%%%%%%%%%%%%%%%%%%%%%%%%%%
\subsection{The result of De Masi, Ferrari, Goldstein and Wick}
\label{subsec-result}
%%%%%%%%%%%%%%%%%%%%%%%%%%%%%%%%%%%%%%%%%%%%%%%%%%%%%%%%%%%

 A main idea in~\cite{DFGW} is to study  the process
$(S_{X_t^\xi} \xi )_{t \geq 0}$ with values in the space $\Nn_0$
of the environment configurations,
 instead of
the random walk $(X_t^\xi)_{t \geq 0}$. This process is called the process
{\it environment viewed from the particle}. It is defined on the probability
space
$(\Omega_\xi,\PP_0^\xi)$, with
$\Omega_\xi=D([0,\infty),\supp (\hat{\xi}))$.
Let
$\PP_\xi$ be its distribution
on the path space $\Xi:=D([0,\infty),\Nn_0)$
(endowed as usual with the Skorohod topology).
A generic
element of $\Xi$ will be denoted by $\underline \xi=(\xi_t)_{t\geq 0}$.
Let us set  $\PP :=\int\Pp_0(d\xi)\PP_\xi$. The {\it environment process}
is the process $(\xi_t)_{t\geq 0}$ defined on the probability space
$(\Xi,\PP)$ with distribution $\PP$.
 This is a continuous--time jump Markov process with initial measure $\Pp_0$
and transition probabilities

$$
\PP( \xi_{s+t}=\xi'\,|\,\xi_s=\xi)
\;=\;
\PP_\xi(\xi_t=\xi')
\;=:\;
p_t(\xi'|\xi)
\qquad\forall \;s,t\geq 0
$$

\noindent with, for any $\xi\in\Ww$,

\begin{equation}
\label{cecenia}
p_t(\xi'| \xi)\;
\;=\;
\begin{cases}
%\sum_{z\,:\, S_z\xi = \xi' }p^\xi _t (z|0)
\;p^\xi _t (x|0)
&  \;\;\;\text{if $\xi' = S_x \xi$ for some $x\in \hat{\xi}$}\;,
\\
0         &\;\;\; \text{otherwise .}
\end{cases}
\end{equation}

\noindent For any time $t\geq 0$, let us introduce the random variable
$X_t:\Xi\to\RR^d$ defined by
\begin{equation}
\label{eq-positions}  X_t(\uxi) \;:=\; \sum_{s\in [0,t]}
\Delta_s(\uxi)\;,
\end{equation}
where
\begin{equation*}
 \Delta_s(\uxi) \;:=\;
\begin{cases}
 x
& \text{if}\;\; \xi_s\;=\;S_x\xi_{s^-}
\\
0 & \text{otherwise}
\end{cases}
\end{equation*}
and  the sum runs over all jump times $s$ for which
$\Delta_s(\uxi)\neq 0$. Note that
 $\{ X_{[s,t]} := X_t - X_s : t > s \geq 0 \}$
defines an antisymmetric additive covariant family of random variables
as defined in~\cite{DFGW}, and $X_t$ has paths in $D([0,\infty),\RR^d)$.
The crucial link to the dynamics of a particle in a fixed environment is now
the following: due to Remark \ref{silenzio}, for any $\xi\in\Ww$,  the
distribution of  the process $(X_t)_{t\geq 0}$ defined  on
$(\Xi,\PP_\xi)$ is equal to the distribution  $\PP_0^\xi$   of the
random walk on $\supp (\hat{\xi})$ (naturally embedded in $\RR^d$)
starting at the origin.
Recalling that $\PP=\int \Pp_0(d\xi)\,\PP_\xi$, this implies
\begin{equation}
\label{eq-link}
\EE_{\Pp_0}\left(\EE_{{\PP}^\xi_0 } \left((X_t^\xi \cdot a)^2 \right)\right)
\;=\;
\EE_{\PP}\left((X_t \cdot a)^2 \right)\;,
\end{equation}
which gives a way to calculate the diffusion matrix $D$
from the distribution $\PP$ on $\Xi$.

\vspace{2mm}

In order to apply Theorem~2.2 of \cite{DFGW}, it is enough to
verify  the following hypothesis:

\vspace{1mm}

\noindent (a)~the environment process is reversible and
ergodic;
\\
(b)~the random variables  $X_{[s,t]}$, $0\leq s<t$  are
in $L^1(\Xi,\PP)$;\\
(c)~the mean forward
velocity exists:
\begin{equation}
\label{eq-meanvelocity}  \varphi(\xi) \;:=\;
L^2\!-\!\lim_{t\downarrow 0} \;\frac{1}{t}\;\EE_{\PP_\xi}(X_t) \;.
\end{equation}
(d)~the
martingale $X_t-\int^t_0ds\,\varphi(\xi_s)$ is in $L^2(\Xi,\PP)$.

\vspace{1mm}

Let us assume  $\rho_{12}<\infty$. Then,  statement (a)
will be proved in Proposition~\ref{prop-ergodicity},
Subsection~\ref{sec-construct}.
The statement (b) follows from
Proposition~\ref{giulietta}. The $L^2$-convergence in (c) will be
proved in Subsection~\ref{sec-mfv}
(Proposition~\ref{prop-integrabilities}), where we also show the
$L^2$-convergence in the following formula defining the mean
square displacement matrix $\psi(\xi)$:
\begin{equation}
\label{eq-meansquarespread}  (a\cdot\psi(\xi)a) \;:=\;
L^2\!-\!\lim_{t\downarrow 0}
\;\frac{1}{t}\;\EE_{\PP_\xi}\left((a\cdot X_t)^2\right)\;.
\end{equation}
 The last
point (d) is a consequence  of Proposition~\ref{giulietta}
assuring that $X_t\in L^2(\Xi,\PP)$ and the fact that
$\int^t_0ds\,\varphi(\xi_s) \in L^2(\Xi,\PP)$, which can be proved
by means of the Cauchy--Schwarz inequality, the stationarity of
$\PP$  following from (a), and the property $\varphi\in
L^2(\Nn_0,\Pp_0)$.

\vspace{2mm}

Once hypothesis (a)-(d) have been verified,
one can invoke \cite[Theorem 2.2 and Remark~4, p.~802]{DFGW} to conclude that
 limit (\ref{eq-Diffmatrix}) exists and that the  rescaled
 random walk  $\underline{Y}^{\xi,\varepsilon}$
 converges weakly in $\Pp_0$--probability to the Brownian
motion  $\underline{W}_D$ with covariance matrix $D$ given by (\ref{eq-Diffmatrix}), and
that $D$ is moreover given by
\begin{equation}
(a\cdot Da)
\; =\;
\EE_{\Pp_0} \bigl(\;(a\cdot \psi a)\;\bigr)
\;-\;
2\,\int^\infty_0dt\,
\left\langle
 \varphi\cdot a
  \,
 , \,e^{t\Ll}\, \varphi\cdot a
\right\rangle _{\Pp_0}\;,
\label{eq-Dform}
\end{equation}
where $\Ll$ is the generator of the environment process and
the integral on the r.h.s. is finite.
Formula (\ref{eq-Dvarformula}) can be deduced from the expressions
of $\Ll$, $\varphi$ and $\psi$ established in the following subsections
(Propositions~\ref{prop-gen} and~\ref{prop-integrabilities}) by using
a known general result  on self-adjoint operators stated in
(\ref{eq-int_sup}) below.

%%%%%%%%%%%%%%%%%%%%%%%%%%%%%%%%%%%%%%%%%%
\subsection{Preliminaries}
\label{sec-preliminaries}

Before starting to prove the above-mentioned  statements (a)-(d),
let us fix some notations and
recall some general facts about jump Markov processes.
In what follows, given a complete separable metric space $Z$
we denote by $\Ff (Z)$ the family of bounded Borel functions on $Z$
and, given a
(not necessarily finite) interval $I\subset\RR$,
we denote by
$D(I,Z)$  the space of right continuous paths $\underline{z}=(z_t)_{t\in I}$,
$z_t\in Z$, having left limits. The path space $D(I,Z)$  is endowed
with the Skorohod  topology \cite{Bil} which is the natural choice for the
study of jump Markov processes.
For a time $s\geq 0$,
the time translation $\t_s$ is defined as

$$\tau_s:D([0,\infty),Z)\to D([0,\infty),Z), \qquad
 (\tau_s\underline{z})_t \;:=\;z_{t+s}\,.
$$

\noindent Moreover, given $0\leq a<b$, we denote by $R_{[a,b]}$ the function
$$
R_{[a,b]}\,:\, D([0,\infty),Z)\rightarrow D([a,b],Z),
\qquad ( R_{[a,b]}\underline{z} )_t\; :=\;
\lim_{\delta \downarrow 0}\;z_{a+b-t-\delta}\,.
$$

\noindent  $R_{[a,b]} \underline{z}$ is the time--reflection of
$(z_t)_{t \in [a,b]}$ w.r.t. the middle point of $[a,b]$, and it
can naturally be extended to paths on $[0,a+b]$.

\vspace{.2cm}

A continuous--time Markov process with path in $D([0,\infty),Z)$
and distribution $\pp$ is called  \emph{stationary} if
$\EE_{\pp} (F)=\EE_{\pp}(F\circ \t_s)$ for all $s\geq 0$
and  for any  bounded Borel function $F$ on $ D([0,\infty),Z)$.
It is called \emph{reversible} if
$ \EE_{\pp} (F)=\EE_{\pp}(F\circ R_{[a,b]})$ for all
$b>a\geq 0$ and any  bounded Borel function $F$ on $ D([0,\infty),Z)$
such that $F(\underline z)$ depends only  on $(z_t)_{t\in [a,b]}$.
Thanks to  the Markov property, one can show that stationarity is
equivalent to

\begin{equation}
\label{bruna1}
\EE_{\pp}\bigl( f(z_0) \bigl)
\;=\;
\EE_{\pp}\bigl( f(z_s) \bigl)\; ,
\qquad
\forall \;s\geq 0\;,\;\; \forall \;f \in\Ff(Z),\,
\end{equation}

\noindent
% where $\Ff(Z)$ is the family of bounded Borel functions on $Z$
while reversibility is equivalent to

\begin{equation}
\label{bruna2}
\EE_{\pp}\bigl( f(z_0) g(z_s)
 \bigl)
\; =\;
\EE_{\pp}\bigl(g(z_0) f(z_s) \bigl)\; ,
\qquad
\forall\; s\geq 0\;,\;\; \forall\; f,g \in \Ff(Z)
\;.
\end{equation}

\noindent In particular, stationarity follows from reversibility. Finally,
the Markov process is called  \emph{(time) ergodic} if $\pp(A)\in\{0,1\}$
whenever $A\in \Bb\bigl(\, D([0,\infty),Z) \bigl)$ is time-shift invariant, i.e.
 $A=\t_s A$ for all $s\geq 0$.
Recall that if  the Markov  process   is stationary then it can be extended
to a Markov process with path space $D(\RR, Z)$  and the resulting
distribution is univocally determined (this follows from  Kolmogorov's
extension theorem and  the regularity of paths). Now stationarity,
reversibility and ergodicity of the  extended process are defined  as above
by means of $\t_s$, $s\in \RR$, and $R_{[a,b]}$, $-\infty<a<b<\infty$. Then
one can check that these properties are preserved  by extension (for what
concerns ergodicity, see in particular \cite[Chapter~15,
p. 96--97]{Ros}). Therefore our definitions coincide with those in
\cite{DFGW}.

\vspace{0,2cm}

All the above definitions  and remarks can be extended in a natural way to
discrete--time Markov processes  (with path space $Z^{\NN}$). Moreover, in
the discrete case, stationarity and reversibility are  equivalent
respectively to (\ref{bruna1}) and (\ref{bruna2}) with $s=1$.

\vspace{2mm}

We conclude this section   recalling  the standard construction
of the continuous--time random walk satisfying conditions (C1) and (C2)
in the Introduction.
  We first note that these  conditions  are meaningful
 for $\Pp_0$--almost all $\xi$
if  $\EE_{\Pp_0}(\lambda_0)<\infty$.  In fact, due to the bound
$\lambda_x(\xi)\leq e^{4\beta}\, e^{|x|}\,\lambda_0(\xi) $, one
can infer  from $\EE_{\Pp_0}(\lambda_0)<\infty$ that
$\lambda_x(\xi) <\infty$ for any $x\in\hat{\xi}$, $\Pp_0$ a.s.. We
note that the condition $\EE_{\Pp_0}(\lambda_0)<\infty$ is
equivalent to $\rho_2<\infty$ due to the following Lemma:

%%%%%%%%%%%%%%%%%%%%%%%%%%%%%%%%%%%%
\begin{lemma}
\label{prop-finite_moment}
For any positive integer $k$,
$\EE_{\Pp_0} ( \lambda_0^{k}) <\infty $ if and only if $\rho_{\,k+1}<\infty$.
\end{lemma}
%%%%%%%%%%%%%%%%%%%%%%%%%%%%%%%%%%%%

\pro Note that for suitable positive constants $c_1,c_2$ one has

$$
c_1\, \sum_{z\in \ZZ^d} \hat{\xi}(C_1+z) e^{-|z|} \;\leq\;
\lambda_0(\xi) \;\leq c_2\; \sum_{z\in \ZZ^d} \hat{\xi}({C_1}+z)
e^{-|z|}\;, \qquad \Pp_0\mbox{-a.s.}\;.
$$

\noindent Next let us expand  the $k$-th
power of these inequalities. By applying Lemma
\ref{simmetria}(iv) and using the stationarity of $\Pp$, one gets
that $\EE_{\Pp_0} ( \lambda_0^{k}) <\infty $  if
$\rho_{\,k+1}<\infty$. Suppose now that  $\EE_{\Pp_0} (
\lambda_0^{k}) <\infty $. Then the above expansion in $k$-th power
implies that $\EE_{\Pp_0} ( \hat{\xi}(C_1) ^k ) <\infty $. Since
due to (\ref{rino}),
$$
\EE_{\Pp_0}(\hat{\xi}(C_1)^k)
\;=\;
\frac{2^d}{\rho} \int _{\Nn} \Pp
(d\xi)\int_{C_{1/2}}\hat\xi (dx)\, \hat{\xi}(C_1+x) ^k
\;\geq\;
\frac{2^d}{\rho}\;\EE_{\Pp}(\hat{\xi}(C_{1/2})^{k+1})\,,
$$
one concludes that $\EE_{\Pp}(\hat{\xi}(C_{1/2})^{k+1})<\infty$,
which is equivalent to $\rho_{k+1}<\infty$.
 \finpro

\vspace{2mm}

 The construction of the continuous--time random walk
 follows standard references
({\sl e.g.} \cite[Chapter 15]{Br} and \cite[Chapter 12]{Kal}) and can be described
roughly as follows:
After arriving at site $y\in\hat{\xi}$, the particle waits an exponential
time with parameter $\lambda_y(\xi)$ and then jumps to  another site
$z\in\hat{\xi}$ with probability

\begin{equation}
\label{eq-p_xy}
p^\xi (z| y )
\;: =\;
\frac{c_{y,z}(\xi)}{\lambda_y (\xi)} \mbox{ . }
\end{equation}

\noindent
More precisely, consider $\xi\in \Nn_0$ such that $0<\lambda_z(\xi)<\infty$
for any $z\in \hat{\xi}$ and  set $\tilde{\Omega}_\xi
:=\bigl(\,\text{supp}(\hat\xi)\,\bigr)^{\NN}$.  A generic path in
$\tilde{\Omega}_\xi$ is denoted by $\bigl(\tilde{X}^\xi_n \bigr)_{n\geq
0}$.  Given $x\in\hat{\xi}$, let $\tilde{\PP}_x^{\xi}$ be the distribution
on $\tilde{\Omega}_\xi$  of a discrete--time random walk on
$\text{supp}(\hat \xi)$ starting in $x$ and having  transition
probabilities $p^\xi (z|y )$.  Let $\bigl(\Theta,\QQ\bigr)$ be  another
probability space  where the random variables $T^\xi_{n,z}$,
$z\in\hat{\xi}$, $n\in \NN$,  are independent and exponentially distributed
with parameter $\lambda_z(\xi)$,  namely   $ \QQ \bigl(\,T^\xi_{z,n} >
t\,)=\exp\bigl(-\lambda_z(\xi)t\bigr)$.  On the probability space
$(\tilde{\Omega}_\xi\times\Theta,
\tilde{\PP}_x^\xi\otimes \QQ)$ define the following functions:
\begin{align}
& R^\xi_0\;:=\;0\,; \;\;\; R^\xi_n:=T_{0,\tilde{X}_0^\xi}^\xi+
T_{1,\tilde{X}_1^\xi}^{\xi} +\cdots+ T^{\xi}_{n-1, \tilde{X}_{n-1}^\xi}
\;\;\;\text{ if }\;\;\; n \geq 1 \;.
\nonumber
\\
& n_\ast^\xi (t)\;:=\;n \;\;\text{ if }\;\; R_n^\xi\leq t < R_{n+1}^\xi\;,
\label{eq-randomtimechange}
\nonumber
\end{align}

\noindent Note that  $n_\ast^\xi (t)$  is well posed for any $t\geq 0$
only if $\lim_{n\uparrow\infty} R_n^\xi =\infty$. If
\begin{equation}\label{pedro}
\tilde{\PP}_x^\xi \otimes\QQ\Bigl(\,
 \lim_{n\uparrow \infty} R_n^\xi =\infty\,\Bigr)
\;=\;1\;,
\end{equation}
 then,  due to \cite[Theorem 15.37]{Br}, the random walk
$( \, \tilde{X}^\xi _{n_\ast^\xi (t) } \, )_{t\geq 0}$, defined
$\tilde{\PP}_x^\xi \otimes\QQ$--almost everywhere, is a jump Markov
process whose distribution  satisfies the infinitesimal conditions (C1)
and (C2).   The condition $\lim_{n\uparrow\infty}R_n^\xi=\infty$    assures
that no \emph{explosion phenomenon}  takes place, notably only finitely many
jumps can occur in finite time intervals. In Appendix  \ref{app-process} we
prove that (\ref{pedro}) is verified if $\rho_2<\infty$.

%%%%%%%%%%%%%%%%%%%%%%%%%%%%%%%%%%%%%%%%%%
\subsection{The environment viewed from the particle}
\label{sec-construct}

  The process environment viewed from the particle and the
environment process have been introduced in
Section~\ref{subsec-result}. Given $t>0$, we write $n_\ast(t)$ for
the function on the path space
$\Xi=D\left([0,\infty),\Nn_0\right)$ associating to each
$\underline{\xi}\in \Xi$ the corresponding number of jumps in the
time interval $[0,t]$. Motivated by further applications, it is
convenient to consider also the discrete--time versions of the
above processes. Consider the discrete-time Markov process
$\bigl(\,S_{\tilde{X}_n^\xi}\xi\,\bigr)_{n\geq 0}$ defined on
 $\bigl(\,\tilde{\Omega}_\xi,\tilde{\PP}_0^\xi\,\bigr)$,
call  $\tilde{\PP}_\xi$ its distribution  on the path space
$\tilde{\Xi}:=\Nn_0^{\NN}$  and denote a generic path in
$\tilde{\Xi}$  by $(\xi_n)_{n\geq 0}$. Such a Markov process  can
be thought of as the environment viewed from the particle
performing the discrete--time random walk with distribution
$\tilde{\PP}^\xi_0$. Let us point out a few properties of the
distribution $\tilde{\PP}_\xi$. First, we remark that due to the
covariant relations
\begin{equation}
\label{eq-covariance}
c_{z,y} (S_x \xi)
\;=\;
c_{z + x, y+ x} (\xi)
\;,
\qquad
\lambda_{y} (S_x \xi )
\;=\;
\lambda_{y+x} ( \xi)
\end{equation}
the
process $ \bigl(\lambda _{\tilde{X}^\xi_n}(\xi)\bigr)_{n\in\NN}$ defined on
$(\tilde{\Omega}_\xi, \tilde{\PP}^\xi_0)$ and the process
$ (\lambda _0(\xi_n))_{n\in\NN}$ defined
on $(\tilde{\Xi},\tilde{\PP}_\xi)$ have the same distribution. Moreover,
due to Remark \ref{silenzio}, if $\xi\in\Ww$, then  the process
$(\zeta_n)_{n\in\NN}$ defined on  $(\tilde{\Xi},\tilde{\PP}_\xi)$ by $\zeta_0=0$ and
$$
\zeta_n
\;=\;
\sum_{k=0}^{n-1} \Delta(\xi_k,\xi_{k+1})
\;, \forall \;n \geq 1\;,
$$
 where $\Delta(\xi,\xi')$ is given by (\ref{eq-delta}), has paths
in $\tilde{\Omega}_\xi$ with distribution $\tilde{\PP}_0^\xi$.
Finally, it is convenient to consider a suitable average of the
distributions $\tilde{\PP}_\xi$. To this aim, let $\Qq_0$ be the
probability measure on $\Nn_0$ defined as

$$
\Qq_0(d\xi)
\;: =\;
\frac{\lambda_0 (\xi) }{\EE_{\Pp_0} (\lambda _0 )}\;\Pp_0(d\xi)\;,
$$

\noindent and set  $\tilde{\PP}:=\int \Qq_0(d\xi) \tilde{\PP}_\xi $.
If $\xi \in \Ww$, the transition probabilities are
$$
p(\xi'| \xi)\; :=\;\tilde{\PP}
\bigl( \xi_{n+1}=\xi'\,|\xi_n=\xi\bigr)
\;=\;
\begin{cases}
 \lambda_0^{-1} (\xi) c_{0,x}(\xi)
&  \;\;\;\text{if $\xi' = S_x \xi$}\;,
\\
0         &\;\;\; \text{otherwise .}
\end{cases}
$$
 Note that,
 due to (\ref{eq-covariance}) and the symmetry of the jump
rates (\ref{eq-rates}),  $\lambda_0(\xi) p(\xi'|\xi)= \lambda_0(\xi') p(\xi|\xi')$.

%%%%%%%%%%%%%%%%%%%%%%%%%%
\begin{prop}
\label{prop-ergodicity}
Let  $\rho_2<\infty$.
Then the process $(\xi_t)_{t\geq 0}$ defined on $(\Xi,\PP)$  is
reversible, i.e.

\begin{equation}\label{napoleone}
\EE_{\PP}\bigl( f(\xi_0) g(\xi_t) \bigr)
\;=\;
\EE_{\PP}
\bigl( g(\xi_0) f(\xi_t) \bigr)
\qquad
\forall\;
f,g \in \Ff(\Nn_0)\;,\;\; \forall\; t>0\;,\;
\end{equation}

\noindent and is {\rm (}time{\rm )}  ergodic if  $\Pp$ is ergodic.
Similarly, the  discrete-time Markov process $(\xi_n)_{n\geq 0}$ defined on
$(\tilde{\Xi},\tilde{\PP} )$ is reversible  and is  {\rm (}time{\rm )} ergodic.
\end{prop}
%%%%%%%%%%%%%%%%%%%%%%%%%%

Having at our disposal Lemma \ref{simmetria}, the proof
follows modifying arguments of {\sl e.g.} \cite{DFGW}.

\vspace{0.2cm}

\pro We give the proof for the continuous--time process, the
discrete--time case being similar. We first verify the symmetric
property $p_t( \xi'|\xi)=p_t (\xi| \xi')$. Actually, thanks to the
construction of the dynamics given in Section
\ref{sec-preliminaries}, one can show that for any positive
integer $n$ and any $\xi=\xi^{(0)}, \xi^{(1)}, \dots ,
\xi^{(n-1)}, \xi^{(n)}= \xi' \in \Nn_0$,
$$
 \PP_\xi
  \bigl( \, n_\ast(t)=n,
   \xi_{R_1}=\xi^{(1)},  \ldots , \xi_{R_n} = \xi^{(n)}
  \bigr)
\;=\;
 \PP_{\xi ' }
  \bigl(  \, n_\ast(t)=n,
   \xi_{R_1}=\xi^{(n-1)} , \ldots ,  \xi_{R_n} = \xi^{(0)}
   \bigr) \mbox{ . }
$$

\noindent where, given $\underline{\xi}\in \Xi$,
 $R_1(\underline{\xi})<R_2(\underline{\xi}) <\dots$ denote the
jump times of the path $\underline{\xi}$.
Next, given $f,g \in \Ff(\Nn_0) $  one gets
by applying Lemma~\ref{simmetria}(i) and using
$p_t( \xi'|\xi)=p_t (\xi| \xi')$
that

\begin{equation} \label{eq-revers}
\int \Pp_0 ( d \xi) \int \hat{\xi} (d x) \, p_t ( S_x \xi| \xi) \;
f (\xi) g ( S_x \xi)
\; =\;
\int \Pp_0 ( d \xi)  \int \hat{\xi}(dx)\,  p_t (  S_x\xi| \xi)\; f (S_x \xi) g
( \xi) \;,
\end{equation}

\noindent  which is equivalent to (\ref{napoleone}). Hence $\PP$ is reversible.
Due to Corollary 5 in \cite[Chapter IV]{Ros}, in order to prove ergodicity
it is enough to show that  $\Pp_0(A)\in\{0,1\}$ if  $A\in \Bb(\Nn_0)$ has the following property: $\PP_\xi(\xi_t\in A)=\chi_A(\xi)$
 for $\Pp_0$--almost all $\xi$. Given such a set $A$, then there  exists a
 Borel subset $B\subset A$ such that
$\Pp_0(A\setminus B)=0$ and
$\PP_\xi(\xi_t\in A)=1$ for any $\xi\in B$. Fix $\xi\in B$ and $x\in \hat{\xi}$, then
$\PP_\xi ( \xi_t = S_x \xi, \xi_t \in A) =
\PP_\xi ( \xi_t = S_x \xi)>0
$ (the last bound follows from the positivity of the jump rates).
Hence $S_x\xi\in A$. Lemma~\ref{simmetria}(iii) implies
that $\Pp_0(A)\in\{0,1\}$,  thus allowing to  conclude the proof.
\finpro

\vspace{0.2cm}

Let  $\Pp$ fulfill the assumption of Proposition
\ref{prop-ergodicity}.
Then,

\begin{equation} \label{eq-Markov_sem}
\quad   (\Tt_t f)(\xi)\;:= \;\EE_{\PP_\xi} \bigl( f(\xi_t) \bigr)
\;=\; \int\hat{\xi}(dx)\; p_t(S_x\xi|\xi)\,f(S_x\xi) \;,\qquad
\Pp_0 \text{ a.s. }
\end{equation}

\noindent defines a strongly continuous contraction
semigroup on $L^2(\Nn_0, \Pp_0) $ (Markov semigroup).  Actually, (i) $\Tt_t:
L^2(\Nn_0,\Pp_0) \to L^2(\Nn_0,\Pp_0)$ is self-adjoint by (\ref{napoleone})
and is  a contraction
by  the Cauchy-Schwarz inequality and the stationarity of $\PP$; (ii)
$\Tt_{t+s} = \Tt_t \Tt_s$ follows from the Markov nature of the process;
(iii) the continuity follows from the following argument: first
observe that it is enough to prove the continuity of $\Tt_t f$ at $t=0$ for
$f\in L^\infty ( \Nn_0, \Pp_0)$, which is obtained from the dominated
convergence theorem and   the estimate  $| (\Tt_t f - f)(\xi)|
\leq 2 \| f \|_{\infty} ( 1 - p^\xi_t ( 0 | 0) )$.

\vspace{0,1cm}

Let us denote by $\Ll$ the generator of the Markov semigroup $( \Tt_t )_{t
\geq 0}$ and by $D(\Ll) \subset L^2(\Nn_0, \Pp_0)$ its domain.

%%%%%%%%%%%%%%%%%%%%%%%%%%
\begin{prop}
\label{prop-gen}
Let $\Pp$ satisfy $\rho_4<\infty$.
Then $\Ll$ is nonpositive and self--adjoint with core
$L^\infty(\Nn_0,\Pp_0)$.
For any $f\in L^\infty(\Nn_0,\Pp_0)$, one has
\begin{equation}
\label{eq-envirogen}
( \Ll f)(\xi)
\;=\;
\int \hat{\xi}(dx)\;
c_{0,x}(\xi)
\;\nabla_x f(\xi)\;,
\qquad \text{for $\Pp_0$-a.e. $\xi$}\;,
\end{equation}
where $\nabla_x f$ is defined in {\rm (\ref{eq-nabla})}, and, moreover,
\begin{equation}
\label{eq-Dirichlet}
{\langle f,  (-\Ll)  f\rangle }_{\Pp_0}
\;=\;
 \frac{1}{2}\;
\int \Pp_0 (d\xi) \int \hat{\xi} (dx ) \;c_{0,x}(\xi)
\;\left(\nabla_x f(\xi)\right)^2
\;.
\end{equation}
\end{prop}
%%%%%%%%%%%%%%%%%%%%%%%%%%

\vspace{.2cm}

\noindent \pro
The self-adjointness of $\Ll$ follows from \cite[Vol.2, Theorem X.1]{RS}.
Actually,
{\rm (i)} $\Ll$ is closed as a generator of a strongly continuous
semigroup~\cite[Vol.2, Chapter X.8]{RS};
{\rm (ii)}  $\Ll$ is symmetric because $\Tt_t$ is self-adjoint;
{\rm (iii)} the spectrum of $\Ll$ is included in $(-\infty,0]$
by contractivity of the semigroup. Note that {\rm (iii)} also implies
that $\Ll$ is non-positive.

\vspace{1mm}

We use the abbreviation $L^p$ for  $L^p=L^p(\Nn_0, \Pp_0)$, $p=2$ or
$\infty$.
For any $f\in L^\infty$, denote by $\Lambda f$ the function
defined by  the r.h.s. of (\ref{eq-envirogen}).
Due to  Lemma~\ref{prop-finite_moment}, $\EE_{\Pp_0}(\lambda _0^{2} )<\infty$
and in particular
$$
\int {\Pp_0}(d \xi)  \bigl|(\Lambda  f)(\xi) \bigr|^2
\;\leq\;
4 \,\|f\|_\infty^2\,
  \EE_{\Pp_0} \bigl( \lambda_0^2 \bigr)<\infty \;,
$$
thus implying that  $\Lambda:L^\infty\to L^2$ is
a well-defined operator.
We claim that
\begin{equation}\label{fame}
L^2-\lim_{t\downarrow 0} \frac{\Tt_t f -f }{t}
\;= \;
\Lambda f\;, \qquad \forall\;  f\in
L^\infty \mbox{ . }
\end{equation}
Note that (\ref{fame}) implies that $ L^\infty \subset D(\Ll)$ and  $\Ll
f=\Lambda f$ for all $f\in L^\infty$.
Since moreover $\Tt_t$ is a contraction and
$\Tt_t \,L^\infty \subset L^\infty$, it then follows
from~\cite[Vol.2, Theorem X.49]{RS} that $L_\infty$ is a core for
$\Ll$ and $\Ll$ is the closure of $\Lambda$.
Finally,
using (\ref{eq-revers}) in the limit $t \to 0$,  by straightforward
computations (\ref{eq-Dirichlet}) can be derived from  (\ref{eq-envirogen}).

\vspace{.2cm}

Let us now prove (\ref{fame}). We assume $\xi\in\Ww$ and we  set, for
$\xi'\neq\xi$,
\begin{equation*}
p_{t,1} ( \xi' | \xi)
\; : = \;
  \PP_\xi ( \xi_t = \xi' , n_\ast(t) = 1 )
   \; = \;
   p_t (\xi'| \xi ) - \PP_\xi ( \xi_t = \xi' , n_\ast(t) \geq 2)
\mbox{ . }
\end{equation*}
Thanks to the construction of the dynamics described in Section
\ref{sec-preliminaries} and due to
 the estimate $1 - e^{-u} \leq u$, $u \geq 0$,
one has for any $x \in \hat{\xi}$ and $x\neq 0$
\begin{equation} \label{eq-cyril}
p_{t,1} ( S_x \xi | \xi)
\; \leq \;
\tilde{\PP}^\xi_0 \otimes \QQ  ( \tilde{X}^\xi_1=x,\, T_{0,0}^\xi \leq t )
\;=\;
p^\xi (x| 0) ( 1 - e^{-\lambda_0(\xi) t } )
   \; \leq \; c_{0, x} (\xi)\,  t
\mbox{ . }
\end{equation}
Let $f \in L^\infty$. In view also of
%(\ref{def-kernel}),
(\ref{eq-Markov_sem}) and $\int \hat{\xi}(dx)\,p_t(S_x\xi | \xi)=1$,
\begin{eqnarray}
\Bigl| \!\!\!\! & \bigl( & \!\!\! \!  \!
\Tt_t f  -  f -t\, \Lambda f \bigr) (\xi) \Bigr|
\;=\;
  \biggl| \int  \hat{\xi}( dx)   \bigl( f(S_x \xi)-f(\xi) \bigr)
   \bigl(p_t( S_x \xi|\xi) - c_{0, x} (\xi ) \,t  \bigr) \biggr|
\nonumber
\\
%\label{eq-veronica}
& &
\nonumber
\\
&
\leq & \!
2\,\| f\|_\infty \,
\biggl(
   \int_{\{x\neq 0\}}
\hat{\xi}( dx)  \bigl( p_{t} ( S_x \xi |\xi) - p_{t,1} ( S_x \xi |\xi) \bigr)
 +
   \int_{\{x\neq 0\}}
\hat{\xi}( dx)   \bigl( - p_{t,1} ( S_x \xi |\xi) + c_{0 , x}( \xi ) \,t   \bigr)
\biggr) \mbox{ .}
\nonumber
\end{eqnarray}
The first integral in second line can be bounded by
$\PP_\xi (n_\ast(t) \geq 2)$.
The second integral equals
$$
-\,\PP_\xi (\, n_{\ast}(t) =1\,)+ \lambda _0 (\xi)\,t
\; = \;
   - 1 + e^{- \lambda_0 (\xi) t} + \lambda _0 (\xi)\,t + \PP_\xi ( n_\ast (t) \geq 2 )
\mbox{ . }
$$
By collecting the above estimates, we get
\begin{multline}
\label{domdom}
\frac{1}{t^2}\;
\EE_{\Pp_0} \bigl[\bigl( \Tt_t f-f -t \Lambda f \bigr)^2\bigr]
 \leq  \frac{32\, \| f\|_\infty^2 }{t^2}\,
   \EE_{\Pp_0}
    \Bigl( \PP^2 _\xi (  n_\ast(t) \geq 2) \Bigr)
      \,+\,\\
         \frac{8 \| f\|_\infty^2 }{t^2} \,\EE_{\Pp_0}
       \Bigl( \,\bigl(- 1 + e^{- \lambda_0 \,t } + \lambda _0 \,t\,\bigr)^2\,
 \Bigr)
 \mbox{ .}
\end{multline}
By using the estimate
$(e^{-u}-1  + u )^2 \leq u^{3}/2$ for $u \geq 0$
and the finiteness of $\EE_{\Pp_0} ( \lambda_0^{3} )$,
it is easy to check that the second term in the r.h.s. tends to zero as $t \to 0$.
In order to bound the first term, we observe that
$$
\PP_\xi \bigl( n_\ast (t) \geq  2 \bigr)
    \, \leq  \,
     \tilde{\PP}_0^\xi \otimes \QQ
        ( T_{0,0}^\xi \leq t,\,T_{1,\tilde{X}^\xi_1}^{\xi} \leq t )
        \,  = \,
        \bigl( 1 - e^{-\lambda_0(\xi) t} \bigr) \int \hat{\xi} (d x)\,
    p (S_x \xi|  \xi )
         \bigl( 1 - e^{-\lambda_0( S_x \xi)t} \bigr)  \mbox{ .}
$$
Due to the estimate  $ 1 - e^{-u} \leq u$, this implies the bound
\begin{equation} \label{eq-rodolf}
 \PP_\xi (  n_\ast(t) \geq 2)
\;\leq \;
t^2\,\lambda_0(\xi)
 \int \hat{\xi} ( d x) \, p( S_x \xi| \xi)  \lambda _0( S_x \xi)
\;=\;
  t^2\,\lambda _0 (\xi)\, \EE_{\tilde{\PP}_\xi} \bigl(\lambda
  _0(\xi_1)\bigr)
\; .
\end{equation}
Due also to the estimate $1 - e^{-u} \leq 1$, it is also true that
\begin{equation}
\label{eq-rodolf2}
 \PP_\xi (  n_\ast(t) \geq 2)
\;\leq \;
t\, \EE_{\tilde{\PP}_\xi} \bigl(\lambda _0(\xi_1)\bigr)  \;.
\end{equation}
By multiplying the last two inequalities,
and  using the stationarity
of $\tilde{\PP}$, one obtains
$$
\frac{1}{t^2} \EE_{\Pp_0}
 \bigl(
  \PP_\xi^2  (  n_\ast(t) \geq 2)
 \bigr)
   \leq   t \,  \EE_{\Pp_0}
   \bigl(
    \lambda _0 (\xi) \bigl[ \EE_{\tilde{\PP}_\xi} \bigl( \lambda _0 ( \xi_1 ) \bigr) \bigr]^2
    \bigr)
\leq  t \,  \EE_{\Pp_0}
   \bigl(
    \lambda _0 (\xi) \EE_{\tilde{\PP}_\xi} \bigl( \lambda _0^{2} ( \xi_1 ) \bigr)
    \bigr)
    \,=\,  t \, \EE_{\Pp_0} \bigl( \lambda_0^{3} \bigr)  \;,
$$
thus implying that the first term on the r.h.s. of (\ref{domdom}) goes to
$0$ as $t\to 0$.
\finpro

%%%%%%%%%%%%%%%%%%%%%%%%%%%%%%%%%%%%%%%%%%%%%%%%%%%%%%%%%%
\subsection{Mean forward velocity and infinitesimal square displacement}
\label{sec-mfv}

%%%%%%%%%%%%%%%%%%%%%%%%%%%%%%%%%%%%%%%%%%%%%%%%%%%%%%%%%%%%%%%%%%%%
\begin{prop}
\label{prop-integrabilities}  Let $\Pp$ satisfy $\rho_{12}<\infty$
and let $\varphi$ be the $\RR^d$-valued function  on $\Nn_0$ and
$\psi$ be the function on $\Nn_0$ with values in the real
symmetric $d\times d$ matrices, respectively defined by

\begin{equation}
\label{eq-meanvelocitydef}
\varphi(\xi)
\;=\;
\int \hat{\xi} (dx)
\;
c_{0,x}(\xi)
\;x
\;,
\qquad
(a\cdot \psi(\xi)a)
\;=\;
\int \hat{\xi} (dx)
\;
c_{0,x}(\xi )
\;(a\cdot x)^2
\;.
\end{equation}

\vspace{.1cm}

\noindent {\rm (i)} $\varphi(\xi)$ is in $L^2(\Nn_0,\Pp_0)$ and is equal to
the mean forward velocity
given by the convergent $L^2$-strong limit {\rm (\ref{eq-meanvelocity})}.

\vspace{.1cm}

\noindent
{\rm (ii)} $(a \cdot \psi(\xi) a)$ is in $L^2(\Nn_0,\Pp_0)$ and is equal to
the infinitesimal mean square displacement
given by the convergent $L^2$-strong limit {\rm (\ref{eq-meansquarespread})}.
\end{prop}
%%%%%%%%%%%%%%%%%%%%%%%%%%%%%%%%%%%%%%%%%%%%%%%%%%%%%%%%%%%%%%%

We point out that $\varphi(\xi)$  and $\psi(\xi)$ are well defined for
$\Pp_0$ almost all $\xi$ since    $\rho _2<\infty$ (see for example the proof of
Lemma \ref{prop-finite_moment}).

\vspace{0.1cm}

\pro
(i) One has
\begin{eqnarray}
\label{eq-claudia}
\nonumber
\frac{1}{t^2} \int {\Pp_0} ( d \xi)
\bigl| \EE_{\PP_\xi} (X_t ) - t \,\varphi (\xi) \bigr|^2
&  \leq &
  \frac{2}{t^2} \int {\Pp_0} ( d \xi)\,
   \bigl| \EE_{\PP_\xi}
    \bigl( X_t \,\chi \bigl( n_\ast (t) =1 \bigr) \bigr)
     - t \,\varphi (\xi)
   \bigr|^2
\\
& &
    \;+\;
     \frac{2}{t^2} \int {\Pp_0} ( d \xi) \,  \bigl| \EE_{\PP_\xi}
     \bigl( X_t\, \chi \bigl( n_\ast (t) \geq 2 \bigr)  \bigr)
   \bigr|^2
\mbox{ . }
\end{eqnarray}
We first show that the first term on the r.h.s. vanishes  as $t \to 0$.
Using the same notation as in the proof of
Proposition~\ref{prop-gen} and invoking (\ref{eq-cyril}),
\begin{equation*}
\EE_{\Pp_0}
 \Bigl(  \Bigl|
  \EE_{\PP_\xi} \bigl( X_t \, \chi \bigl( n_\ast (t) =1 \bigr) \bigr)
- t \, \varphi (\xi)
  \Bigr|^2
 \Bigr)
\; =\;
   \EE_{\Pp_0}
   \biggl(
    \biggl|
     \int _{\{x\not=0\}}\hat{\xi} ( dx) \bigl( p_{t,1} ( S_x \xi | \xi ) - t \, c_{0 ,x} (\xi ) \bigr) \, x
    \biggr|^2
  \biggr)
 \end{equation*}
 is bounded according to the Cauchy-Schwarz inequality by
\begin{equation}
\label{eq-inequ}   \EE_{\Pp_0} \biggl(
 \int_{ \{x\not=0\} } \hat{\xi} ( d x)
   \bigl( -  p_{t,1} ( S_x \xi | \xi ) + t\,  c_{0, x} (\xi ) \bigr)
\;  \int _{\{y\not = 0\}} \hat{\xi} ( dy) \bigl( -  p_{t,1} (
S_y \xi | \xi ) + t \, c_{0, y} (\xi ) \bigr) | y|^2 \biggr)
\mbox{ . }
\end{equation}
Let us denote by $I_1(\xi)$ and $I_2(\xi)$ the (non negative) integrals over
$\hat{\xi}(dx)$ and $\hat{\xi}(dy)$ respectively.
Using the identities of the proof of Proposition~\ref{prop-gen}, the
inequality  $0 \leq -1 + e^{-u} + u \leq u^2$, $u \geq 0$, and
(\ref{eq-rodolf}), we deduce
$$
I_1(\xi)\;=\;
-1+e^{-t\lambda_0(\xi)}+t\lambda_0(\xi)+\PP_\xi(n_\ast(t)\geq 2)
\;\leq\;
t^2\lambda_0(\xi)^2+t^2\lambda_0(\xi)\;\EE_{\tilde{\PP}_\xi}(\lambda_0(\xi_1))
\;.
$$
\noindent Moreover, $I_2(\xi)\leq t
\int \hat{\xi} ( dy)  \, c_{0, y} (\xi )  | y|^2$. Hence
(\ref{eq-inequ}) is bounded by
$$
t^3\,\left(
\EE_{\Pp_0}
\biggl(\lambda^2_0(\xi)
\int \hat{\xi}(dy)\,c_{0,y}(\xi)|y|^2\biggr)
+
\EE_{\Pp_0}
\biggl(
\lambda_0(\xi)\EE_{\tilde{\PP}_\xi}(\lambda_0(\xi_1))
\int \hat{\xi}(dy)\,c_{0,y}(\xi)|y|^2\biggr)
\right)
\;.
$$
 As long as $\rho_4<\infty$,
the first expression can be  bounded by applying Lemma~\ref{simmetria}(iv)
(see the  argument leading to Lemma~\ref{prop-finite_moment}).  A short
calculation shows that the second expression equals
$$
 \int \Pp_0(d\xi)\; \int\hat{\xi}(dx)\,c_{0,x}(\xi)\,
\int\hat{\xi}(dz)\,c_{x,z}(\xi)\,
\int\hat{\xi}(dy)\,c_{0,y}(\xi)\,|y|^2
$$
and is therefore bounded if  $\rho_4<\infty$ (again by means of Lemma
\ref{simmetria}(iv)). Resuming
the results obtained so far, one gets
\begin{equation} \label{eq-clarysse}
\frac{1}{t^2} \, \int {\Pp_0} ( d \xi) \;
 \bigl|
  \EE_{\PP_\xi} \bigl( X_t
\, \chi \bigl( n_\ast (t) =1 \bigr) \bigr)  - t \, \varphi (\xi)
  \bigr|^2
  \;=\; {\mathcal O} ( t )
\mbox{ . }
\end{equation}

We now turn to the second term in (\ref{eq-claudia}).
By Proposition~\ref{giulietta},
$\EE_{\Pp_0} ( \EE_{\PP_\xi} ( | X_t |^\gamma) )  < \infty$ as long as
$0<\gamma<\kappa-3$ whenever
$\rho_\kappa<\infty$ for $\kappa$ integer.
By applying twice the H{\"o}lder inequality, if $\g>2$,
\begin{equation*}
\EE_{\Pp_0}
 \Bigl(
  \bigl| \EE_{\PP_\xi} \bigl( X_t
\,\chi \bigl( n_\ast (t) \geq 2 \bigr)  \bigr) \bigr|^2
 \Bigr)
 \;\leq\;
   \Bigl(
    \EE_{\Pp_0} \Bigl( \EE_{\PP_\xi} \bigl( \bigl| X_t
\bigr|^\gamma  \bigr) \Bigr)
   \Bigr)^{\frac{2}{\gamma} }
   \Bigl(
     \EE_{\Pp_0}
      \Bigl(
       \PP_\xi \bigl( n_\ast (t ) \geq 2 \bigr)^{\frac{2 \gamma - 2}{\gamma - 2}}
      \Bigr)
    \Bigr)^{1-\frac{2}{\gamma} } \mbox{ . }
\end{equation*}
Let us take (\ref{eq-rodolf2}) to the power $\gamma/(\gamma - 2)$,
multiply the result
by (\ref{eq-rodolf}).
This yields
\begin{equation*}
\EE_{\Pp_0}
 \Bigl(
  \PP_\xi \bigl( n_\ast (t ) \geq 2 \bigr)^{\frac{2 \gamma - 2}{\gamma - 2}}
  \Bigr)
\;\leq \;
 t^{\frac{3 \gamma -4}{ \gamma - 2}} \,
  \EE_{\Pp_0}
   \Bigl(
    \lambda_0 (\xi) \,
     \EE_{\tilde{\PP}_\xi} \Bigl( \lambda_0^{\frac{2 \gamma - 2}{\gamma - 2}} (\xi_1 ) \Bigr)
   \Bigl)
     \;=\;
      t^{\frac{3 \gamma -4}{ \gamma - 2}}  \,\EE_{\Pp_0}  \Bigl(  \lambda_0^{\frac{3 \gamma -4}{\gamma - 2} } \Bigr)
      \mbox{ . }
\end{equation*}
Hence, by Lemma~\ref{prop-finite_moment},
if $\rho_\kappa <\infty$
is satisfied for integer
$\kappa> (4 \gamma - 6)/(\gamma -2)$ and $\gamma<\kappa-3$, there is a finite
constant $C>0$ such that
\begin{equation} \label{eq-madarine}
\EE_{\Pp_0}
 \Bigl(
  \bigl| \EE_{\PP_\xi} \bigl( X_t
\,\chi \bigl( n_\ast (t) \geq 2 \bigr)  \bigr) \bigr|^2
 \Bigr)
  \leq C \, t^{\frac{3 \gamma -4}{\gamma }}
\mbox{ . }
\end{equation}
One concludes the proof by
choosing $\gamma > 4$ and by combining (\ref{eq-claudia}),
(\ref{eq-clarysse}) and (\ref{eq-madarine}), as long as $\kappa>7$.

\vspace{.2cm}

\noindent (ii) One follows the same strategy.
The first term in the equation corresponding to
(\ref{eq-claudia}) can be dealt with in exactly the same way.
In the argument for the second term, $|X_t|$ is replaced by $|X_t|^2$ so
that one needs $2\gamma<\kappa-3$, hence $\kappa>11$.
\finpro

%%%%%%%%%%%%%%%%%%%%%%%%%%%%%%%%%%%%%%%%%%%%%%%%%%%%%%%%
\subsection{Proof of Theorem~\ref{theo-DFGW}} \label{sec-proofDFGW}
%%%%%%%%%%%%%%%%%%%%%%%%%%%%%%%%%%%%%%%%%%%%%%%%%%%%%%%

 Since all conditions (a)-(d)  of Subsection
\ref{subsec-result}
 have been checked in
the preceding subsections, as already pointed out, one can invoke
\cite[Theorem 2.2]{DFGW}  to conclude that the limit
(\ref{eq-Diffmatrix}) exists and that the rescaled
 random walk  $\underline{Y}^{\xi,\varepsilon}$
  converges weakly in $\Pp_0$--probability to the Brownian
motion  $\underline{W}_D$. We can now also
derive  the variational formula (\ref{eq-Dvarformula}) from
the general expression  (\ref{eq-Dform}). Let us first quote
some general results concerning self--adjoint
operators. Let $(\Omega,\mu)$ be a probability space and
 denote by  $\langle \,.\, ,\, .\, \rangle_{\mu}$ and by $\|.\|_{\mu}$ the
 scalar product and the norm on  $\Hh = L^2 ( \Omega,\mu)$.  Let  $\Ll
 :D(\Ll ) \to \Hh $ be a nonpositive self--adjoint operator  with (dense)
 domain $D(\Ll)\subset \Hh$ and assume $\Cc \subset D (\Ll) $ is a core of
 $\Ll$.  The space  $\Hh_{1}$ is the completion of $D( | \Ll|^{1/2} )
\cap ( {\rm Ker} (\Ll) )^\bot$ under the norm $\| f \|_{1} :
= \bigl\| |\Ll|^{1/2} f \bigr\|_{\mu}$ for $ f\in D( | \Ll|^{1/2}
)$, while the dual $\Hh_{-1}$ of $\Hh_1$ under $\langle \,.\, ,\,
.\, \rangle_{\mu}$ can  be identified with the completion of $D(|
\Ll|^{-1/2})= {\rm Ran}( | \Ll|^{1/2})$  under the $\| .
\|_{-1}$-norm defined as  $\| \varphi \|_{-1}  : = \bigl\|
|\Ll|^{-1/2 } \varphi \bigr\|_{\mu}$ for $ \varphi \in
 D( | \Ll|^{-1/2} )$.
Given  $\varphi \in \Hh\cap \Hh_{-1}$, the dual norm $\| \varphi \|_{-1}$ admits several useful characterizations:
\begin{equation}\label{vonnegut}
\| \varphi \|_{-1}^2
  \;  = \;
  \sup_{ f \in \Hh_{1}\cap \Hh } \frac{ | \langle \varphi , f \rangle_{\mu}|^2
}{ \| f \|_{1}^2 }
   \;= \;
\sup_{ f \in \Cc \cap ( {\rm Ker} (\Ll) )^\bot } \frac{ | \langle \varphi , f
\rangle_{\mu}|^2 }{ \| f \|_{1}^2 }   \;,
\end{equation}
where the last identity results from the fact that $\Cc$ is a core for $\Ll$.
Moreover, the identity
\begin{equation}\label{vonnegut2}
\|\varphi \|_{-1} ^2
\;  =\;
\sup _{ f \in \Cc }
 \Bigl( 2\, \langle \varphi , f\rangle_{\mu} -
\langle f, (-\Ll)  f\rangle _{\mu} \Bigr)
\end{equation}
is obtained  by using the nonlinearity in $f$ of the expression in
the r.h.s. of (\ref{vonnegut2}) and observing that
 $\varphi\in  ( {\rm Ker}( \Ll) )^\bot$.
 Finally, it follows from spectral calculus that
\begin{equation}
\label{vonnegut1}
 \| \varphi \|_{-1}^2
\;= \;
\int_0^\infty d t \, \langle \varphi  , e^{t\Ll }  \varphi \rangle_{\mu}
\mbox{ . }
\end{equation}
In what follows, we extend the definition of $\|\cdot\|_{-1}$ to the
whole space $\Hh$ by setting $ \| \varphi \|_{-1}:=\infty$ whenever
 $\varphi\in \Hh$ and $\varphi\not\in\Hh_{-1}$.
Thanks to  this choice,
identities (\ref{vonnegut}), (\ref{vonnegut2}) and  (\ref{vonnegut1})
are true for all $\varphi\in \Hh$.

\vspace{.2cm}

Invoking (\ref{vonnegut2}) and (\ref{vonnegut1}), one obtains
\begin{equation} \label{eq-int_sup}
\int^\infty_0dt \left\langle
 \varphi\cdot a
\, , \,e^{t\Ll}\, \varphi\cdot a
\right\rangle _{\Pp_0}
\;  =\;
\sup_{f\in L^\infty(\Nn_0,\Pp_0)}
\;
\Big(\, 2\,\langle  \varphi\cdot a\,,\, f \rangle_{\Pp_0}
\;-\;\langle f \, , \, (-\Ll) f\rangle_{\Pp_0}
\Big)
\;.
\end{equation}
Using (\ref{eq-Dirichlet}), (\ref{eq-meanvelocitydef}) and Lemma
\ref{simmetria}(i), a short calculation
starting from (\ref{eq-Dform})
yields
(\ref{eq-Dvarformula}).

%%%%%%%%%%%%%%%%%%%%%%%%%%%%%%%%%%%%%%%%%%
\section{Bound by cut-off on the transition rates}
\label{sec-bounds}

This section and the next ones are devoted to the proof of
Theorem \ref{theo-Mott}(ii).  In particular, we assume that $\hat\Pp$,
$\Pp$ and $\nu$ satisfy the conditions of Theorem \ref{theo-Mott}(ii)
although many partial results are
true under much weaker conditions.
The variational formula (\ref{eq-Dvarformula}) is particularly suited
in order to derive bounds on the diffusion matrix $D$.
For example, due to the monotonicity of the jump rates
$c_{x,y}(\xi)$ in the inverse temperature $\beta$, one deduces
that the diffusion matrix is non-increasing function of $\beta$.
The aim of this section is to obtain more quantitative bounds.

\vspace{0.2cm}

Given an energy  $0\leq  E_c \leq  1$,
we define the map $\Phi_{c}:\, \Nn\to
\hat\Nn := \Nn(\RR^d)$ as follows:

\begin{equation}\label{trieste}
\bigr( \Phi_{c}(\xi) \bigl) ( {A} )
\;:=\;
\xi ( A \times [-E_c, E_c ])\;, \qquad {A} \in \Bb( \RR^d)\;.
\end{equation}
Note that
$\hat{\Pp}^{c}:=  \Pp \circ   \Phi_{c}^{-1}$ is the
distribution of a point process on $\RR^d$ with finite intensity
$\rho_{c}:= \EE_{\hat{\Pp}^{c}} (\,\hat{\xi} (C_1)\,)
\leq \EE_\Pp (\,\hat{\xi} ( C_1 )\,) = \rho$,
and in general
\begin{equation}
\label{follia}
\EE_{\hat{\Pp}^{c}} (\,\hat{\xi} (C_1)^\kappa\,)
\;\leq\;  \rho_\kappa\,
\,,\qquad
\forall\; \k>0\,.
\end{equation}

\noindent In what follows, we assume that $\rho_{c}>0$.
It can readily be checked that $\hat{\Pp}^{c}$ is an ergodic SSPP on $\RR^d$.
We write $\hat{\Pp}^{c}_0$ for the Palm distribution associated to
$\hat{\Pp}^{c}$. Note that the distribution $\hat{\Pp}^{c}$ is obtained
from $\hat\Pp$ by  $\delta_c$--thinning with $\delta_c:=\nu([ -E_c,E_c])$.
Thus, $\rho_{c} = \d_c\, \rho$.
The relation between the Palm distributions
$\Pp_0$ and $\hat{\Pp}^{c}_0$ is described in the following lemma.

%%%%%%%%%%%%%%%%%%%%%%%%%%%%%%%%%%%%%%%%%%%%%%%%%%%%%
\begin{lemma}
\label{domenica_mbo}
For any Borel set $A\in \Bb(\hat{\Nn}_0)$ one has
$
\hat{\Pp}_0^{c}(A)=\rho\, \rho_{c}^{-1}
\;\Pp_0(\, |E_0|\leq E_c, \;
 \Phi_c(\xi)
\in A\,)\;.
$

\end{lemma}
%%%%%%%%%%%%%%%%%%%%%%%%%%%%%%%%%%%%%%%%%%%%%%%%%%%%%

\pro
The assertion is proven by comparing the two following identities obtained from
(\ref{rino}):
$$
\hat{\Pp}_0^{c}(A)\,=\,\frac{1}{\rho_{c}} \int_{\hat \Nn}
\hat{\Pp}^{c} (d\hat{\xi})\int _{C_1} \hat{\xi}(dx) \chi_A(S_x
\hat{\xi} )\;,
$$
\begin{eqnarray}
\Pp_0(\, |E_0|\leq E_c, \;
 \Phi_{c} (\xi)
\in A\,)
& = &
\frac{1}{\rho } \int_{\Nn}
\Pp (d \xi)\int _{C_1} \hat{\xi}(dx)\, \chi(|E_x|\leq E_c)
\,\chi_A(\Phi_{c} (S_x \xi))
\nonumber
\\
& = &
\nonumber
\frac{1}{\rho } \int_{\Nn}
\Pp (d \xi)\int _{C_1} \bigl(\Phi_{c} (\xi)\bigr)(dx)
\, \chi_A \bigl(S_{x} (\Phi_{c} ( \xi)) \bigr)\;.
\end{eqnarray}

\finpro

%%%%%%%%%%%%%%%%%%%%%%%%%%%%%%%%%%%%%%%%%%
\begin{prop}
\label{marcopantani}
Fix a distance  $r_c>0$ and an energy
$0\leq  E_c\leq 1$ and let $\hat{\Pp}^c_0$
be as above.
Moreover,  define
\begin{equation}
\label{sfruttamento}
\varphi _c (\hat{\xi})
\;:=\;
\int \hat{\xi} (dx)
\;
\hat{c}_{0,x}
\;x
\;,
\qquad
(a\cdot \psi_c (\hat{\xi})a)
\;:=\;
\int \hat{\xi} (dx)
\;
\hat{c}_{0,x}\;(a\cdot x)^2
\end{equation}
as functions on $\hat{\Nn}_0$.
where $\hat{c}_{0,x}:=\chi(|x|\leq r_c) $.
Then the diffusion matrix $D$ for the process $(X_t^\xi)_{t \geq 0}$
in {\rm Theorem \ref{theo-DFGW}}
admits the following lower bound
$$
D\;\geq \;  \frac{\rho_c}{\rho}\,
e^{- r_c-4\,\beta\,E_c}
\; D  _c(r_c, E_c)
\;,
$$
where
\begin{equation}
\label{eq-Dcformula}
(a\cdot D  _c(r_c,E_c) \, a)
\; := \;
\EE_{\hat{\Pp}_0^c} \Big(\,(a\cdot \psi_c a)\,\Big)
\;-\;
2\,\int^\infty_0dt\;
\left\langle
 \varphi _c\cdot a
\,  , \,e^{t \Ll_c}\, \varphi_c\cdot a
\right\rangle_{\hat{\Pp}^c_0}
\;,
\end{equation}
and $\Ll_c$ is the unique self--adjoint operator on
$L^2(\hat{\Nn}_0,\hat{\Pp}^c_0)$ such that
\begin{equation}
(\Ll_cf)(\hat{\xi})
\;=\;
\int \hat{\xi} (dx) \;\hat{c}_{0,x} \nabla_x f (\hat{\xi})\;, \qquad
\forall \;f \in L^\infty (\hat{\Nn}_0,\hat{\Pp}_0^c)\,.
\end{equation}
%Moreover, if $\Pp$ is the distribution of a {\rm PPP}, then
%\begin{equation}
%\label{eq-rescaling}
% D_c( r_c, E_c)
%\;= \;
%r_c^2 \,
%D_c( 1, E_c')
%\qquad
%\mbox{ where }\;\; \rho_c(E_c')\;=\;   r_c^d\, \rho_c(E_c)
%\;.
%\end{equation}
\end{prop}
%%%%%%%%%%%%%%%%%%%%%%%%%%%%%%%%%%%%%%%%%%

One can prove by the same arguments used in the proof of  Proposition
\ref{prop-gen} that $\Ll_c$ is well-defined and
self--adjoint.   Let $\hat{\PP}^c$ and $\hat{\PP}^c_{\hat{\xi}}$
be the  probability measures on
the path space  $\hat{\Xi}:=D(\,[0,\infty),\hat{\Nn}_0\,)$ associated to the
Markov process with generator  $ \Ll_c$  and initial distribution
$\hat{\Pp}_0^c$ and $\d_{\hat{\xi}}$, respectively, with $\hat{\xi}\in \hat{\Nn}_0$.
One can prove
that  these Markov processes are well--defined (in particular,
$\hat{\PP}^c_{\hat{\xi}}$ is well--defined for $\hat{\Pp}^c_0$--almost
all $\hat{\xi}$)
and exhibit a realization as jump processes by means of the same arguments
used in Section
\ref{sec-preliminaries}  (note that, for a suitable positive
constant $c$, $\int \hat{\xi}(dx) \hat{c}_{0,x}\leq c\, \lambda_0(\xi)$ for any
$\xi \in \Nn_0$, thus allowing to exclude explosion phenomena from the
results of Appendix~\ref{app-process}).  Finally, given
$\underline{\hat{\xi}}\in\hat{\Xi}$,  $X_t(\underline{\hat{\xi}})$ is
defined as in (\ref{eq-positions}).

\vspace{0.2cm}

\pro Note that
\begin{equation*}
\label{uffa}
c_{0,x}(\xi )
\;\geq\;
e^{- r_c-4\,\beta\,E_c}\;
\tilde{c}_{0,x}(\xi )\;,
\end{equation*}
where
$$
 \quad
\tilde{c}_{x,y}(\xi)
\;:=\;
\chi\bigl(
|E_x|\leq E_c, \;|E_y|\leq E_c,\;|x-y|\leq r_c  )
\quad , \quad x,y \in \hat{\xi}\;.
$$
Then (\ref{eq-Dvarformula}) implies that
$
(a\cdot Da)
\;\geq\;
 e^{- r_c-4\,\beta\,E_c}\;g( a)
$,
where
\begin{equation*}
g(a) \;:=\;
  \inf_{f\in L^\infty(\Nn_0,\Pp_0)}
     \int \Pp_0 (d \xi  )\,\int \hat{\xi} (dx )\;
       \tilde{c}_{0,x}(\xi)\;
      \Big( a\cdot x\,+\,\nabla_x f(\xi)\, \Big)^2\;\geq 0\;.
\end{equation*}
By the same arguments used in the proof of
Proposition  \ref{prop-gen} one can show that there is a unique self--adjoint
operator $\tilde{\Ll}$ on $L^2(\Nn_0,\Pp_0)$ such that
\begin{equation*}
\label{bambi}
(\tilde{\Ll}f)(\xi)
\;:=\;
\int \hat{\xi}(dx )\;
\tilde{c}_{0,x}(\xi)\; \nabla_x f(\xi)\;,
\qquad
\forall\;
 f\in L^\infty(\Nn_0,\Pp_0) .
\end{equation*}
Moreover, $L^\infty(\Nn_0,\Pp_0) $ is a core of $\tilde{\Ll}$ and
\begin{equation}
\label{giroditalia}
{\langle f , ( -\tilde{\Ll}) f\rangle }_{\Pp_0}
\;=\;
\frac{1}{2}\;
\int \Pp_0 (d\xi) \int \hat{\xi} (dx ) \;\tilde{c}_{0,x}(\xi)
\;\left(\nabla_x f(\xi)\right)^2\;,
\qquad \forall \;f\in L^\infty(\Nn_0,\Pp_0)
\;.
\end{equation}
Next let us introduce the functions
\begin{equation*}
\tilde{\varphi}(\xi)
\;=\;
\int \hat{\xi} (dx )
\;
\tilde{c}_{0,x}(\xi)
\; x\;
,\qquad
(a\cdot \tilde{\psi}(\xi)a)
\;=\;
\int \hat{\xi} (dx )
\;
\tilde{c}_{0,x}(\xi)
\;(a\cdot x)^2
\;.
\end{equation*}
Then we obtain by means of straightforward computations and
the identities  (\ref{vonnegut2}), (\ref{vonnegut1}) and (\ref{giroditalia})
  that
\begin{equation*}
\begin{split}
 g(a)
\;& =  \;
\EE_{\Pp_0}
  \Big(\,(a\cdot \tilde{\psi} a)\,\Big)-2\;
\sup_{f \in L^\infty(\Nn_0,\Pp_0)}\Bigl(
 2 \,\left \langle \tilde{\varphi}\cdot a, f\right
\rangle _{\Pp_0} -\langle f, (-\tilde{\Ll} )f \rangle
_{\Pp_0} \Bigr)\\
 &=\;
 \EE_{\Pp_0} \Big(\;(a\cdot \tilde{\psi} a)\;\Big)
\;-\;
2\,\int^\infty_0 dt\;
\left\langle
 \tilde{\varphi}\cdot a \, ,\,  e^{t\tilde{\Ll}} \, \tilde{\varphi}\cdot a
\right\rangle _{\Pp_0 }.
\end{split}
\end{equation*}
At this point, in order to get (\ref{eq-Dcformula}), it is enough to show that
$$
\EE_{\Pp_0} \Big(\;(a\cdot \tilde{\psi} a)\;\Big)
=\delta_c\,
 \EE_{\hat{\Pp}^c_0} \Big(\;(a\cdot \psi_c a)\;\Big)\;,
$$
and
$$
\left\langle
\tilde{ \varphi}\cdot a \,  , \,e^{t\tilde{\Ll}}\,\tilde{\varphi}\cdot a
\right\rangle _{\Pp_0}
 =\delta_c
\left\langle
 \varphi_c\cdot a \, , \,e^{t \Ll_c}\,\varphi_c\cdot a
\right\rangle _{\hat{\Pp}_0^c}\;.
$$
This can be derived from Lemma \ref{domenica_mbo} and the following
identities, where $\Phi_{c}$ is defined by (\ref{trieste}):
\begin{eqnarray*}
\tilde{\psi}  & = &
\chi(|E_0|\leq E_c)\,\psi_c \circ \Phi_{c}
\;, \\
 \tilde{\varphi}  & = &
\chi(|E_0|\leq E_c)\, \varphi _c \circ \Phi_{c}\;,\\
\tilde{\Ll}(f\circ\Phi_{c} )&  = & \chi(|E_0|\leq E_c) \, (\Ll_cf)\circ \Phi_{c}
\;.
\end{eqnarray*}
\finpro

%%%%%%%%%%%%%%%%%%%%%%%%%%%%%%%%%%%%%%%%%%
\section{Periodic approximants and resistor networks}
\label{sec-approx}

In this section, we compare $D_c(r_c,E_c) $ to the diffusion
coefficient of adequately defined periodic approximants, which
then in turn can be calculated as the conductance of a random
resistor network as in \cite{DFGW}. There have been numerous works
on periodic approximants; a recent one containing further
references is \cite{Owh}.

%%%%%%%%%%%%%%%%%%%%%%%%%%%%%%%%%%%%%%%%%%
\subsection{Random walk on a periodized medium}
\label{subsec-periodize}

Let us choose a given direction in $\RR^d$, say, the direction parallel
to the axis of the first coordinate.
Given a  fixed configuration $\hat{\xi} \in\hat{\Nn}$ and $N>r_c $,
we define the following subsets of $\RR^d$
\begin{align*}
&  Q_N^{\hat{\xi}}\;:= \;\text{supp}(\hat{\xi}) \cap
\check{C}_{2N} \;,\qquad \Gamma^{\pm} _N\;:=\; \ZZ^{d}  \cap \{ x:
x^{(1)} = \pm N, \; |x^{(j)}|<N \text{ for }j=2,\dots, d\}\,,
\\
& \overline{\Vv}^{ {\hat{\xi}}}_N
\;:=\;
Q_N^{ {\hat{\xi}}} \cup \Gamma_N^+\cup \Gamma _N^-
\;,
\qquad
B^{\hat{\xi}\pm}_N\;:=\;Q_N^{\hat{\xi}}\cap B^\pm_N\;,
\end{align*}
where   $\check{C}_{2N}:=(-N,N)^d$, $B^-_N:=\{x\in
\check{C}_{2N}\,:\,x^{(1)}\in(-N,-N+r_c]\}$ and $B^+_N:=\{x\in
\check{C}_{2N}\,:\,x^{(1)}\in[N-r_c,N)\}$.

\vspace*{0.2cm}

Next let us introduce a graph    $(\overline{\Vv}^{
{\hat{\xi}}}_N,\overline{\Ee}^{ {\hat{\xi}}}_N)$ with set of vertices
$\overline{\Vv}^{ {\hat{\xi}}}_N$ and set of edges  $\overline{\Ee}^{
{\hat{\xi}}}_N$. Two vertices  $x, y \in  Q_N^{ {\hat{\xi}}}$ are connected
by a non-oriented edge  $\{x,y\}\in \overline{\Ee}^{ {\hat{\xi}}}_N$ if and
only if $|x-y|\leq r_c$; moreover, all
vertices  $x \in  B_N^{ {\hat{\xi}}+}$ (respectively $x\in
B_N^{\hat{\xi}-}$)   are connected to all $y \in \Gamma_N^{+}$ (respectively
$y\in \Gamma_N^{-}$)  by an edge $\{x,y\}\in \overline{\Ee}^{
{\hat{\xi}}}_N$ and the points of $\Gamma^\pm_N$ are not connected between
themselves.

\vspace*{0.2cm}

We now define another graph
$( {\Vv} ^{ {\hat{\xi}}}_N, {\Ee}^{ {\hat{\xi}}}_N)$ obtained from
$(\overline{\Vv}^{ {\hat{\xi}}}_N,\overline{\Ee}^{ {\hat{\xi}}}_N)$
by identifying the vertices
$$
 x_{-} \;=\;(-N,x^{(2)},\dots, x^{(d)} )
\qquad \text{and} \qquad x_{+}\;=\; (N, x^{(2)},\dots, x^{(d)} )
\;.
$$
Let us write $\pi:\, \overline{\Vv}_N^{\hat{\xi}}\to
{\Vv}_N^{\hat{\xi}}$ for the identification map on the sets of
vertices. Hence $\pi ( \Gamma_N^{-}) = \pi ( \Gamma_N^{+} )$ and
$\pi$ restricted to $Q^{\hat{\xi}}_N$ is the identity map. The set
${\Vv}^{\hat{\xi}}_N = \pi ( \overline{\Vv}_N^{\hat{\xi}})$
represents the medium periodized along the first coordinate.  A
vertex $y \in \pi(\Gamma_N^-)$ is connected to all vertices $x \in
B_N^{\hat{\xi}+} \cup B_N^{\hat{\xi}-}$ by an edge of
${\Ee}^{{\hat{\xi}}}_N$. \vspace{.2cm}

Now a continuous--time random walk with state space
$\Vv _N^{\hat{\xi}}$ and infinitesimal generator $\Ll_N^{\hat{\xi}}$ is
given by
$$
\bigl( \Ll_N^{\hat{\xi}} f \bigr) (x)
\;=\;
\sum _{ y\in {\Vv}^{\hat{\xi}}_N \,:\,
\{x,y\}\in  {\Ee}^{\hat{\xi}}_N } c(\{x,y\})
\bigl( \,f(y)-f(x)\,\bigr)
\; ,
\qquad \forall\;
x\in  {\Vv}^{\hat{\xi}}_N
\;,
$$
where the bond-dependent transition rates $c(\{x,y\})$
 are defined for any $\{ x, y\} \in  {\Ee}^{\hat{\xi}}_N$ by
\begin{equation}\label{impero}
c(\{x,y\})
\;=\;
\begin{cases}
1 & \;\;\text{ if }  x,y\in Q_N^{\hat{\xi}}\;,\\
\frac{1}{|\G_N^{-} |} &  \;\;\text{ if } x \in \pi ( \Gamma_N^{-} )
\mbox{ or } y\in \pi ( \Gamma_N^{-} )\; .
\end{cases}
\end{equation}
Clearly the generator $\Ll_N^{\hat{\xi}}$ is symmetric w.r.t.
 the uniform distribution $m_N^{\hat{\xi}}$ on
${\Vv}_N^{\hat{\xi}} $ given by
$$
m_N^{\hat{\xi}}
\;=\;
\frac{1}{\bigl| {\Vv}^{\hat{\xi}}_N\bigr|}\;
\sum_{x\in {\Vv}^{\hat{\xi}}_N}
\;\delta_x
\;.
$$
Hence the Markov process with generator $\Ll_N^{\hat{\xi}}$ and
initial distribution $m_N^{\hat{\xi}}$ is reversible.
Note that it is not ergodic, however,
if there are more than one cluster (equivalence class of edges).
In the latter case, the ergodic measures are the
uniform distributions on a given cluster and this is sufficient for
the present purposes.

\vspace*{0.2cm}

We write ${\PP}^{ {\hat{\xi}}} _N$ (respectively ${\PP}^{ {\hat{\xi}}}_{N,x}$)
  for the probability on the path space $\Omega^{ {\hat{\xi}}}_N=
D\bigl([0,\infty), {\Vv} _N^{ {\hat{\xi}}}\bigr)$  associated
to the random walk
with initial distribution $m_N^{ {\hat{\xi}}} $ (respectively $\delta_x$)
and generator $\Ll_N^{ {\hat{\xi}}}$.

\vspace{.2cm}

Let us introduce an antisymmetric function
$d_1(x,y)$ on ${\Vv}_N^{ {\hat{\xi}}}$ such that
$$
d_1(x,y)\;= \;
\begin{cases}
y^{(1)} -x^{(1)}   &\text{ if } x,y\in Q_N^{ {\hat{\xi}}}\;,\\
y^{(1)} +N & \text{ if } y\in Q_N^{ {\hat{\xi}}},\; y^{(1)} <0,\; x\in \pi( \G_N^-)
\;,\\
y^{(1)} -N & \text{ if } y\in Q_N^{ {\hat{\xi}}} , \;y^{(1)} >0 ,\; x\in \pi(\G_N^-) \;.
\end{cases}
$$
Finally, given $t\geq 0$, we define the random variable
$$
X^{(1) { {\hat{\xi}}} }_{N,t}(\underline{\omega})
\;=\;
\sum_{s\in [0,t]\,:\,\omega_s\neq \omega_{s-}}
d_1(\omega_{s-},\omega_{s})\;,
$$
where $(\omega_s)_{s\geq 0} \in \Omega^{ {\hat{\xi}}}_N$.
It is the sum of position  increments along the first coordinate axis
for  all jumps occurring in the time interval $[0,t]$.
Clearly, $X^{(1) { {\hat{\xi}}} }_{N,t}$ gives rise to a time-covariant and
antisymmetric family so that, as in Section~\ref{sec-viewed},
\cite[Theorem 2.2]{DFGW} can be used in order to deduce the following
result.

%%%%%%%%%%%%%%%%%%%%%%%%%%
\begin{prop}
Given $N\in\NN$,  $N>r_c$,  and $\hat{\xi}\in \hat{\Nn}$
$$
\lim_{t\uparrow\infty}
\;\frac{1}{t} \;\EE _{\PP^{ {\hat{\xi}}}_N}\bigl(
(X^{(1){ {\hat{\xi}}} }_{N,t})^2\bigr)
\;=\;
D_N^{ {\hat{\xi}}}
\;,
$$
\noindent where the diffusion coefficient
  $D_N^{ {\hat{\xi}}}$ is finite and given by

\begin{equation}
\label{eq-periodicdiff}
D_N^{ {\hat{\xi}}}
\;=\;
m_N^{\hat{\xi}}
\bigl(\psi_N^{ {\hat{\xi}}}\bigr)
\;-\;
2\,
\int_0^\infty dt\;
\langle \varphi_N^{\hat{\xi}}
 \,, \, e^{t\Ll_N^{\hat{\xi}} }\, \varphi_N^{ {\hat{\xi}}}
\rangle _{m_N^{ {\hat{\xi}}} }
\;,
\end{equation}

\noindent with $\psi_{N}^{ {\hat{\xi}}}$, $\varphi_{N}^{ {\hat{\xi}}}$  (scalar)
functions on
${\Vv} _N^{ {\hat{\xi}}}$  defined as
\begin{equation}
\label{eq-periodicdef}
\psi_{N}^{ {\hat{\xi}}}(x)
\;=\;
\sum_{y\,:\,\{y,x\}\in \Ee_N^{ {\hat{\xi}}}   } c(\{x,y\})\;d_1(x,y)^2
\;,
\qquad
\varphi_{N}^{ {\hat{\xi}}} (x)
\;=\;
\sum_{y\,:\,\{y,x\}\in \Ee_N^{ {\hat{\xi}}}    } c(\{x,y\})\;d_1(x,y)
\;.
\end{equation}
\end{prop}
%%%%%%%%%%%%%%%%%%%%%%%%%%

\vspace{.2cm}

%%%%%%%%%%%%%%%%%%%%%%%%%%%%%%%%%%%%%%%%%%
\subsection{Link to periodized medium}
\label{subsec-linkperiodize}

Here we show that the diffusion matrix  (\ref{eq-Dcformula}) can
be bounded from  below in terms of   the average of the
diffusion coefficient associated to the  periodized random media.
Our proof follows the arguments of \cite[Prop. 4.13]{DFGW}, but
additional technical problems are related to the randomness of
geometry (absence of any lattice structure) and possible (albeit
integrable) singularities of the mean forward velocity and
infinitesimal mean square displacement.

%%%%%%%%%%%%%%%%%%%%%%%%%%
\begin{prop}
\label{odissea}
Suppose that for $1\leq p \leq 8$
\begin{equation}
\label{eq-Lpassump}
\lim_{N\uparrow \infty}
\frac{ \rho _c \,\ell(C_{2N})}{{\hat{\xi}}(C_{2N})+a_{2N}}
\;=\;1
\qquad \text{ in } \;\; L^p(\,\hat{\Nn}, \hat{\Pp}^c )
\;,
\end{equation}
where $\rho_c:=\EE_{\hat{\Pp}^c} (\hat{\xi}(C_1)) $ and
$a_{2N}:=|\G_N^{\pm}|=(2N-1)^{d-1}$. Then, for any $t>0$,
\begin{align}
&
\lim _{N\uparrow\infty}
\EE_{\hat{\Pp}^c}\Bigl( m_N^{\hat{\xi}} \bigl(\psi_N^{\hat{\xi}}\bigr)
\Bigr) \;=\;
\EE_{\hat{\Pp}^c_0} \bigl( \psi_c^{(11)} \bigr)   \;,
\label{forza1}
\\
& \lim _{N\uparrow \infty} \EE_{\hat{\Pp}^c }
\left(
\langle \varphi_N^{\hat{\xi}}\,,\, e^{t\Ll_N^{\hat{\xi}}}\, \varphi_N^{\hat{\xi}}
\rangle _{m_N^{\hat{\xi}}}
\right)
\;=\;
\left\langle  \varphi_c^{(1)}
\,, \,e^{t\Ll_c } \,\varphi_c^{(1)}  \right\rangle_{\hat{\Pp}^c_0}
\label{forza2}\;,
\end{align}

\noindent where $\psi_c^{(11)}$ and
$\varphi_c^{(1)}$ are the first diagonal matrix element  of
the matrix $\psi_c$ and
the first  component of  the vector $\varphi_c$
 introduced in {\rm (\ref{sfruttamento})}, respectively.
\end{prop}
%%%%%%%%%%%%%%%%%%%%%%%%%%

Since  $D_c(r_c,E_c)$ is given by (\ref{eq-Dcformula}) and is a
multiple of the identity ({\it cf.} Remark~\ref{virgilio}), the
identities (\ref{forza1}) and (\ref{forza2}) combined with Fatou's
Lemma immediately imply:

%%%%%%%%%%%%%%%%%%%%%%%%%%
\begin{coro}
Under the same hypothesis as above,

\begin{equation}
\label{kitano}
D_c(r_c ,E_c )
\;\geq\;
\left(
\limsup_{N\uparrow\infty}
\EE_{\hat{\Pp}^c} \bigl(D_N^{\hat{\xi}}\bigr)
\right)\,
{\bf 1}_d
\;,
\end{equation}
\noindent where  ${\bf 1}_d$ is the $d\times d$ identity matrix.
\end{coro}
%%%%%%%%%%%%%%%%%%%%%%%%%%

\vspace{.2cm}

Before giving the proof, let us  comment on
its assumptions. In Section \ref{sec-randomized}  we will show that
condition  (\ref{eq-Lpassump}) is always satisfied.
Due to  (\ref{follia}),
 $\rho_p <\infty $ implies
$\EE_{\hat{\Pp}^c}(\,{\hat{\xi}} (C_1)^p\,)<\infty$ for any $p >0$.
As $\hat{\Pp}^c$ is ergodic, this implies the following ergodic theorem, an
extension of \cite[Theorem 10.2]{DVJ}.
We recall that a convex averaging sequence of sets
 $\{A_n\}$ in $\RR^d$  is a sequence of convex sets   such that
$A_n\subset A_{n+1}$ and $A_n$ contains a ball of radius $r_n$ with $r_n\to \infty$ as $n\to \infty$.

%%%%%%%%%%%%%%%%%%%%%%%%%%
\begin{lemma}
\label{eternauta}
Suppose that  $\rho_p<\infty$,
$p \geq 1$. Then,
given a convex averaging sequence of Borel sets $\{A_n\}$ in $\RR^d$,
$$
\frac{{\hat{\xi}}(A_n)}{\rho_c\, \ell(A_n)}\;\to\; 1
\;\;\;
\text{ in }\;\; L^p (\,\hat{\Nn}, \hat{\Pp}^c )
\;,
\qquad\text{ and }\qquad
\frac{{\hat{\xi}} (A_n)}{\rho_c\, \ell(A_n)}\;\to\; 1
 \qquad   \hat{\Pp}^c \text{-a.s.}\;.
$$
\end{lemma}
%%%%%%%%%%%%%%%%%%%%%%%%%%

We will also need a bound on
$\EE_{\hat{\Pp}^c}(({\hat{\xi}} (A_n)/\ell(A_n))^p)$, uniformly in $n$,  for a sequence of sets
that does not satisfy the assumptions
of Lemma~\ref{eternauta}. To this aim we note that, given a Borel set
$B\subset\RR^d$ which is a union of
$k$ non-overlapping cubes of side $1$, one has

\begin{equation}
 \label{sabato} \EE_{\hat{\Pp}^c}\bigl(\,\bigl( \hat{\xi}
(B)/k\bigr) ^p\,\bigr) \;\leq\;
\EE_{\hat{\Pp}^c}\bigl(\,{\hat{\xi}} ( C_1)^p \bigr) \;\leq\;
\rho_p\,,\qquad \forall \;p\geq 1\,.
\end{equation}
%where $\overline{C}_1$ is the closed cube $[-\frac{1}{2},\frac{1}{2}]^d$.
 This follows from  the stationarity of $\hat{\Pp}^c$ and
the convexity of the function $f(x)=x^p$, $x\geq 0$.

\vspace{.2cm}

\noindent {\bf Proof of Proposition \ref{odissea}}. Without loss
of generality, we assume $r_c=1$. Note that, since $\hat\Pp$
is stationary with  finite intensity $\rho_1$, one has $\Pp$-a.s.
$\hat\xi \left( \partial C_k\right)=0$ for all $ k\in\NN$.
 In what follows we hence may assume $\xi$ to be as such,
thus allowing to simplify notation since
 $C_{2N}\cap\text{supp}(\hat\xi)
 =\check{C}_{2N}\cap\text{supp}(\hat\xi)$.
 A key observation in order
to prove (\ref{forza1}) and (\ref{forza2}) is the following
identity, valid for any nonnegative measurable function $h$
defined on $\hat{\Nn}_0$. It follows easily from (\ref{aida}):
\begin{equation}
\label{chiara}
\EE_{\hat{\Pp}^c}
\left( \int_B {\hat{\xi}} (dx) h(S_x{\hat{\xi}}) \right)
\;=\;
\rho_c\,\ell(B) \, \EE_{\hat{\Pp}^c_0 }(h),
\qquad \forall \;\;B\in \Bb(\RR^d).
\end{equation}
From this identity we can deduce  for any $h\in L^2(\Nn_0,\hat{\Pp}^c_0)$ that
\begin{equation}
\label{cile}
\lim_{N\uparrow\infty} \EE_{\hat{\Pp}^c}
\left(\frac{1}{ {\hat{\xi}}(C_{2N})+a_{2N}}
\int _{C_{2 N-2} } {\hat{\xi}}(dx) h(S_x {\hat{\xi}} ) \right)
\;=\;\EE_{\hat{\Pp}_0^c}(h)\;.
\end{equation}
In fact, due to (\ref{chiara}), it is enough to show that
\begin{equation}
\label{romania1}
\EE_{\hat{\Pp}^c} \left( \,\Bigl(
   \frac{1}{{\hat{\xi}} (C_{2N}) +a_{2N} }-\frac{1}{\rho_c \ell
   (C_{2N-2})}\Bigr)\,
 \int _{C_{2 N-2} } {\hat{\xi}} (dx) h (S_x{\hat{\xi}})
\right)\;\downarrow \;0\;,\qquad
 \text{ as } \;N\;\uparrow \;\infty\;.
\end{equation}
By applying twice the Cauchy-Schwarz inequality and by invoking
 (\ref{chiara}), we obtain
\begin{eqnarray*}
& & \Bigl( \text{l.h.s. of } (\ref{romania1}) \Bigr)^2
\\
& & \;\;\;
 \leq\; \EE_{\hat{\Pp}^c} \left( \,\Bigl( \frac{ \rho_c \,\ell(C_{2N-2}) }{
{\hat{\xi}} (C_{2N}) +a_{2N}}- 1 \Bigr)^2\frac{{\hat{\xi}} (C_{2N-2} ) }{
\rho^2_c\,  \ell(C_{2N-2}) ^2 }\right)
\EE_{\hat{\Pp}^c}\left(\frac{1}{\hat{\xi}(C_{2N-2})}  \Bigl(
  \int _{C_{2 N-2} } {\hat{\xi}} (dx) h (S_x{\hat{\xi}})
  \Bigr)^2\right)
\\
& & \;\;\;\leq\;
\EE_{\hat{\Pp}^c} \left( \,\Bigl( \frac{ \rho_c \,\ell(C_{2N-2}) }{
{\hat{\xi}} (C_{2N}) +a_{2N}}- 1 \Bigr)^2\frac{{\hat{\xi}} (C_{2N-2} ) }{ \rho_c\,
 \ell(C_{2N-2})}  \right)
\EE_{\hat{\Pp}^c_0 }(h^2 )
\;.
\end{eqnarray*}
At this point, (\ref{romania1})
follows by applying the   Cauchy-Schwarz inequality to the first
expectation above and then applying (\ref{sabato})  and the limit
(\ref{eq-Lpassump}) for $p=4$.

Let now $h_N^{\hat{\xi}} $ be a function on ${\Vv}_N^{\hat{\xi}}$ such that
for some constant $c>0$ independent of $N$
$$
  \quad |h_N^{\hat{\xi}} (x)| \; \leq\; c\;\begin{cases}
{\hat{\xi}} (\overline{ B_1(x)} ) & \text{ if } x\in Q_N^{\hat{\xi}},\\
\frac{|B_N^{\hat{\xi}}|}{a_{2N}} & \text{ otherwise },
\end{cases}
$$
where $B_N^{\hat{\xi}}=B_N^{{\hat{\xi}} -}\cup B_N^{{\hat{\xi}}
+}$ and $\overline{B_1(x)}$ is the closed unit  ball centered in
$x$. Note that $\psi_N^{\hat{\xi}}$ and $\varphi_N^{\hat{\xi}}$
satisfy this inequality. We claim that the mean boundary
contribution vanishes in the limit:
\begin{equation}\label{iacopo27}
\lim _{N\uparrow \infty} \EE_{\hat{\Pp}^c} \Bigl(
\frac{1}{{\hat{\xi}}(C_{2N})+a_{2N}} \sum _{x\in \Vv_N^{\hat{\xi}}\setminus
Q_{N-1}^{\hat{\xi}}} |h_N^{\hat{\xi}}(x) |^p
\Bigr) \;=\;
0\;,
\qquad  \text{ for } 1\leq p\leq 4.
\end{equation}
In fact, the sum in (\ref{iacopo27}) can be bounded by
\begin{equation}
\label{iacopo27p}   c^p\,
a_{2N}\frac{|B_N^{\hat{\xi}}|^p}{a_{2N}^p} \;+\;c^p\, \sum _{x\in
Q_N^{\hat{\xi}}\setminus Q_{N-1}^{\hat{\xi}}} \left( {\hat{\xi}} (
\overline{B_1(x)}) \right)^p \;.
\end{equation}
By the Cauchy-Schwarz inequality
$$
\EE_{\hat{\Pp}^c}
\Bigl(\frac{a_{2N}}{{\hat{\xi}}(C_{2N})+a_{2N}}\;
\frac{|B_N^{\hat{\xi}}|^p}{a_{2N}^p}\Bigr)
\;\leq\;
\EE_{\hat{\Pp}^c} ^\frac{1}{2}
\Bigl(\frac{a^2_{2N}}{({\hat{\xi}}(C_{2N})+a_{2N})^2}\Bigr)
\,\EE_{\hat{\Pp}^c}^\frac{1}{2}
\Bigl(\frac{|B_N^{\hat{\xi}}|^{2p}}{a_{2N}^{2p}}\Bigr)
\;.
$$
The first factor on the r.h.s.
is negligible as $N\uparrow\infty$ because of the limit
(\ref{eq-Lpassump}) for $p=2$, while
the second factor is bounded, uniformly in $N$, because of  (\ref{sabato}).
For the second summand in (\ref{iacopo27p}), we use
 twice the Cauchy-Schwarz inequality
and invoke (\ref{chiara})   to deduce

\begin{equation*}
\begin{split}
\EE_{\hat{\Pp}^c}
 \biggl(\frac{1}{{\hat{\xi}}(C_{2N})+a_{2N}}
& \sum _{x\in Q_N^{\hat{\xi}}\setminus Q_{N-1}^{\hat{\xi}}} \Bigl(
{\hat{\xi}} \bigl( \overline{B_1(x)} \bigr)  \Bigr)^p \biggr)
\quad \leq \quad \EE_{\hat{\Pp}^c}^{\frac{1}{2}} \biggl(
\frac{{\hat{\xi}}(C_{2N}\setminus C_{2N-2}) }{({\hat{\xi}}
(C_{2N})+a_{2N})^2} \biggr)
\\
&  \times \EE_{\hat{\Pp}^c}^{\frac{1}{2} }
 \biggl(\frac{1}{{ \hat{\xi}}(C_{2N}\setminus C_{2N-2})    }
\Bigl( \sum _{x\in Q_N^{\hat{\xi}}\setminus Q_{N-1}^{\hat{\xi}}}
 \Bigl( {\hat{\xi}} \bigl( \overline{B_1(x)} \bigr)  \Bigr)^p \Bigr)^2\biggr)
\\
 & \leq \biggl(\rho_c\,\ell( C_{2N}\setminus C_{2N-2})
 \EE_{\hat{\Pp}^c}
\biggl( \frac{{\hat{\xi}}(C_{2N}\setminus C_{2N-2}) }{({\hat{\xi}}
(C_{2N})+a_{2N})^2} \biggr) \biggr)^{\frac{1}{2}} \;
\EE_{\hat{\Pp}^c_0}^{\frac{1}{2}} \biggl( \Bigl(
\hat{\xi}\bigl(\overline{B_1(0)} \bigr) \Bigr)^{2p}\biggr) \;.
\end{split}
\end{equation*}
The last factor is bounded by hypothesis, the first one converges to $0$ as
$N\uparrow\infty$ because of Lemma~\ref{eternauta} and (\ref{eq-Lpassump}).

\vspace{0.2cm}

In order to prove (\ref{forza1}) observe that
$ \psi_c^{(11)}(S_x {\hat{\xi}} )= \psi_N^{\hat{\xi}} (x) $ if $x\in Q^{\hat{\xi}}_{N-1}$. Therefore  we
 can write
$$
m_N^{\hat{\xi}} (\psi_N^{\hat{\xi}})
\;=\;
\frac{1}{{\hat{\xi}} (C_{2N}) +a_{2N} }
\int_{C_{2N-2}}{\hat{\xi}}(dx) \psi_c^{(11)} (S_x {\hat{\xi}} )
\;+\;
\frac{1}{{\hat{\xi}} (C_{2N})+a_{2N}}
\sum_{x\in\Vv^{\hat{\xi}}_N\setminus C_{2N-2} } \psi_N^{\hat{\xi}}(x)
\;.
$$
Now  (\ref{forza1}) follows easily from (\ref{cile})
 and  (\ref{iacopo27}) with $h_N^{\hat{\xi}}: = \psi_N^{\hat{\xi}} $.
Note that by the same arguments one can prove
\begin{equation}\label{ricamo}
\lim _{N\uparrow\infty} \EE_{\hat{\Pp}^c }
\Bigl\{m_N^{\hat{\xi}}\bigl[\;|\varphi_N^{\hat{\xi}}(x)|^p\; \bigr] \Bigr\}
\;=\;
\EE _{\hat{\Pp}_0^c } ( |\varphi_c^{(1)} |^p )\;<\;\infty\;,
\qquad 1\leq p \leq 4\;,
\end{equation}
which will be useful below.

\vspace{0.2cm}

In order to prove (\ref{forza2}),  we fix $0<\a<1$ and
set $M = 2N-2 [N^\alpha]$, where
$[N^\alpha ]$ denotes the integer part of $N^\alpha$. Moreover, we define the hitting
times

\begin{equation}
\label{tempo1}
\t^{\hat{\xi}}_N(\underline{\omega})
\;=\;
\inf
\left\{s\geq 0\,:\, \omega_s \not \in C_{2N-2}\right\}\;,
\qquad
\underline{\omega}
\;=\; (\omega_s)_{s\geq 0}\in\Omega_N^{\hat{\xi}}=D([0,\infty),
\Vv _N ^{\hat{\xi}})
\; .
\end{equation}
Recall the definitions of the distribution $\hat{\PP}^c_{\hat{\xi}}$,
$\PP^{\hat{\xi}}_{N,x}$ and  $\PP^{\hat{\xi}}_{N}$ given in Sections
\ref{sec-bounds} and \ref{subsec-periodize}.
 Thanks to the identity $(e^{t\Ll_N^{\hat{\xi}}}\varphi_N^{\hat{\xi}})
(x)=\EE_{\PP_{N,x}^{\hat{\xi}}}
\bigl(\varphi_N^{\hat{\xi}} (\omega_t) \bigr)$, we can write

$$
\EE_{\hat{\Pp}^c }
\Bigl(
\langle \varphi_N^{\hat{\xi}}\,, \, e^{t\Ll_N^{\hat{\xi}}}\, \varphi_N^{\hat{\xi}}
\rangle _{m_N^{\hat{\xi}}}
\Bigr)
\;=\;
\EE_{\hat{\Pp}^c } \bigl( A^{\hat{\xi}}_{1,N}+
A^{\hat{\xi}}_{2,N}+A^{\hat{\xi}}_{3,N} \bigr)
\;,$$

\noindent where

\begin{align*}
& A^{\hat{\xi}}_{1,N}\;=\; m_N^{\hat{\xi}} \Bigl( \chi(x\not \in C_M)\,
\varphi_N^{\hat{\xi}}(x) \, \EE_{\PP_{N,x}^{\hat{\xi}}}
\bigl(\varphi_N^{\hat{\xi}}(\omega_t)\bigr)\Bigr)
\;,
\\
& A^{\hat{\xi}}_{2,N}\;=\;
 m_N^{\hat{\xi}} \Bigl( \chi(x\in C_M)\,
\varphi_N^{\hat{\xi}}(x) \, \EE_{\PP_{N,x}^{\hat{\xi}}}  \bigl(\, \chi(\t_N^{\hat{\xi}}\leq t
)\,\varphi_N^{\hat{\xi}}(\omega_t)\bigr)\Bigr)
\;,
\\
&  A^{\hat{\xi}}_{3,N} \;=\;
 m_N^{\hat{\xi}} \Bigl( \chi(x\in C_M)\,
\varphi_N^{\hat{\xi}}(x) \, \EE_{\PP_{N,x}^{\hat{\xi}}}  \bigl(\,\chi(\t_N^{\hat{\xi}}> t )\,
\varphi_N^{\hat{\xi}}(\omega_t)\bigr)\Bigr)
\;.
\end{align*}

\noindent Then (\ref{forza2}) follows from
\begin{equation}\label{rondini}
 \lim _{N \uparrow \infty}  \EE_{\hat{\Pp}^c } \bigl(A^{\hat{\xi}}_{1,N}\bigr)
\;=\;0
\;,\qquad
 \lim _{N \uparrow \infty}  \EE_{\hat{\Pp}^c } \bigl(A^{\hat{\xi}}_{2,N}\bigr)
\;=\;0
\;, \qquad
 \lim _{N \uparrow \infty}  \EE_{\hat{\Pp}^c } \bigl( A^{\hat{\xi}}_{3,N}\bigr)
\;=\;
\langle  \varphi_c^{(1)}
\,, \,e^{t\Ll_c }\, \,\varphi_c^{(1)}  \rangle_{\hat{\Pp}^c_0}
\;.
\end{equation}

Let us first prove the first limit in (\ref{rondini}).
By several applications of Cauchy-Schwarz inequality and due to
the identity $\PP_N^{\hat{\xi}}=\int m_N^{\hat{\xi}} (dx ) \PP_{N,x}^{\hat{\xi}}$
we get
\begin{equation*}\label{jirihafame}
\begin{split}
\bigl|\EE_{\hat{\Pp}^c } \bigl( A^{\hat{\xi}}_{1,N}\bigr)\bigr|
& \leq
\EE_{\hat{\Pp}^c }^{\frac{1}{2}} \Bigl\{m_N^{\hat{\xi}}\bigl(\,
\Vv_N^{\hat{\xi}} \setminus C_M\,\bigr) \Bigr\}
\;
\EE_{\hat{\Pp}^c }^\frac{1}{2} \Bigl\{ m_N^{\hat{\xi}} \bigl[\;\varphi_N^{\hat{\xi}}(x)^2
\EE_{\PP_{N,x}^{\hat{\xi}}}  \bigl(\varphi_N^{\hat{\xi}}(\omega_t)^2 \bigr)\;\bigr]
\Bigr\}
\\
& \leq
\EE_{\hat{\Pp}^c }^{\frac{1}{2}}   \Bigl\{m_N^{\hat{\xi}}\bigl(\,
\Vv_N^{\hat{\xi}} \setminus C_M\,\bigr)    \Bigr\}\;
 \EE_{\hat{\Pp}^c }^{\frac{1}{4}} \Bigl\{ m_N^{\hat{\xi}} \bigl[\;\varphi_N^{\hat{\xi}}(x)^4\;\bigr] \Bigr\}
\;
\EE_{\hat{\Pp}^c }^{\frac{1}{4}} \Bigl\{ \EE_{\PP_N^{\hat{\xi}}  }  \bigl(\varphi_N^{\hat{\xi}}(\omega_t)^4
\bigr)\;\bigr] \Bigr\}
\\
& =
\EE_{\hat{\Pp}^c }^\frac{1}{2} \Bigl\{m_N^{\hat{\xi}}\bigl(\,
\Vv_N^{\hat{\xi}} \setminus C_M\,\bigr)
\Bigr\}\;
\EE_{\hat{\Pp}^c }^{\frac{1}{2}}\Bigl\{m_N^{\hat{\xi}}\bigl[\;\varphi_N^{\hat{\xi}}(x)^4\;
\bigr] \Bigr\}
\;,
\end{split}
\end{equation*}
\noindent where
the last identity follows from the stationarity  of $\Ll _N^{\hat{\xi}}$
w.r.t. $m_N^{\hat{\xi}}$. Due to  the dominated convergence
theorem, the first expectation  on the r.h.s.  goes to
$0$, while the second expectation is bounded due to (\ref{ricamo}).

\vspace{0.2cm}

In order to  prove the second limit in   (\ref{rondini}), we apply twice the
Cauchy-Schwarz inequality in order to obtain  the bound
$\EE_{\hat{\Pp}^c } \bigl( A^{\hat{\xi}}_{2,N} \bigr)$ by
 \begin{equation}
\label{tempio}
\EE_{\hat{\Pp}^c }^{\frac{1}{2}}
\Bigl\{ m_N^{\hat{\xi}}  \bigl[\;\varphi_N^{\hat{\xi}}(x)^2\;\bigr]
 \Bigr\}
\EE_{\hat{\Pp}^c }^{\frac{1}{4}} \Bigl\{ \EE_{\PP_N^{\hat{\xi}}  } \bigl[
\,\varphi_N^{\hat{\xi}}(\omega_t)^4 \;\bigr]
 \Bigr\}
\EE_{\hat{\Pp}^c }^{\frac{1}{4}} \Bigl\{
m_N^{\hat{\xi}}  \bigl[
\chi(x\in C_M)\, \PP_{N,x}^{\hat{\xi}} (\t_N^{\hat{\xi}}\leq t)
\bigr] \Bigr\}
\;.
\end{equation}
Again, because of stationarity and (\ref{ricamo}), the first two
factors on the  r.h.s. are bounded while the last one converges to $0$ due to
Lemma~\ref{coraggio} below.

\vspace{0.2cm}

Finally we prove the last limit in  (\ref{rondini}).
To this aim, given $\underline{ {\hat{\xi}} }\in \hat \Xi =D([0,\infty),\hat{\Nn}_0)$ and $x\in
C_M$,
we set

\begin{equation}
\label{tempo2}
\t_{N,x} (\underline{{\hat{\xi}}}) =
\inf
\left\{ s\geq 0\,:\, x+X_s(\underline{{\hat{\xi}} })\not\in C_{2N-2}
\right\}
\;,
\end{equation}

\noindent where $X_s(\underline{{\hat{\xi}} })$ is defined as in
(\ref{eq-positions}).
Note that for $x\in C_M\cap \text{supp}({\hat{\xi}}) $,

$$
\varphi_N^{\hat{\xi}}(x)
\;=\;
\varphi_c^{(1)} (S_x {\hat{\xi}} )
\;,
\qquad
\EE_{\PP_{N,x}^{\hat{\xi}}}  \bigl(\,\chi(\t_N^{\hat{\xi}}> t )\,
\varphi_N^{\hat{\xi}}(\omega_t)\bigr)
\;=\;
\EE_{\hat{\PP}^c_{S_x {\hat{\xi}}} }
\bigl(\,\chi( \t_{N,x} >t  )\,\varphi_c^{(1)} ({\hat{\xi}}_t )
\,\bigr)
\;.
$$

\noindent Therefore

$$
\EE_{\hat{\Pp}^c }\bigl(\,A_{3,N}^{\hat{\xi}}\,\bigr)
\;= \;
\EE_{\hat{\Pp}^c }\Bigl\{
m_N^{\hat{\xi}} \Bigl[ \chi(x\in C_M)\,
\varphi_c^{(1)} (S_x {\hat{\xi}} )\,\EE_{\hat{\PP}^c_{S_x {\hat{\xi}} }}\bigl(\,\chi(\t_{N,x}>t)\,\varphi_c^{(1)}
({\hat{\xi}}_t)
\,\bigr)\Bigr]\Bigr\}\,.
$$

\noindent
On the other hand, by
applying the Cauchy-Schwarz inequality as in (\ref{tempio})  and due to Lemma
\ref{coraggio},  we obtain
$$
\lim_{N\uparrow\infty}\EE_{\hat{\Pp}^c }\Bigl\{
m_N^{\hat{\xi}} \Bigl[ \chi(x\in C_M)\,
|\varphi_c^{(1)} (S_x {\hat{\xi}} )|\,\EE_{\hat{\PP}^c_{S_x {\hat{\xi}} }}\bigl(\,\chi(\t_{N,x}\leq t)
\,|\varphi_c^{(1)} ({\hat{\xi}}_t) | \,\bigr)\Bigr]\Bigr\}
\;=\;0
\;.
$$

\noindent The last two identities imply
\begin{equation}
\label{urca}
\lim_{N\uparrow\infty} \EE_{\hat{\Pp}^c }\bigl(\,A_{3,N}^{\hat{\xi}}\,\bigr)
\;=\; \lim_{N\uparrow\infty}
\EE_{\hat{\Pp}^c }\Bigl\{
m_N^{\hat{\xi}} \Bigl[ \chi(x\in C_M)\,
\varphi_c^{(1)} (S_x {\hat{\xi}} )\,\EE_{\hat{\PP}^c _{S_x {\hat{\xi}} }}
 \bigl( \,\varphi_c^{(1)} ({\hat{\xi}}_t)\bigr)
\Bigr]\Bigr\}
\;.
\end{equation}

\noindent
Observe now that (\ref{cile}) remains valid if the integral is performed on
$C_M$ in place of $C_{2N-2}$ (the arguments used in the proof there  work also in this case) and  the function $h({\hat{\xi}})$ is
defined as
$$
 h({\hat{\xi}} )
\;=\;
\varphi_c^{(1)} ({\hat{\xi}})\,\EE_{\hat{\PP}^c_{{\hat{\xi}}}}\bigl( \,\varphi_c^{(1)} ({\hat{\xi}}_t)\bigr)
\;=\;
\varphi_c^{(1)} ({\hat{\xi}}) \bigl( e^{t\Ll_c }\varphi_c^{(1)} \bigr)({\hat{\xi}})
\;.
$$
Note that $h\in L^2(\hat{\Nn}_0,\hat{\Pp}^c_0)$.
Therefore we can conclude that the r.h.s. of (\ref{urca}) is
equal to $\langle  \varphi_c^{(1)} \,,\,e^{t\Ll_c } \,\varphi_c^{(1)}  \rangle_{\hat{\Pp}^c_0}$.
\finpro

%%%%%%%%%%%%%%%%%%%%%%%
\begin{lemma}
\label{coraggio}
Let $\tau_N^{\hat{\xi}}$ and $\tau_{N,x} $ be defined as in
{\rm (\ref{tempo1})}
and {\rm (\ref{tempo2})}, and let $M=2N-2[N^\alpha]$. Then

\begin{align}
& \lim_{N\uparrow\infty}
\; \EE_{\hat{\Pp}^c}
\Bigl\{ m_N^{\hat{\xi}}\Bigl[\chi (x\in C_M)\,
 \PP_{N,x}^{\hat{\xi}}
\bigl(\, \t_N^{\hat{\xi}}\leq t\, \bigr)\Bigr]\Bigr\}
\;=\;0\;,
\label{cardi}
\\
& \lim_{N\uparrow\infty}
\; \EE_{\hat{\Pp}^c}\Bigl\{m_N^{\hat{\xi}}\Bigl[\chi (x\in C_M)\,
 \,\hat{\PP}^c_{S_x{\hat{\xi}}}\bigl(\, \t_{N,x}\leq t\,\bigr)\Bigr]\Bigr\}
\;=\;0
\;.
\label{viole}
\end{align}

\end{lemma}
%%%%%%%%%%%%%%%%%%%%%%%

\pro
One can check by a coupling argument
that the two expectations in (\ref{cardi}) and  (\ref{viole}) coincide:
for each $N\in\NN_+$,
$\hat{\xi}\in\hat{\Nn}$ and $ x\in C_M\cap \text{supp}
(\hat{\xi})$,
one can define a probability
 measure  $\mu$ on $\Omega_N^{\hat{\xi}} \times \hat{\Xi}$ such that
$$
\mu(A\times  {\hat{\Xi}})
\;=\;
\PP_{N,x}^{\hat{\xi}}(A)\;,
\;\;\;\;\;\;
\mu(\Omega_N^{\hat{\xi}}
  \times B)\;=\; \hat{\PP}^c_{S_x{\hat{\xi}}}(B)\;,
\qquad
\forall\; A\in\Bb(\Omega_N^{\hat{\xi}} )\;,\;\;\;
\forall\; B\in \Bb({\hat{\Xi}})\;,
$$
and such that, $\mu$ almost surely,
$\t_N^{\hat{\xi}}(\underline{\omega})
=\t_{N,x}(\underline{{\hat{\xi}}})$ and
$\omega_s=x+X_s(\underline{\hat{\xi}})$ for any
$0\leq s < \t_N^{\hat{\xi}}$. Such a coupling $\mu$ implies
$ \hat{\PP}^c_{S_x{\hat{\xi}}}\bigl(\,\t_{N,x}\leq t \,\bigr)=
 \PP_{N,x}^{\hat{\xi}} \bigl(\,\t^{\hat{\xi}} _N\leq t\, \bigr)$.
Thus we need to prove only (\ref{cardi}). Moreover,
 without loss of generality, we assume $r_c=1$.

\vspace{.1cm}

To this aim let us cover
$C_{2N-2}\setminus C_M$ by disjoint cubes $C_{1,i}$ of side $1$, $i \in I$, so that
$
C_{2N-2}\setminus C_M = \cup_{i\in I} C_{1,i}
$ (the boundaries of these cubes
are  suitably chosen for them to be disjoint).
Finally, given a positive integer $n$, we set
$$
I^n_\ast
\;=\;
\{ (l_1,\dots, l_n)\in I^n\,:\, l_j\not=l_k \; \text{ if }
j\not = k \}
\;.
$$
 For paths $\underline \omega$ such that $\tau _N^{\hat{\xi}}
(\underline \omega ) <\infty$, let us define $k=
k(\underline{\omega})$ as the number of different cubes $C_{1,i}$,
$i\in I$, visited by the particle in the time interval
$[0,\t_N^{\hat{\xi}} (\underline{\omega}) \,)$ and moreover we
define  by induction $(i_1,\dots, i_k)\in I^k_*$,   $(x_1,
\dots ,x_k)\in \bigl(C_{2N-2}\setminus C_M\bigr)^k $ with $x_j\in
C_{1,i_j}$  $\forall j:\, 1\leq j\leq k$, and $(t_1,\dots, t_k)$
as follows: Let  $x_1$ be the first point reached in
$C_{2N-2}\setminus C_M$ and $t_1$ be the  time spent in $x_1$
before jumping
 away. The index $i_1$ is characterized
 by the requirement that  $x_1\in C_{1,i_1}$.
 Suppose now that   $i_1,\dots,i_j$,
 $x_1,\dots, x_j$ and  $t_1,\dots, t_j$ have been  defined and that $j<k$.
Then  $x_{j+1}$ is the first point
 in $C_{2N-2}\setminus \bigl(C_M\cup C_{1,i_1}
 \cup \dots \cup C_{1,i_j} \bigr)$ visited during the time interval
 $[0,\t_N^{\hat{\xi}}(\underline{\omega})\,)$ and
$t_{j+1}$ is the time spent at $x_{j+1}$ during such
a first visit.  Moreover, $i_{j+1}$ is such that
$x_{j+1}\in C_{1,i_{j+1}} $.

\vspace{.2cm}

Now let  $T_i^{\hat{\xi}}$, $i\in I$ and
${\hat{\xi}} \in\hat{\Nn}$, be a family of
independent
  exponential random variables (all independent from
the above random objects) and such that $T_i^{\hat{\xi}}$ has  parameter
 ${\hat{\xi}} \bigl(\,\widetilde{C}_{1,i}\,)$, where
$$
\widetilde{C}_{1,i}
\;=\;
\{y \in \RR^d \,:\, \text{dist}(y,C_{1,i}) \leq 1 \,\}.
$$
Since, given   $\hat{\xi}$, $k$  and $(x_1, \dots, x_k)$,
 $t_j $ ($1\leq j \leq k$) are independent
exponential variables  and $t_j$ has parameter  non  larger than
${\hat{\xi}}\bigl(\, \widetilde{C}_{1, i_j} \,\bigr)$ and since
$k\geq k_{\min}:= [N^\a]-1 $, we obtain

\begin{eqnarray}
\label{cosmo}
& & \!\!\!\!\!\! \EE_{\hat{\Pp}^c }
\Bigl\{ m_N^{\hat{\xi}}\Bigl[\chi (x\in C_M)\,
\PP_{N,x}^{\hat{\xi}}
\bigl(\, \t_N^{\hat{\xi}}\leq t\, \bigr)\Bigr]\Bigr\}
\nonumber
\\
& &\;=\;
\sum _{n= k_{\min}}^{|I|}\sum_{{\underline{l}\in I^n_\ast}}
\EE_{\hat{\Pp}^c}
\Bigl\{  m_N^{\hat{\xi}}\Bigl[\chi (x\in C_M)
\sum _{{\underline y}\in {\prod}_{j=1}^n C_{1,l_j} \cap {\Vv} _N^{ {\hat{\xi}}} }
\PP_{N,x}^{\hat{\xi}}
\bigl(\, \t_N^{\hat{\xi}}\leq t,\, k=n, x_l=y_l,1\leq l \leq n
 \bigr)\Bigr]\Bigr\} \nonumber
\\
& & \; \leq\;
\sum _{n= k_{\min}}^{|I|}\sum_{{\underline{l}\in I^n_\ast}}
\EE_{\hat{\Pp}^c }
\Bigl\{   m_N^{\hat{\xi}}\Bigl[\chi (x\in C_M)\, \PP _{N,x}^{\hat{\xi}}
\bigl( k=n, i_1=l_1,\dots, i_n=l_n\  )\Bigr]
\\
& & \qquad\qquad\qquad\qquad\qquad\qquad\qquad\qquad\qquad
\qquad\qquad\qquad\qquad
\times \text{ Prob }\bigl( T_{l_1}^{\hat{\xi}}+
\cdots+T_{l_n}^{\hat{\xi}}\leq t)\Bigr\} \nonumber
%& \;\leq \;
%\sum _{n=k_{\min}}^{|I|}{\sum}_{\underline{a}\in I^n_\ast}
%\EE_{\hat{\Pp}_0 } \Bigl\{{\vartheta}_{a_1,\dots, a_n}({\hat{\xi}})
%  \PP\bigl( T_{a_1}^{\hat{\xi}}+T_{a_2}^{\hat{\xi}}+\cdots+T_{a_n}^{\hat{\xi}}\leq t)\Bigr\}
\end{eqnarray}

\noindent where the last inequality follows from  the bound

\begin{equation*}
\PP_{N,x}^{\hat{\xi}}
\bigl(\, \t_N^{\hat{\xi}}\leq t\,|\, k=n, x_1=y_1,\dots, x_n=y_n
 \bigr)
\leq
 \text{ Prob }\bigl( T_{l_1}^{\hat{\xi}}+
\cdots+T_{l_n}^{\hat{\xi}}\leq t)\,.
\end{equation*}

  In order to estimate the probability in the r.h.s., we use an
argument similar that of the proof of Proposition \ref{giulietta}
in Appendix~\ref{app-LPestimate}. Let us define $ m:=
\EE_{\hat{\Pp}^c} \bigl( \, {\hat{\xi}} (\widetilde{C}_1 )\,
\bigr)$, where $\widetilde{C}_1 = \{ y \in \RR^d : \text{dist}
(y,C_1) \leq 1\}$. Given $\k>0$ and $\underline{l} \in I^n_\ast$
as above, we define $\Aa = \Aa(\k, \underline{l}) $ as follows

$$
\Aa\; =\;\Bigl\{\,
{\hat{\xi}} \in \hat{\Nn}\,:\, \,
\bigl|\,\bigl\{
j\,:\, 1\leq j\leq n \text{ and } {\hat{\xi}}
\bigl(\widetilde{C}_{1,l_j}\bigr)>\k\, m \,
\bigr\} \bigr| > \frac{n}{2} \,
\Bigr\}.
$$
Then, by the  Chebyshev inequality and the stationarity of $\hat{\Pp}^c$,
$$
\hat{\Pp}^c\bigl(\,\Aa\,\bigr)
\;\leq\;
\frac{2}{n}
\;
\EE_{\hat{\Pp}^c}
\Bigl(
\bigl|\,\bigl\{
j\,:\, 1\leq j\leq n \text{ and } {\hat{\xi}} \bigl(\widetilde{C}_{1,l_j}\bigr)>\k\, m \,
\bigr\} \bigr|
\bigr)
\;\leq\;
2 \,\hat{\Pp}^c \bigl(\, {\hat{\xi}} (  \widetilde{C}_1 )>\k\,m \, \bigr)\;\to \; 0
\;,
$$
as $ \k \to  \infty$.
Note that  the complement $\Aa^c$ of $\Aa$ can be written  as
$$
\Aa^c
\;=\;
\Bigl\{\,
{\hat{\xi}} \in \hat{\Nn}\,:\, \,
\bigl|\,\bigl\{
j\,:\, 1\leq j\leq n \text{ and } {\hat{\xi}}
\bigl(\widetilde{C} _{1,l_j}\bigr) \leq \k\, m \,
\bigr\} \bigr| \geq \bigl[\,\frac{n}{2}\,\bigr]_\ast \,
\Bigr\}
\;,
$$
where $[n/2]_\ast$ is defined as $n/2$ for   $n$  even and as
$(n+1)/2$ for $n$ odd. If ${\hat{\xi}} \in \Aa^c$ then at least
$\bigl[\,\frac{n}{2}\,\bigr] _\ast $ of the exponential variables
$T_{l_1}^{\hat{\xi}}$, \dots ,$T_{l_n}^{\hat{\xi}}$ have parameter
non larger than $\k\,m$. Then, by a coupling argument  ({\sl e.g.}
Appendix~\ref{app-LPestimate}),
 we get for all ${\hat{\xi}} \in \Aa^c$
$$
 \text{Prob}
\bigl( T_{l_1}^{\hat{\xi}}+\cdots+T_{l_n}^{\hat{\xi}}\leq t)
\;\leq\;
e^{-\k \,m t}\sum _{r=[n/2]_\ast }^\infty
\frac{ (\k \,m\,t)^r}{r!}
\;=:\;
\phi(\k,\,n)\;.
$$
Due to the above estimates and since $n\geq k_{\min}= [N^\a]-1$, we get
$$
\text{r.h.s. of (\ref{cosmo}) }
\;\leq\;
 2\; \hat{\Pp}^c \bigl(\, {\hat{\xi}} (  \widetilde{C}_1 )>\k\,m \, \bigr)
\;+\;\phi(\k,  N^\a)\;.
$$
The lemma follows by taking first the limit
 $N\uparrow \infty$ and then the limit $\k \uparrow \infty$.
\finpro

%%%%%%%%%%%%%%%%%%%%%%%%%%%%%%%%%%%%%%%%%%
\subsection{Random resistor networks}
\label{subsec-networks}

We conclude this section by pointing out that the diffusion coefficient
$D_N^{\hat{\xi}}$ of the periodized medium can be expressed in terms of the
effective conductance of the graph
$(\overline{\Vv}^{\hat{\xi}}_N,\overline{\Ee}^{\hat{\xi}}_N)$   when
assigning suitable bond conductances. More precisely, consider the
electrical network given by the graph
$(\overline{\Vv}^{\hat{\xi}}_N,\overline{\Ee}^{\hat{\xi}}_N)$ where the bond
$\{x,y\}\in \overline{\Ee}_N^{\hat{\xi}}$ has conductivity $c(\{\pi(x),
\pi(y)\} )$ with $c(\{ \cdot,\cdot\})$ defined in (\ref{impero}).  Then,
the effective conductance $G^{\hat{\xi}}_N$ of this network is defined as
the current flowing from $\G_N^-$ to $\G_N^+$ when a unit potential
difference between  $\G_N^-$ to $\G_N^+$   is imposed. It can be calculated
from Ohm's law and the Kirchhoff rule as follows.  Let  the electrical
potential $V(x)$ vanish on the left border $\Gamma_N^{-}$, be equal to $1$
on the right border $\Gamma_N^{+}$, and satisfy:
$$
\sum_{y\,:\,\{y,x\}\in \overline{\Ee}_N^{\hat{\xi}} } c(\{\pi(x),\pi(y)\}) \;
\bigl(V(y)-V(x)\bigr)\;=\;0
\;\;\;\;\mbox{ for any }\;\;\;\;x\in Q_N^{\hat{\xi}}
\;.
$$
 \noindent Then the effective conductance is given by the current
flowing through the surfaces $\{ x \in [-N,N]^d : x^{(1)} = \pm N \}$:
\begin{equation}
\label{eq-conductance}
G^{\hat{\xi}}_N
\;=\;\sum _{x\in B_N^{\hat{\xi} -}} V(x)
\;=\;\sum _{x\in B_N^{\hat{\xi} +}}  \bigl( 1- V(x) \bigr)\,.
\end{equation}

\noindent By a well-known analogy it is linked to the diffusion
coefficient $D_N^{\hat{\xi}}$
(see {\sl e.g.} \cite[Proposition 4.15]{DFGW} for a similar proof):

%%%%%%%%%%%%%%%%%%%%%%%%%%
\begin{prop}
\label{prop-resistors} One has

\begin{equation}
\label{eq-resistorlink}
D^{\hat{\xi}}_N
\;=\;
\frac{8\,N^2}{|\overline{\Vv}^{\hat{\xi}}_N|}\;G^{\hat{\xi}}_N
\;.
\end{equation}
\end{prop}
%%%%%%%%%%%%%%%%%%%%%%%%%%

%%%%%%%%%%%%%%%%%%%%%%%%%%%%%%%%%%%%%%%%%%
\section{Percolation estimates}
\label{sec-randomized}

Let us set $\Ff_r:=\Ff_{\RR ^d \setminus C_r}$ and  recall that
$\rho_c=\rho\, \delta_c$ with $\delta_c=\nu([-E_c,E_c])$.

%%%%%%%%%%%%%%%%%%%%%%%%%%%%%%%%%%%%%%%%%%
\subsection{Point density estimates}
\label{sec-auxillary}

Here we show how the ergodic properties of Lemma~\ref{eternauta}
combined with the hypothesis (H1) or (H2) imply (\ref{eq-Lpassump}).

%%%%%%%%%%%%%%%%%%%%%%%%%%
\begin{prop}
\label{prop-partdens}  Suppose that $\rho_8<\infty$ and that the
hypothesis {\rm (H1)} or {\rm (H2)} holds. For  $1\leq p\leq 8$,

\begin{equation}
\label{eq-Lpassump2}
\lim_{N\uparrow \infty}  \;
\frac{ \rho _c \,\ell(C_{N})}{{\hat{\xi}}(C_{N})+a_{N}}
\;=\;1
\qquad \text{ in } \;\; L^p(\,\hat{\Nn}, \hat{\Pp}^c )
\;,
\end{equation}

\noindent where  $a_N=(N-1)^{d-1}$.
\end{prop}
%%%%%%%%%%%%%%%%%%%%%%%%%%

We will first prove the following criterion.

%%%%%%%%%%%%%%%%%%%%%%%%%%
\begin{lemma}\label{cri_eindhoven}
Property {\rm (\ref{eq-Lpassump2})} holds if one has,
for some $0<\rho'<\rho$,

\begin{equation}
\label{eq-Lpassump3}
\lim_{N\uparrow \infty}
\;N^p\;
\;\hat{\Pp}\left(
\hat{\xi}(C_N)\leq \rho'\,N^d\right)
\;=\;
0\;.
\end{equation}

\end{lemma}
%%%%%%%%%%%%%%%%%%%%%%%%%%

\pro We first check that (\ref{eq-Lpassump3}) implies that, for some
$0<\rho''<\rho'\delta_c$,

\begin{equation}
\label{eq-Lpassump4}
\lim_{N\uparrow \infty}
\;N^p
\;\hat{\Pp}^c\left(
\hat{\xi}(C_N)\leq \rho''\,N^d\right)
\;=\;
0\;.
\end{equation}

\noindent If $\delta_c=1$, this is clearly true so let us
suppose that $0<\delta_c<1$. Set $\tilde{\delta}_c=1-\delta_c$. If
${\cal C}^k_j$ denotes the binomial coefficient, we have

\begin{eqnarray*}
\hat{\Pp}^c\left(
\hat{\xi}(C_N)\leq \rho''N^d\right)
& = &
\sum_{k=0}^{[\rho''N^d]}
\hat{\Pp}(\hat{\xi}({C}_N)=k)
\;+
\sum_{k=[\rho''N^d]+1}^\infty
\hat{\Pp}(\hat{\xi}({C}_N)=k)
\sum_{j=k-[\rho''N^d]}^k
{\cal C}^k_j\,\tilde{\delta}_c^j\delta_c^{k-j}
\\
& \leq &
\sum_{k=0}^{[\rho'N^d]}
\hat{\Pp}(\hat{\xi}(C_N)=k)
\;+\;
\sup_{k> [\rho' N^d ]} \,
\sum_{j=k-[\rho''N^d]}^k
{\cal C}^k_j\,\tilde{\delta}_c^j\delta_c^{k-j}
%\sum_{j=[\rho'N^d]+1-[\rho''N^d]}^{[\rho'N^d]}
%{\cal C}^{[\rho'N^d]}_j
%\,\tilde{\delta}_c^j\delta_c^{[\rho'N^d]-j}
\\
& \leq &
\hat{\Pp}\left(
\hat{\xi}(C_N)\leq \rho'N^d\right)
\;+\;
\exp(-c[\rho'N^d](\delta_c-\rho''/\rho')^2)
\;,
\end{eqnarray*}

\noindent where the last inequality, given $\rho''<\delta_c\rho'$,
follows from a standard large deviation type estimate for Bernoulli
variables with some $c>0$. Multiplying by $N^p$, (\ref{eq-Lpassump3}) thus
implies (\ref{eq-Lpassump4}).

\vspace{.2cm}

Now set $A_N=\{\hat{\xi}\,:\,\hat{\xi}(C_N)\leq\rho''N^d\}$. Then, for some
$c'>0$ independent of $N$,

$$
f_N(\hat{\xi})
\;:=\;
\left|\frac{\rho_c\,\ell(C_N)}{\hat{\xi}(C_N)+a_N}-1\right|^p
\;\leq\;
c'\,\rho_c^p\,N^p\,\chi_{A_N}(\hat{\xi})
\;+\;
f_N(\hat{\xi})\,\chi_{A^c_N}(\hat{\xi})
\;.
$$

\noindent Integrating w.r.t. $\hat{\Pp}^c$, the first term vanishes in the
limit $N\uparrow\infty$ because of (\ref{eq-Lpassump4}). For the second, let
us first note that Lemma~\ref{eternauta} implies that
$\lim_{N\uparrow 0}f_N\chi_{A^c_N}= 0$ holds $\hat{\Pp}^c$-a.s.. Furthermore,
$|f_N\chi_{A^c_N}|\leq c''<\infty$ uniformly in $N$ so that the dominated
convergence theorem assures that
$\lim_{N\uparrow 0}\EE_{\hat{\Pp}^c}(f_N\chi_{A^c_N})= 0$.
\finpro

\vspace{.2cm}

%%%%%%%%%%%%%%%%%%%%%%%%%%

\vspace{.2cm}

\noindent {\bf Proof of Proposition~\ref{prop-partdens}}. Due to
Lemma \ref{cri_eindhoven} we only need to show that (\ref{eq-Lpassump3})
is satisfied for some $\rho'<\rho$. This is trivially true if (H1) holds.
Hence let us consider the
case where (H2) holds. This implies

\begin{equation}
\label{eq-mixing2}
\left| \EE_{\hat{\Pp}}(f\,|\, \Ff_{r_2})-\EE_{\hat{\Pp}}(f) \right|
\;\leq \;
\|f \|_\infty
\,r_1^d r_2^{d-1}\, h(r_2-r_1)
\;,
\qquad
\hat{\Pp}\mbox{-a.s.}\;,
\end{equation}

\noindent where $f$ is a bounded
 $\Ff_{C_{r_1}}$--measurable function.

\vspace{2mm}

Let  $C_{1}^{i}$ denote the unit  cube centered at $i\in
\ZZ^d$ and $\check{C}_1^i$ be the interior of $C_{1}^{i}$. Let
$I_N \subset \ZZ^d$ be such that $C_N = \cup_{i \in I_N} C_1^{i}$
and $\check{C}_1^{i} \cap \check{C}_1^{j} = \emptyset$ if $i \not=
j$. Hence $|I_N|=N^d$. Given $M>0$, set  $\tilde{Y}_i(\hat{\xi})=
\min\{\hat{\xi}(\check{C}_{1}^{i}),\frac{M}{2}\}$ and
$Y_i=\tilde{Y}_i-\EE_{\hat{\Pp}}(\tilde{Y}_i)$. Note that
 $Y_i$ is centered,  $\Ff_{C^{i}_1}$--measurable and
$\|Y_i\|_\infty\leq M$. We choose $M$ large enough so that
$\tilde{\rho}:= \EE_{\hat{\Pp}}(\tilde{Y}_i)>\rho'$ which is possible because
 $\lim_{M\uparrow\infty}\EE_{\hat{\Pp}}(\tilde{Y}_i)=\rho> \rho'$.
Now

$$
\left\{\hat{\xi}(C_N)\leq\rho'N^d\right\}
\;\subset \;
\left\{\sum_{i\in I_N} \tilde{Y}_i(\hat{\xi})\leq \rho' N^d
%Y_i(\hat{\xi}) \geq(\tilde{\rho}-\rho')N^d
\right\}
\;\subset\;
\left\{\bigl|\sum_{i\in I_N}Y_i(\hat{\xi})\bigr|
\geq(\tilde{\rho}-\rho')N^d\right\}
\;.
$$

\noindent Hence it is sufficient to show that, for $a>0$,

\begin{equation}\label{lafine}
\lim _{N\uparrow\infty}
N^p\; \hat{\Pp}
\left( \bigl|\sum _{i\in I_N} Y_i\bigr|\geq a N^d\right)
\;=\;0
\;.
\end{equation}

\noindent By the Chebyshev inequality, one has for any even $q\in\NN$:
\begin{equation}\label{sispera}
\hat{\Pp}
\left( \bigl|\sum _{i\in I_N} Y_i\bigr|\geq a N^d\right)
\;\leq \;
 \frac{1}{a^q\,N^{dq}}
\;
\sum _{i_1,\ldots,i_q\in I_N}
\EE_{\hat{\Pp}}
\left(
Y_{i_1}\cdots Y_{i_q}
\right)
\;.
\end{equation}
 We will now bound the sum in the r.h.s. of (\ref{sispera}).
Let us define the norm $\|x\|=\max\{ |x^{(k)}|\,:\, 1\leq k\leq d
\}$ on $\RR^d$ (recall that $x^{(k)}$ is the $k$th component of
$x$) and introduce the notation $\underline{i}=(i_1,\ldots,i_q)$,
$\underline{I}_N = (I_N)^{q}$, and $r_j (\underline{i}) = \min \{
\| i_j- i_k \|\, :\, k=1,\ldots , q, k\not= j\}$. If $r_1(
\underline{i})=\ldots = r_N(\underline{i})=0$, i.e., if each point
appears at least twice in $(i_1,\ldots, i_q)$, then use the bound
$\EE_{\hat{\Pp}} (Y_{i_1}\cdots Y_{i_q}) \leq M^q$. The number of
$\underline{i}\in \underline{I}_N$ satisfying this property is at
most $cN^{dq/2}$ (here and below $c$ is a varying constant depending
only on $d$ and on $q$).
 Suppose now that, say, $r_1(\underline{i}) = r
\geq 1$. Then the open cubes $\check{C}_1^{i_2}, \ldots ,
\check{C}_1^{i_q}$ are contained in $A : = \RR^d - C_{2r-1}^{i_1}$
and thus $Y_{i_2}, \ldots , Y_{i_q}$ are $\Ff_{A}$-measurable.
Using conditional expectation, (\ref{eq-mixing2}) and the fact
that $Y_{i_1}$ is centered and $\|Y_{i_j} \|_\infty \leq M$,

$$
\EE_{\hat{\Pp}}
(Y_{i_1}\cdots Y_{i_q})
 \leq
M^{q-1}\;
\EE _{\hat{\Pp} }\bigl(
\bigl|
\EE _{\hat{\Pp} }(
Y_{i_1}|\Ff_{A} )
\bigr| \bigr)
\;\leq \;
M^q\,h(2r -2 )\, (2r -1)^{d-1}
\;.
$$
Note that $\EE_{\hat{\Pp}} (Y_{i_1}\cdots Y_{i_q})$ is
invariant under permutations of the indices
$i_1,\ldots, i_q$. Hence

\begin{eqnarray*}
\sum _{\underline{i} \in \underline{I}_N} \EE_{\hat{\Pp}} (Y_{i_1}\cdots
Y_{i_q})
& \leq  &
c\,
\sum_{r=0}^N \; \,\sum_{\underline{i} \in \underline{K}_N(r)}
\EE_{\hat{\Pp}}
(Y_{i_1}\cdots Y_{i_q})
\\
& \leq  &
  c\,M^q
   N^{d q/2}
  + c\,M^q  \sum_{r=1}^N h(2 r -2)\, (2r -1)^{d-1}\,|\underline{K}_N(r)|\;,
\end{eqnarray*}
where $\underline{K}_N(r)=\{ \underline{i} \in \underline{I}_N  :
  r_1(\underline{i})=r ,    r_2(\underline{i})\leq r,   \ldots,
  r_q(\underline{i}) \leq r \}$. One has

\begin{equation}
\label{eq-correct}
|\underline{K}_N(r)|  \;\leq \;c \,N^{dq/2} r^{dq/2-1}\;.
\end{equation}

\noindent
In fact, on the set of points  $i_1,\ldots,i_q$
(treated as distinguishable) let us define a graph structure by connecting
two points $i \sim j$ with a bond whenever $\| i-j \|\leq r$. We call
$\Gg(\underline i)$ the resulting graph.
Note that each connected component of $\Gg(\underline i)$ has
cardinality at least $2$ whenever    $\underline i\in
\underline{K}_N(r)$, therefore
$\Gg(\underline i)$ has at most $q/2$  connected components.
We claim  that, given $1\leq l\leq q/2$,

\begin{equation}\label{noia}
\left|\{ \underline i\in  \underline{K}_N(r)\,:\,
\text{$\Gg(\underline i)$ has $l$ connected components} \}\right|
\;\leq\;
 c (N/r)^{dl} r^{dq-1}\;.
\end{equation}

In order to prove (\ref{noia}), suppose that the connected component
containing $i_1$ has cardinality $k_1$, while the other components
have cardinality $k_2,\dots , k_l$ respectively.
Each component can be built by first choosing one of its  point  in
$I_N$ (there are $N^d$ possible choices), then its
neighboring points w.r.t. $\sim$ (for each such neighboring point there
are at most $c r^d$ possible choices) and then iteratively adding
neighboring points w.r.t. $\sim$. Therefore,  the  $j$-th component can be
built in at most  $c N^d r^{d(k_j-1)}$ ways. If  $j=1$, since $i_1$
has a neighboring point at distance exactly $r$, the upper
bound can be improved by
$c N^d r^{d-1+d(k_1-2)}$. Summing over all possible $k_1,\dots ,k_l$ such that
$k_1+\ldots + k_l=q$, one gets (\ref{noia}). Since $r\leq N$,
(\ref{noia}) implies
$$
|\underline{K}_N(r)|
\;\leq\;
\sum _{l=1}^{q/2} c  (N/r)^{dl} r^{dq-1}
\;\leq\;
c  (N/r)^{dq/2 } r^{dq-1}
\;,
$$

\noindent thus concluding the proof of  (\ref{eq-correct}).  It implies

\begin{equation}
\label{eq-bound_sum_I_N}
\sum _{\underline{i} \in \underline{I}_N} \EE_{\hat{\Pp}} (Y_{i_1}\cdots
Y_{i_q})
\leq
M^q N^{d q/2} \left( 1 + c' \sum_{r=1}^\infty h(2 r -2)\, r^{d q/2+d-2}
\right)\;.
\end{equation}
Provided that $d q > 2 p$ and the sum over $r$ converges, that is, if
$dq/2 -d \leq 8$, we get the result (\ref{eq-Lpassump3})
by combining (\ref{lafine}), (\ref{sispera}), and (\ref{eq-bound_sum_I_N}). Choosing for $q$
the smallest even integer larger than $16/d$,
(\ref{eq-Lpassump3}) is true for $1 \leq p \leq 8$ and
$dq/2-d \leq 8$ as required.
\finpro

%%%%%%%%%%%%%%%%%%%%%%%%%%%%%%%%%%%%%%%%%%
\subsection{Domination}
\label{sec-domination}

Due to Proposition~\ref{prop-partdens}, we may apply the results of
Section~\ref{sec-approx} so that combining with Proposition~\ref{domenica_mbo}

\begin{equation}
\label{eq-collect}
D
\;\geq\;
\nu([-E_c,E_c])
\;
e^{-r_c-4\beta E_c}
\;
\limsup_{N\to\infty}
\;
\EE_{\hat{\Pp}^c}
\left(
\frac{8\,N^2}{|\overline{\Vv}^{\hat{\xi}}_{N}|}\;
  G^{\hat{\xi}}_{N}
\right)
\;.
\end{equation}

\noindent In order to bound the conductance $G^{\hat{\xi}}_N$ for
$N\gg r_c$ from below, we will discretize  the space $\RR^d$ using cubes of
appropriate size and spacing.  Given $r_2\geq r_1>0$, let us then consider
the following functions on $\hat{\Nn}$:

\begin{equation}
\label{eq-sitevar}
\sigma_j(\hat{\xi})
\;:=\;
\chi\bigl(\, \hat{\xi}(C_{r_1}+r_2 \,j)>0
\,\bigr) \;,\qquad j\in\ZZ^d \;.
\end{equation}

\noindent They form a random field $\Sigma=(\sigma_j)_{j\in\ZZ^d}$ on the
probability space $(\hat{\Nn},\hat{\Pp}^c)$.
If $\hat{\Pp}$ is a PPP, the $\sigma_j$ are independent random
variables. For a process with finite range correlations, this independence
can also be assured by an adequate choice of $r_1$ and $r_2$, but in general
the $\sigma_j$ are correlated. The side length $r_1$ and spacing $r_2$
are going to be chosen of order $\Oo(r_c)$ in such a way that all
points of neighboring cubes have an euclidian distance less than $r_c$ and
they are thus connected by an edge of the graph $(\overline{\Vv}^{
{\hat{\xi}}}_N,\overline{\Ee}^{ {\hat{\xi}}}_N)$.

\vspace{.2cm}

Next note that the $\sigma_j$ take values in $\{0,1\}$. We shall
consider the associated site percolation problem with bonds
between nearest neighbors only \cite{Gri}. For this purpose, we
shall compare $\Sigma$ with a random field
$Z^p=(z^p_j)_{j\in\ZZ^d}$ of independent and identically
distributed random variables with Prob$(z^p_j=1)=p$ and
Prob$(z^p_j=0)=1-p$. In this independent case, it is well-known
that there is a critical probability $p_c(d)\in (0,1)$ such that,
if $p>p_c(d)$, there is almost surely a unique infinite cluster,
while for $p< p_c(d)$ there is almost surely none \cite{Gri}. We
will need somewhat finer estimates for the super-critical regime.
 Let $|.|$ denote the Euclidean norm in $\RR^d$. A left-right
crossing (LR-crossing) with length $k-1$ of $C_{2N}$ of a
configuration $(z^p_j)_{j\in\ZZ^d}$ is a sequence of distinct
points $y_1, \dots, y_k$ in $C_{2N}\cap \ZZ^d$ such that
$|y_i-y_{i+1}|=1$ for $1\leq i<k$, $z^p_{y_i}=1$ for $1\leq i \leq
k$, $y_1^{(1)}=-N$, $y_k^{(1)}=N$, $-N<y_i^{(1)}<N$ for $1<i<k$,
and finally $y_i^{(s)}=y_j^{(s)}$ for any $s\geq 3$ and for $1\leq
i<j\leq k$. Two crossings are called disjoint if all the involved
$y_j$'s are  distinct. In the same way, one defines disjoint
LR-crossings for  $(\sigma_j)_{j\in\ZZ^d}$. Note that this
definition of LR-crossings for $d\geq 3$ uses LR-crossings in
$2$-dimensional slices only. For the random field $Z^p$, the
techniques of \cite[Section~2.6 and 11.3]{Gri} transposed to site
percolation imply that, if $p>p_c(2)$, there are positive
constants $a=a(p) $, $b=b(p)$,   and $c=c(p)$ such that for all
$N\in \NN_+$

\begin{equation}
\label{eq-LR}   \quad \text{Prob} \bigl( Z^p \text{ has less than
$b N^{d-1}$ disjoint LR--crossings in } C_{2N}\bigr) \;\leq\; c\,
e^{-a\, N}\,.
\end{equation}

In order to transpose this result on $Z^{p}$ to one for
$\Sigma$, we will use the concept of stochastic dominance
\cite[Section 7.4]{Gri}. One writes $\Sigma\geq_{\mbox{\rm\tiny st}}Z^{p}$
whenever

\begin{equation}
\label{eq-domination}   \EE_{\hat{\Pp}^c}(f(\Sigma)) \;\geq\;
\EE_{\mbox{\tiny\rm Prob}}(f(Z^p)) \;,
\end{equation}

\noindent for any bounded, increasing, measurable function
 $f:\{0,1\}^{\ZZ^d}\to\RR$ (recall that a function is increasing if
$f((z_j)_{j\in\ZZ^d}) \geq f((z'_j)_{j\in\ZZ^d})$ whenever
$z_j\geq z'_j$ for all $j\in\ZZ^d$). As the event on the l.h.s. of
(\ref{eq-LR}) is decreasing, $\Sigma\geq_{\mbox{\rm\tiny
st}}Z^{p}$ with $p>p_c(2)$ implies that for all $N\in \NN_+$

\begin{equation}
\label{eq-LR2}
\hat{\Pp}^c \bigl( (\sigma_j)_{j\in\ZZ^d} \text{
has  less than $b N^{d-1}$ disjoint LR--crossings in }
C_{2N}\bigr) \;\leq\; c\,  e^{-a\, N}\,.
\end{equation}

\noindent Moreover, let us call the configurations $\hat{\xi}$
in the set on the l.h.s. $N$-{\it bad},
those in the complementary set $N$-{\sl good}.
For every $N$--good configuration $\hat{\xi}$, let us fix
a set of at least $b N^{d-1}$ disjoint LR--crossings in $C_{2N}$
 for $\bigl( \sigma_j(\hat{\xi}) \bigr)_{j\in\ZZ^d} $
 and denote it $\Cc_N(\hat{\xi})$.
Given an LR--crossing $\gamma $ in $C_{2N}$, we write
$L(\gamma)$ for its length.
Note that, since  the LR--crossings are self--avoiding,
$L(\g)=|\text{supp}(\g) |-1$ for all $\g\in\Cc_N(\hat{\xi})$. Moreover,
since that paths in $\Cc_N(\hat{\xi} )$ are disjoint and  have support in
$C_{2N}\cap \ZZ^d$,
$ \sum _{\g\in \Cc_N (\hat{\xi})} |\text{supp}(\g)   |\leq (2N+1)^d $. The above
estimates imply that $\sum_{\gamma\in\Cc_N(\hat{\xi} )} L(\gamma)
\leq (2N+1)^d\leq (4N)^d $.
 In particular, due to Jensen inequality, for any  $N$--good
configuration $\hat{\xi}$,

\begin{equation}
\label{picco}
\sum _{\g\in \Cc_N (\hat{\xi})} \frac{1}{L(\gamma)}
\;\geq\;
\frac{|\Cc_N(\hat{\xi} )|^2 }{ \sum_{\gamma\in\Cc_N(\hat{\xi} )}
L(\gamma) }
\;\geq\;
\frac{b^2\, N^{d-2}}{4^d}
\;.
\end{equation}

\noindent This will allow us to prove a lower bound on
(\ref{eq-collect}). Hence we need the following criterion for domination.

%%%%%%%%%%%%%%%%%%%%%%%%%%
\begin{lemma}
\label{lem-criteria}
$\Sigma\geq_{\mbox{\rm\tiny st}}Z^{p}$ holds with $r_1=r$, $r_2=2r$
if $\hat{\Pp}$ and $r>0$ satisfy the following:
There exists $\rho'>0$ such that

\begin{equation}
\label{eq-cond2}
r^d\,\nu([-E_c,E_c])
\;\geq\;
-\frac{\ln(p/2)}{\rho'}
\;,
\end{equation}
and
\begin{equation}
\label{eq-cond3}
\hat\Pp \Bigl( \hat{\xi}( C_{r} )<\rho' r^d \,\bigl| \Ff_{2\,r}
\Bigr)
\;\leq \;
1\,-\,\frac{3\,p}{2} \;,\qquad \hat \Pp\text{--a.s. }.
\end{equation}
\end{lemma}
%%%%%%%%%%%%%%%%%%%%%%%%%%

\pro The proof is based on the following criterion \cite[Section 7.4]{Gri}:
if for any finite subset $J$
of $\ZZ^d$, $i\in \ZZ^d\setminus J$ and $z_j\in\{0,1\}$ for
$j\in J$ satisfying
$\hat{\Pp}^c (\sigma_j=z_j\;\,\forall j\in J)>0$, one has

\begin{equation}
\label{eq-needed}
\hat{\Pp}^c(\sigma_i=1\,|\, \sigma_j=z_j \;\; \forall\; j\in J)
\;\geq \;
p\,,
\end{equation}

\noindent then $\Sigma\geq_{\mbox{\rm\tiny st}}Z^{p}$.
Hence let $J,i , z_j$ be as above and set
$\tilde{\delta}_c:=1-\delta_c$ and
$J_0:=\{j\in J\,:\, z_j=0\}$ as well as
$J_1:=\{j\in J\,:\, z_j=1\}$.
Moreover, given $\underline k\in \NN^{J_0}$ and $\underline s\in \NN_{+}
^{J_1} $, let
$$
W(\underline k, \underline s)
\;:=\;
\bigl\{\,\hat{\xi}\in\hat{\Nn}\,:\,
 \hat{\xi}(C_r+2r j )=k_j\;\;\;
\forall j\in J_0,\;\;
\hat{\xi}(C_r+2r j )=s_j\;\;\;
\forall j\in J_1\;
\bigr\}\;.
$$
Then
\begin{eqnarray}
& &\!\!\!\!\!\!\!\!\!\!
\hat{\Pp}^c  (\sigma_i=0 \,
\, | \, \sigma_j=z_j \;\;\; \forall j\in J)
\nonumber
\\
& & \;=\;
\frac{
    \sum _{\underline k\in {\NN}^{J_0} }
      \sum_{\underline s \in \NN_{+}^{J_1} }
        \sum _{n\in \NN}
         \hat{\Pp} \bigl( \hat{\xi}(C_r + 2 r i)=n\,,\,
         W(\underline k, \underline s) \bigr)\,
    \tilde{\delta}_c^n \prod _{j\in J_0} \tilde{\delta}_c^{k_j}
      \prod _{j\in J_1} (1-\tilde{\delta}_c^{s_j})}
{
 \sum _{\underline k\in {\NN}^{J_0} }
        \sum_{\underline s \in \NN_{+}^{J_1} }
       \hat{\Pp} \bigl(
          W(\underline k, \underline s) \bigr)\,
        \prod _{j\in J_0} \tilde{\delta}_c^{k_j}
     \prod _{j\in J_1} (1-\tilde{\delta}_c^{s_j})
}
\nonumber
\;.
\end{eqnarray}

\noindent Within this, we can, moreover, replace
$$
   \hat{\Pp}\bigl( \hat{\xi}(C_r + 2r i)=n\,,\,
         W(\underline k, \underline s) \bigr)
\;=\;
   \hat{\Pp} \bigl( \hat{\xi}(C_r + 2r i)=n\,|\,
         W(\underline k, \underline s) \bigr)
\; \hat{\Pp}\bigl(  W(\underline k, \underline s) \bigr)
\;.
$$

\noindent Finally, note that
$W(\underline k, \underline s)\in \Ff_{A}$
where  $A=\RR^d\setminus \bigl(C_{2 r} +2r i  \bigr)$.
As $\tilde{\delta}_c\leq e^{-\delta_c}$,
we obtain the following bound
$$
 \sum _{n\in \NN}
     \hat{\Pp} \bigl( \hat{\xi}(C_r + 2r i)=n\,|\,
          W(\underline k, \underline s) \bigr) \tilde{\delta}_c^n
\;\leq\;
 \hat{\Pp}\bigl( \hat{\xi} (C_r + 2r i) <\rho' r^d \,|\,
 W(\underline k, \underline s) \bigr)
\;+\;
e^{-\delta_c \rho' r^d }\;.
$$

\noindent Due  the
stationarity of $\hat{\Pp}$,
(\ref{eq-cond2}) and (\ref{eq-cond3}) imply (\ref{eq-needed}).
\finpro

%%%%%%%%%%%%%%%%%%%%%%%%%%%%%%%%%%%%%%%%%%
\subsection{Proof of Theorem~\ref{theo-Mott}(ii)}
\label{sec-proofMott}
We fix $p>p_c(2)$ and $\rho'<\rho$. Then,
given $E_c$, we
choose $r_c$ such that (\ref{eq-cond2}) is satisfied, {\it i.e.}
$r_c=c(E_c^{\a+1})^{-1/d}$ for some constant $c$. As
$r_c\uparrow\infty$ in the limit of low temperature, we can next check
that the condition (\ref{eq-cond3}) also holds.
This is trivial for a process
with a uniform lower bound (\ref{eq-uniformbound}) on the point density. For
a mixing point process satisfying  (\ref{eq-mixing}), one has

$$
\hat\Pp \Bigl( \hat{\xi}( C_{r} )<\rho ' r^d \,\bigl| \Ff_{2r}  \Bigr)
\;\leq\;
\hat\Pp \Bigl( \hat{\xi}( C_{r} )<\rho ' r ^d \Bigr)
\;+\;
 r^d\,(2r)^{d-1} \, h(r) \;,
\qquad
\hat{\Pp}-\mbox{a.s.}\,.
$$

\noindent Due to the hypothesis on $h$, the second term converges to $0$ in
the limit $r\uparrow\infty$. If $\rho'<\rho$, the first one can be bounded
by the Chebychev inequality:

$$
\hat{\Pp}(\hat{\xi}(C_r)\leq \rho'\,r^d)
\;\leq\;
{\hat{\Pp}}
\left(
\left|\frac{\hat{\xi}(C_r)}{\ell(C_r)}-\rho\right|>
\rho-\rho'\right)
\;\leq\;
\frac{1}{\rho-\rho'}
\int \hat{\Pp}(d\xi)\,
\left|\frac{\hat{\xi}(C_r)}{\ell(C_r)}-\rho\right|
\;.
$$

\noindent By Lemma~\ref{eternauta}, the expression on the r.h.s. can be made
arbitrarily small by choosing $r$ sufficiently large, thus implying that
(\ref{eq-cond3}) is satisfied for $r$ sufficiently large.
In conclusion, due to Lemma \ref{lem-criteria},
(\ref{eq-LR2}) holds for $r$ large enough, {\it i.e.} tempurature low enough.
We fix such a value  $r$ satisfying (\ref{eq-LR2}) and call it $r_p$.

\vspace{.2cm}

Consider the variables $(\sigma_j)_{j\in\ZZ^d}$ defined for
$r_1=r_p, r_2=2r_p$  and
choose $r_c=(d+8)^{\frac{1}{2}}r_p$. This assures that, if neighboring sites
$j$ and $j'$
 in $\ZZ^d$ have $\sigma_j(\hat{\xi})=\sigma_{j'}(\hat{\xi})=1$,
then $C_{r_p}+2j r_p  $ and $C_{r_p}+2 j' r_p $ contain each a point
and these points are
separated by a distance less than $r_c$.
Two neighboring sites $j$ and $j'$ in $\ZZ^d$ such that
$\sigma_j(\hat{\xi})=\sigma_{j'}(\hat{\xi})=1$  define a bond of the site percolation
problem. To such a bond one can associate (at least) two points
$x \in \supp\, \hat{\xi} \cap ( C_{r_p}+2j r_p )$ and
$y \in \supp \,\hat{\xi} \cap (C_{r_p}+2j' r_p ) $ separated by a distance
less than $r_c$.
Given $N$ integer, we define
$$
\hat{N}:= \max \bigl\{
n\in \NN\,:\, C_{r_p}+2 r_p j \subset C_{2[r_p N]}, \;\;\forall
j\in C_{2n} \cap \ZZ^d
\bigr\}\;.
$$
Note that $\hat{N}=\Oo(N)$. If $j,j'\in C_{2\hat{N}}\cap\ZZ^d$, then the
above associated points $x$ and $y$ are linked by an edge of the graph
$(\overline{\Vv}^{{\hat{\xi}}}_{[r_pN]},
\overline{\Ee}^{ {\hat{\xi}}}_{[r_pN]})$ defined in
section~\ref{subsec-periodize}.
Each LR-crossing of $C_{2\hat{N}}$ for
the site percolation problem gives in a natural way
a connected path of
edges of the graph $(\overline{\Vv}^{{\hat{\xi}}}_{[r_pN]},
\overline{\Ee}^{ {\hat{\xi}}}_{[r_pN]})$ which
connects the boundary faces
$\Gamma^{\pm}_N$.

\vspace{.2cm}

For a $\hat{N}$--good configuration $\hat{\xi}$, we
now bound the conductance $G^{\hat{\xi}}_{[r_pN]}$ from below.
For this purpose, let us
consider the random resistor network with vertices
$Q_{[r_pN]}^{\hat{\xi}}\cup\{\hat{\Gamma}^+_N,\hat{\Gamma}^-_N\}$ where unit
conductances are put on all
edges in $\overline{\Ee}^{\hat{\xi}}_{[r_pN]}$
with vertices in $Q_{[r_pN]}^{\hat{\xi}}$ as well as between the two added
boundary points
$\hat{\Gamma}^\pm_N$ and all points of $B^{{\hat{\xi}}\pm}_{[r_pN]}$.
This new network is
obtained from the one of Section~\ref{subsec-periodize} upon placing
superconducting wires between all vertices of $\Gamma^+_{[r_pN]}$ and
$\Gamma^-_{[r_pN]}$
so that they can be identified with a single point  $\hat{\Gamma}_N^+$
and $\hat{\Gamma}_N^-$.
The conductance $g^{\hat{\xi}}_N$ of this new network
(defined as the current flowing from $\hat{\Gamma}^-_N$ to
$\hat{\Gamma}^+_N$  when a
unit potential difference is imposed between these two points) is precisely
equal to $G^{\hat{\xi}}_{[r_pN]}$ because all points of
$\Gamma^\pm_{[r_pN]}$ have the same
potential ($0$ or $1$ respectively) and each has links to all points of
$B^{{\hat{\xi}}\pm}_{[r_pN]}$ with equal conductances summing up to $1$.

\vspace{.2cm}

In order to bound $g^{\hat{\xi}}_N$ from below, we now invoke
Rayleigh's monotonicity law which states that eliminating links
({\it i.e.} conductances) from the network always lowers its conductance.
%For each LR--crossing in $\CC_{\hat{N} }(\hat{\xi} )$ we can find..
%We cut all the links not belonging to the LR-crossings of the good
%configuration. As the length of each crossing is bounded above by $N/b$, its
%conductance of is bounded below by $b/N$.
For a given $\hat{N}$-good configuration $\hat{\xi}$,
we cut all links but those belonging to the
family of disjoint paths associated to
$\Cc_{\hat{N}}(\hat{\xi})$. Each of these paths $\gamma$
connecting $\hat{\Gamma}^+_N$ and
$\hat{\Gamma}^-_N$ is self-avoiding and hence
has a conductance bounded below by $1/L(\gamma)$.
As all the paths of $\Cc_{\hat{N}}(\hat{\xi})$ are
disjoint and they are connecting $\hat{\Gamma}^+_N$ and $\hat{\Gamma}^-_N$ in
parallel, $g^{\hat{\xi}}_N$ is the sum of the conductances of all paths and
it follows from (\ref{picco})
that $g^{\hat{\xi}}_N\geq c(b) N^{d-2}$ for some positive constant
$c(b)$ depending on $b$.
We therefore deduce that

$$
\EE_{\hat{\Pp}^c}
\left(
\frac{[r_pN]^2  }{|\overline{\Vv}^{\hat{\xi}}_{[r_pN]}|}
 G_{[r_pN]}^{\hat{\xi}}
\right)
\;\geq\;c(b)\;
\EE_{\hat{\Pp}^c}
\left(
\frac{[r_pN]^2}{|\overline{\Vv}^{\hat{\xi}}_{[r_pN]}|}\; N^{d-2}
\chi({\hat{\xi}}\mbox{ is }
\;\;\hat{N}\mbox{--good}\;)
\right)
\;.
$$

\noindent Due to (\ref{eq-LR2}) and
%Lemma~\ref{eternauta}
Proposition \ref{prop-partdens}  the r.h.s.
converges to a positive value.

\vspace{.2cm}

Combining this with the estimate (\ref{eq-collect}) we obtain
$$
D
\;\geq\;
C\;\nu([-E_c,E_c])
\;
e^{-r_c-4\beta E_c}
\;
\geq\;
C'\,E_c^{1+\alpha}\,
\exp(-cE_c^{-\frac{\a+1}{d}} -4\beta E_c)
\;,
$$

\noindent where $C$ and $C'$ are positive constants.
Optimizing the exponent leads to
 $E_c=c'\,\beta^{-\frac{d}{\a+1+d}}$  which completes the proof.
\finpro

%%%%%%%%%%%%%%%%%%%%%%%%%%%%%%%%%%%%%%%%%%%%%%%%%%%%%%%%%%%%%%%%%%
\appendix

%%%%%%%%%%%%%%%%%%%%%%%%%%%%%%%%%%%%%%%%%%%%%%%%%%%%%%%%%%%%%%%%%%

%%%%%%%%%%%%%%%%%%%%%%%%%%%%%%%%%%%%%%%%%%%%%%%%%%%%%%%%%%%%%%%%%%
\section{Proof that the random walk is well-defined}
\label{app-process}

%%%%%%%%%%%%%%%%%%%%%%%%%%%%%%%%%%%%%%%%%%%%%%%%%%%%%%%%
\begin{prop}
\label{prop_no_explosion}    Let $\Pp$ be ergodic with
$\rho_2<\infty$.   Then for $\Pp_0$--almost all  $\xi\in\Nn_0$ and
for all $x\in\hat{\xi}$, there exists a unique probability measure
$\PP_x^\xi$ on  $\Omega_\xi=D
([0,\infty),\text{supp}(\hat\xi)) $ of a  continuous--time
random walk starting at $x$ whose transition probabilities $p_{t}^\xi
(y|x) :=\PP^\xi_x ( X_{s+t}^\xi=y| X_{s}^\xi = x )$, $x,y \in
\hat{\xi}$, $t \geq 0, s> 0$  
satisfy the infinitesimal conditions {\rm (C1)} and {\rm
(C2)}.
\end{prop}
%%%%%%%%%%%%%%%%%%%%%%%%%%%%%%%%%%%%%%%%%%%%%%%%%%%%%%%%

\noindent
\pro
The uniqueness follows from \cite[Chapter 15]{Br}. In order to prove existence,
due to the construction described in Section \ref{sec-preliminaries},
 we only need to prove (\ref{pedro}) for $\Pp_0$--almost
all  $\xi$ and for any $x\in\hat{\xi}$.
According to~\cite[Prop. 15.43]{Br}, condition (\ref{pedro}) is implied by the
following one:

\begin{equation}\label{raoul}
\tilde{\PP}_x^\xi \Bigl(\, \sum _{n=0}^\infty \frac{1}{\lambda_{\tilde{X}_n^\xi}(\xi)}
=\infty\,\Bigr)
\;=\;1 \;.
\end{equation}

\noindent Due to the identity

$$
\tilde{\PP}_0^\xi
\Bigl(\, \sum _{n=1}^\infty
\frac{1}{\lambda_{\tilde{X}_n^\xi}(\xi)}=\infty\;\bigl|\; \tilde{X}_1^\xi
 = x\,\Bigr)
\;=\;
\tilde{\PP}_x^\xi \Bigl(\, \sum _{n=0}^\infty
\frac{1}{\lambda_{\tilde{X}_n^\xi}(\xi)}=
\infty\,\Bigr)\;,
\qquad \forall \; x\in \hat{\xi} \mbox{ , }
$$

\noindent the proof will be completed if
we can show (\ref{raoul}) for $x=0$ and $\Pp_0$--almost all $\xi$  and, in
particular, if we can show

$$
\tilde{\PP}\bigl(\sum_{n=0}^\infty \frac{1}{\lambda_0(\xi_n)}=\infty\bigr)
\;=\;
\int \Qq_0(d\xi)\; \tilde{\PP}_0^\xi
\Bigl(\, \sum _{n=0}^\infty \frac{1}{\lambda_{\tilde{X}_n^\xi}(\xi)}
=\infty\,\Bigr)=1\;,
$$
where the distributions $\tilde{\PP}$, $\tilde{\PP}_0^\xi$, and $\Qq_0$
are defined in Section~\ref{sec-construct}.
Due to   Proposition  \ref{prop-ergodicity},  $\tilde{\PP}$
is ergodic and therefore, according to ergodic theory
(see \cite[Chapter IV]{Ros}),
$$
\lim_{N \uparrow \infty} \frac{1}{N}\sum_{n=0}^N \frac{1}{\lambda _0(\xi_n)}
\;=\;\EE_{\Qq_0} \Bigl(\frac{1}{\lambda _0 }\Bigr)
\;=\; \frac{1}{\EE_{\Pp_0} (\lambda _0)}\;,
 \qquad  \tilde{\PP}\text{-almost surely},
$$

\noindent thus allowing to conclude the proof.
\finpro

%%%%%%%%%%%%%%%%%%%%%%%%%%%%%%%%%%%
\begin{rem}
{\rm Explosions are excluded if  $\sup_{x \in \hat{\xi} } \lambda_x (\xi) <
\infty$ (in such a case (\ref{raoul}) is always true),  but this simple
criterion is typically not satisfied in our case. For instance, for a PPP

\begin{equation*}
\sup_{x \in \hat{\xi}} \lambda_x (\xi)
 \; \geq \;
   e^{-4 \beta} \sup_{x \in \hat{\xi}} \sum_{y \in \hat{\xi}, | y-x| \leq 1} e^{-| x-y|}
   \; \geq \;
    e^{-4 \beta-1} \sup_{x \in \hat{\xi}} \hat{\xi} ( C_1 + x ) \; = \; \infty
\;,
\qquad \Pp_0{\text{-a.s.}} \mbox{ . }
\end{equation*}
}
\end{rem}
%%%%%%%%%%%%%%%%%%%%%%%%%%%%%%%%%%%

%%%%%%%%%%%%%%%%%%%%%%%%%%%%%%%%%%%
\section{Proof of Lemma \ref{simmetria} }
\label{app-prooflemma}

Note that
the statements (ii)  and (iii) of Lemma \ref{simmetria}
are proved
in~\cite[Corollary 1.2.11 and Theorem 1.3.9]{FKAS} in dimension $d=1$.
The proof below is valid for any    dimension  $d$.

\vspace{0,1cm}

\noindent
{\bf Proof of Lemma \ref{simmetria}.} (i)
Let  $h(\xi,\xi') := k(\xi,\xi') - k(\xi', \xi)$.
By the definition (\ref{rino}) of the Palm distribution $\Pp_0$,
 $\forall N>0$,  $\forall A\in \Bb(\RR^d)$ and for any non negative measurable
function $f$
\begin{equation}\label{mattone}
\int \Pp_0 (d\xi) \int _A \hat{\xi}(dx) f(\xi, S_x \xi)=
\frac{1}{\rho N^d} \int \Pp(d\xi) \int _{C_N} \hat{\xi}(dy) \int _{A+y}
\hat{\xi} (dx) f(S_y\xi, S_x\xi)\;.
\end{equation}
The  antisymmetry of $h(\xi,\xi')$ and the identity above imply
\begin{equation} \label{eq-proof_lemma1}
\int \Pp_0 ( d \xi) \int_{\RR^d} \hat{\xi} ( dx)\, h(\xi,S_x \xi)
 =
\frac{1}{\rho N^d}  \int \Pp ( d \xi)  \int_{C_N} \hat{\xi} ( d y)
 \int_{\RR^d \setminus C_N} \hat{\xi} ( dx) \,
   h (S_y \xi,S_{x} \xi) \, .
\end{equation}
Let us split the last integral into two integrals over $\RR^d \setminus C_{N+ \sqrt{N}}$ and
over $ C_{N+ \sqrt{N}} \setminus C_N$.
Using (\ref{mattone}) again,

\begin{eqnarray*}
& \displaystyle
\frac{1}{\rho N^d}  \left| \int  \Pp ( d \xi)  \int_{C_N} \hat{\xi} ( d y)
 \int_{\RR^d \setminus C_{N+ \sqrt{N}}} \hat{\xi} ( dx)
  h (S_y \xi,S_{x} \xi)  \right|
\\
& \displaystyle
\;\;\;\;\;\leq \;
 \int  \Pp_0 ( d \xi) \int_{\RR^d \setminus C_{\sqrt{N}} } \hat{\xi} (dx)
 \bigl( | k( \xi,S_{x} \xi)|  + | k(S_{x} \xi, \xi) | \bigr)\;,
\end{eqnarray*}
which  converges to zero as $N \to \infty$  by the dominated convergence theorem.
The same holds for
$$
\frac{1}{\rho N^d}  \left| \int  \Pp ( d \xi)  \int_{C_N} \hat{\xi} ( d y)
 \int_{ C_{N+ \sqrt{N}}  \setminus C_N } \hat{\xi} ( dx) \,
  h (S_y \xi,S_x \xi) \right|\;,
$$
since, due to (\ref{mattone}),  it can be bounded by
\begin{eqnarray*}
& \displaystyle
\frac{1}{\rho N^d} \int  \Pp ( d \xi) \int_{ C_{N+ \sqrt{N}} \setminus C_N} \hat{\xi} ( dx)
  \int_{\RR^d} \hat{\xi} ( d y)
   \bigl( | k(S_y \xi,S_{x} \xi)| + |k(S_{x} \xi, S_y \xi) | \bigr)
\\
& \displaystyle
=
\frac{(N+ \sqrt{N})^d - N^d}{N^d} \int  \Pp_0 ( d \xi)  \int_{\RR^d} \hat{\xi} ( d y)
   \bigl( | k(S_y \xi, \xi)| + |k( \xi, S_y \xi) | \bigr)
\;.
\end{eqnarray*}
Letting $N \to \infty$ in (\ref{eq-proof_lemma1}) leads to the result.

\vspace{0.1cm}

\noindent  (ii)
Since  $\Gamma \in \Bb (\Nn)$ is translation invariant,
one has $\chi_{\Gamma_0}(S_x \xi) =
\chi_{\Gamma} (\xi)$ for all
$\xi \in \Nn$ and $ x\in \hat{\xi}$.
The above remark together with  (\ref{rino}) gives
\begin{equation*}
\Pp_0 ( \Gamma_0 )
\;=\;
\frac{1}{\rho} \int \Pp (d\xi)
\int _{C_1} \hat{\xi}(dx) {\chi}_{\Gamma_0}(S_x\xi)
\;=\;
\frac{1}{\rho}   \int_{\Gamma} {\Pp} ( d \xi) \,
\hat{\xi} (C_1) \mbox{ . }
\end{equation*}
Comparing with (\ref{gaetano}), this yields
$\Pp_0 ( \Gamma_0 ) =1$ if $\Pp ( \Gamma ) =1$. Reciprocally, always due to
(\ref{gaetano}),  if
$\Pp_0 ( \Gamma_0 ) =1$, one gets $\hat{\xi} (C_1) = 0$ for $\Pp$--almost
all  $\xi \in \Nn \setminus \Gamma$,
and by translation invariance $\xi = 0$  for
 $\Pp$--almost
all  $\xi \in \Nn \setminus \Gamma$, thus implying that $\Pp(\Gamma)=1$.
\vspace{0.1cm}

\noindent (iii)
Let us suppose that    $\Pp_0 (A)=\Pp_0(B) >0$
and  set $\Gamma:=\bigcup_{x\in\RR^d} S_x B$. This is a
 translation-invariant Borel subset of $\Nn$ (see Lemma \ref{prop-Borel}) and
$B \subset \Gamma\cap \Nn_0 \subset A$. In particular,
 $\Pp(\Gamma) \in \{ 0, 1\}$
by the ergodicity of $\Pp$.
Since
$\chi _B(S_y\xi)\leq \chi_{\Gamma}(\xi)$ for all $\xi\in\Nn$ and
$y\in\RR^d$, it follows from (\ref{rino})  that
$$
\Pp_0 ( B)
\;=\;
\frac{1}{\rho} \int_{\Nn} \Pp ( d \xi)
\int_{C_1 } \hat{\xi} (dy)\, \chi _B (S_y \xi)
\; \leq\;
\frac{1}{\rho} \int_\Gamma \Pp (\xi) \hat{\xi}\, ( C_1 )
 \;.
$$
Therefore,  $\Pp(\Gamma)=0$ would imply that $\Pp_0(B)=0$,
in contradiction with our assumption. Thus $\Pp(\Gamma)=1$.
But
$ \Gamma \cap \Nn_0\subset A$, therefore the statement follows from (ii).

\vspace{0.1cm}

\noindent (iv)
The thesis follows by  observing that  (\ref{rino}) implies
$$
\EE_{\Pp_0} \bigl(\,\prod _{j=1}^k \hat{\xi}(A_j)\,\bigr) \,=\,
\frac{1}{\rho} \,\int_{\Nn} \Pp(d\xi) \int_{C_1} \hat{\xi}(dx)
\prod _{j=1}^k \hat{\xi}(A_j+x)\,\leq\, \frac{1}{\rho} \int_{\Nn} \Pp(d\xi)
\hat{\xi}(C_1) \prod _{j=1}^k \hat{ \xi}(\tilde{A}_j)
$$
and by applying
the estimate
$a_1\cdots a_{k+1}\leq c(k+1)
\, (a_1^{k+1}+\cdots +a_{k+1}^{k+1})$, $a_1,\dots,a_{k+1}\geq 0$.
\finpro

\vspace{0.1cm}

%%%%%%%%%%%%%%%%%%%%%%%%%%%%%%%%%%%%%%%%%%%%%%%%%%%%%%%%
\begin{lemma}
\label{prop-Borel}
Let $A \in \Bb(\Nn_0)$. Then
$\bigcup_{x\in\RR^d} S_x A \in \Bb(\Nn)$.
\end{lemma}
%%%%%%%%%%%%%%%%%%%%%%%%%%%%%%%%%%%%%%%%%%%%%%%%

\pro
Let us introduce the following lexicographic ordering on $\RR^d$:
$x \prec y$ if and only if either  $\;|x|<|y|$ or
$|x|=|y|$ and there is $k$, $1\leq k \leq d$, such that
 $x^{(k)} < y^{(k)}$ and
$ x^{(l)} = y^{(l)} $ for $ l<k $
(here $x^{(k)}$ is the $k$-th component of the vector $x$).
Given $\hat{\xi} \in \hat{\Nn}$, one can then order
the support of $\hat{\xi}$ according to $\prec$:
$$
\supp(\hat{\xi})
\;=\;
\begin{cases}
\{ y_1 (\hat{\xi}), y_2 (\hat{\xi}),\dots , y_N  (\hat{\xi}) \} & \text{ if }
N:=  {\hat{\xi}}(\RR^d) <\infty \;,  \\
\{ y_j (\hat{\xi})\}_{j\in \NN_+} & \text{ otherwise },
\end{cases}
$$
where $y_j\prec y_k  $ whenever  $j<k$.
For any $n\in\NN$,
let $x_n  : \hat{\Nn}\to \RR^d $ then be defined as
$$
x_n(\hat{\xi})
\;=\;
\begin{cases}
 y_n (\hat{\xi}) & \text{ if }  n\leq  \hat{\xi} (\RR^d)\;, \\
 y_N (\hat{\xi}) & \text{ if } n > N:=  \hat{\xi} (\RR^d) \;.
\end{cases}
$$
Using an adequate family of finite disjoint covers of $\RR^d$ and the fact
that $\hat{\xi}\in\hat{\Nn}
\mapsto \hat{\xi}(B)$ is a Borel function for every Borel set
$B\subset\RR^d$, one can verify that $x_n$ is a Borel function
for each $n$. Moreover,
$\supp ( \hat{\xi} ) =\{ x_n (\hat{\xi})\,:\, n \in\NN\}$
for all $\hat{\xi} \in \hat{\Nn}$.

\vspace{.2cm}

Due to the definition of the Borel sets in $\Nn$ and
$\hat{\Nn}$, the map
$\pi: \Nn  \to \hat{\Nn}$ given by $\pi(\xi)=\hat{\xi}$
is Borel, and by \cite[Section 6.1]{MKM} the function
$F: \RR^d\times \Nn  \to \Nn$ given by
$F(x,\xi)=S_x \xi$ is even continuous. Hence we conclude that

$$
H_n: \Nn  \to \Nn_0 \; ,\qquad
H_n(\xi)\;:=\;
F\bigl(\,  x_n (\hat \xi),\xi \,\bigr )
\;=\;
S_{x_n (\hat \xi)}\xi
\;,
$$

\noindent is a Borel function.
Its restriction $\hat{H}_n:\Nn_0 \to \Nn_0$ is then also a Borel function.
%Note that $H^{-1}_1(\{\xi\})$ is similar to a Voronoi cell in $\xi$
%of the point at the origin.
Now given a Borel subset $A$ of $\Nn_0$, we conclude that
$\Phi(A):= \bigcup_{n=1}^\infty \hat H _n^{-1}(A) $ is a Borel subset
in $\Nn_0$. One can check that
$$
\Phi(A)
\;=\;
\{ \xi\,:\, \xi
\;=\;
S_x \eta
\;\text{ for some }
\eta\in A
\text{ and } x\in \hat \eta\,\}.
$$
Since $\Nn_0$ is a Borel subset of $\Nn$, it follows that
$\Phi (A)$ is a Borel subset of $\Nn$ as is
$H_1^{-1}\bigl(\,\Phi(A)\,\bigr)$ since $H_1$ is a Borel function.
The identity
$$
H_1^{-1}\bigl(\,\Phi(A)\,\bigr)
\;=\;
\bigcup_{x\in \RR^d}  S_x A
\mbox{ , }
$$
now completes the proof.
\finpro

%%%%%%%%%%%%%%%%%%%%%%%%%%%%%%%%%%%%%%%%%%%%%%%%%%%%%%%%%%
\section{Proof of Proposition~\ref{giulietta} }
\label{app-LPestimate}
%%%%%%%%%%%%%%%%%%%%%%%%%%%%%%%%%%%%%%%%%%%%%%%%%%%%%

%%%%%%%%%%%%%%%%%%%%%%%%%%%%%%%%%%%%%%%%%%%%%%%%%%%%%%

{\bf Proof of Proposition~\ref{giulietta}.}
Due to the construction of the dynamics given in Section
\ref{sec-preliminaries},
$$
\EE_{\Pp_0} \EE_{\PP_0^\xi} \left( |X_t^\xi|^\g \right) \,=\,
\EE_{\Pp_0}\,\EE_{\tilde{\PP}_0^\xi\otimes \QQ}\bigl(\,
|\tilde{X}^\xi _{n_\ast^\xi (t) }|^\gamma\,\bigr)\,.
$$
Let $p,q>1$ be such that $1/p+1/q=1$. Due to the H{\"o}lder inequality,
\begin{eqnarray}
& & \EE_{\Pp_0}\EE_{\tilde{\PP}_0^\xi\otimes \QQ}\bigl(
|\tilde{X}^\xi _{n_\ast^\xi (t) }|^\gamma\bigr)
\;=\; \sum_{n=1}^\infty \EE_{\Pp_0}
\EE_{\tilde{\PP}_0^\xi\otimes \QQ }
 \Bigl( |\tilde{X}_n^\xi |^\gamma \,\chi \bigl( n^\xi_\ast (t) \geq 1 \bigr)
  \chi \bigl( n^\xi_\ast (t) = n \bigr) \Bigr)
\nonumber
\\
& & \;\;\;\;\;\;\;\;\;\;\leq\; \sum_{n=1}^\infty
 \biggl( \EE_{\Pp_0 }
  \EE_{\tilde{\PP}_0^\xi\otimes\QQ}
  \Bigl(  | \tilde{X}_n^\xi |^{\gamma\, q}
 \chi \bigl( n^\xi _\ast (t) \geq 1 \bigr) \Bigr)
 \biggr)^{\frac{1}{q}}
 \biggl(
 \EE_{\Pp_0}\bigl(
\text{\small{
$\tilde{\PP}_0^\xi\otimes \QQ$
}}( n^\xi_\ast (t) = n )
 \bigr)\biggr)^{\frac{1}{p}}  .
\nonumber
\end{eqnarray}
Clearly, $n^\xi_\ast (t) \geq 1$ means
$T^\xi_{0,\tilde{X}_0^\xi} \leq t$. It then follows from the estimate $1- e^{-u} \leq u$, $u \geq 0$, that
\begin{equation}\label{eq-pietro}
\EE_{\tilde{\PP}_0^\xi\otimes \QQ}\bigl( | \tilde{X}_n^\xi |^{\gamma  \, q} \, \chi \bigl( n^\xi_\ast (t) \geq 1 \bigr) \bigr)
  \; = \;
\bigl( 1 - e^{- \lambda_0 (\xi) t} \bigr) \EE_{\tilde{\PP}_0^\xi}
  \bigl( | \tilde{X}_n^\xi |^{\gamma\, q} \bigr)
   \; \leq \; \lambda_0 (\xi) t \; \EE_{\tilde{\PP}_0^\xi}
  \bigl( | \tilde{X}_n^\xi |^{\gamma\, q} \bigr)
\mbox{ . }
\end{equation}
We then obtain
\begin{equation} \label{eq-sofia}
\EE_{\Pp_0}\,\EE_{\PP_0^\xi} \bigl( | X_t^\xi |^\gamma \bigr)
\; \leq\;
  C
\sum_{n=1}^\infty
 \biggl(
  \int {\Qq_0} (d \xi) \, \EE_{\tilde{\PP}_0^\xi} \Bigl( | \tilde{X}_n^\xi |^{\gamma\, q} \Bigr)
 \biggr)^{1/q}
 \biggl(
  \int {\Pp_0} (d \xi) \,\tilde{\PP}_0^\xi\otimes \QQ
 \Bigl( n^\xi_\ast (t) = n \Bigr)
 \biggr)^{1/p}\;,
\end{equation}
with $C= [ t\, \EE_{\Pp_0} ( \lambda_0 ) ]^{1/q}$.
We claim that there is a (time-independent) constant $C' >0$ such that
\begin{equation} \label{eq-elena}
\int {\Qq_0} (d \xi) \, \EE_{\tilde{\PP}_0^\xi}
\bigl(  | \tilde{X}_n^\xi |^{\gamma \, q} \bigr)
\; \leq \;
C' \,n^{\gamma \,q}
\mbox{ . }
\end{equation}
To show this, let us note first that, given $\tilde{X}_0^\xi = 0$,
by another application of  the H{\"o}lder inequality,
\begin{equation*}
 \bigl| \tilde{X}_n^\xi \bigr|^{\gamma \, q}
\; =\;
 \biggl|  \sum_{m=0}^{n-1}
\bigl( \tilde{X}_{m+1}^\xi -\tilde{X}_{m}^\xi \bigr) \biggr|^{\gamma\,q}
\; \leq \;
n^{\gamma\,q-1}
     \sum_{m=0}^{n-1} \bigl| \tilde{X}_{m+1}^\xi - \tilde{X}_{m}^\xi  \bigr|^{\gamma\,q}
\;,
\end{equation*}
where it has been assumed that $\gamma \,q> 1$.
One can derive from the stationarity of $\tilde{\PP}$ and Remark \ref{silenzio}
 that
\begin{equation*}
\int {\Qq_0} (d \xi) \,
 \EE_{\tilde{\PP}_0^\xi}
   \bigl( \bigl|\tilde{X}_{n+1}^\xi -\tilde{X}_{n}^\xi  \bigr|^{\gamma \, q} \bigr)
 \; =  \;
\int {\Qq_0} (d \xi) \,
  \EE_{\tilde{\PP}_0^\xi}
   \bigl( \bigl|\tilde{X}_1^\xi  \bigr|^{\gamma \, q} \bigr)
   \; : =\; C'\;.
\end{equation*}
for any $n \in \NN$.
One concludes the proof of (\ref{eq-elena}) by checking that $C'$ is finite.
Actually, by (\ref{eq-p_xy}), $\EE_{\Pp_0} ( \lambda_0 ) \, C'$ is equal to
\begin{equation*}
\int {\Pp_0} (d \xi)
  \int \hat{\xi} ( dx)  \, c_{0, x} (\xi )  | x|^{\gamma \, q} \,\leq\,
c\, \int {\Pp_0} (d \xi)
  \int \hat{\xi} ( dx)\,  e^{-\frac{|x|}{2}}
\;,
\end{equation*}
for a suitable constant $c$.
The r.h.s. can be bounded by means of  Lemma \ref{simmetria}(iv)  and the
same argument leading to Lemma \ref{prop-finite_moment}.

\vspace{.2cm}

In view of (\ref{eq-sofia}) and (\ref{eq-elena}),
the proposition will be proved if we can
show that the expectation \text{$\EE_{\Pp_0} ( \tilde{\PP}_0^\xi\otimes \QQ ( n^\xi_\ast (t) = n ))$}
converges to zero more rapidly than $n^{- (\gamma+1)p}$ as
$n \to \infty$.
Let us fix $0 < \alpha <1$. We will  show that, if $l>0$ is such that
$\EE_{\Pp_0} ( \lambda_0^{l+1} ) < \infty$, then
\begin{equation} \label{eq-esmalda}
\EE_{\Pp_0} \bigl( \tilde{\PP}_0^\xi\otimes\QQ \bigl( n^\xi_\ast (t ) = n \bigr)  \bigr)
\; =\;
  \Oo ( n^{-\alpha l} ) \mbox{ . }
\end{equation}

To this end, let us first make a general observation.
Let $\lambda>0$ and let $T_1, \dots , T_k$ be
independent exponential variables on some probability space $(\Omega,\mu)$,
with parameters $\lambda_1, \dots, \lambda_k \leq \lambda$.
Define  the random variables $T_j^\prime: = (\lambda_j / \lambda) T_j$,
 $j=1, \dots ,k$. These
are independent identically distributed  exponential
variables  with parameter $\lambda$. As $T_j^\prime \leq T_j$, this shows that
\begin{equation}\label{baviera}
\mu \bigl( T_1 +\dots +T_k \leq t\bigr)
\;\leq\;
\mu \bigl( T_1^\prime +\dots +T_k^\prime \leq t\bigr)
\;=\;
e^{-\lambda t }\sum_{j=0}^\infty
\frac{(\lambda  t)^{j+k}}{(j+k)!}
\;\leq\;
\frac{(\lambda  t) ^k}{k!}
\mbox{ . }
\end{equation}

In order to proceed, for all $\xi \in \Nn_0$, let us set
$B_n^\xi :=
\bigl\{ x\in \hat{\xi} \,:\, \lambda _x(\xi) \leq n^{\a} \bigr\} $
as well as
\begin{equation*}
A_n^\xi \;: =\;
 \Bigl\{
  \bigl(\tilde{X}_k^\xi)_{k\geq 0} \in \tilde{\Omega}_\xi \,:\,
   \;\exists \;J \subset I_n \,,\, | J | > \frac{n}{2}\, ,\,
     \tilde{X}_j^\xi\in B_n^\xi \;\,\,\forall\; j \in J
 \Bigr\}
\end{equation*}
where $I_n :=\{ 0,\dots,n-1\}$  and $|J|$ is the cardinality of $J$.
We write $\tilde{\PP}_0^\xi\otimes \QQ \bigl( n^\xi_\ast (t ) = n  \bigr)
=  g_n(\xi) + h_n(\xi)$ with
$$
g_n (\xi)
\;: =\;
  \tilde{\PP}_0^\xi\otimes\QQ
   \Bigl( \bigl\{ n^\xi_\ast (t) = n \bigr\} \cap  A^\xi_n \Bigr)
\;,
\qquad
h_n (\xi )
\;: = \;
    \tilde{\PP}_0^\xi\otimes \QQ
\Bigl( \bigl\{ n^\xi_\ast (t) = n \bigr\} \cap (A^\xi_n)^c \Bigr)
 \mbox{ . }
$$
We first  estimate $g_n$. Obviously $ \{ n^\xi_\ast (t) = n \}$ is contained
 in
$\{ \sum_{j \in J} T_{j,\tilde{X}^\xi_j}^{\xi} \leq t \}$.
As a result,
\begin{eqnarray*}
g_n (\xi)
& \leq &
  \sum_{J \subset I_n, |J| > n/2} \;
   \sum_{x_0,\ldots, x_{n-1} \in \hat{\xi}}
\chi \bigl(  x_j \in B_n^\xi\;\,\forall\;j \in J \bigr)\,
    \chi \bigl(  x_i \notin B_n^\xi\;\,\forall\;i \in I_n \setminus J \bigr)
\\
 & &\;\;\;\;\;
      \tilde{\PP}_0^\xi
       \bigl(
        \tilde{X}_0^\xi = x_0 , \ldots, \tilde{X}_{n-1}^\xi = x_{n-1}
         \bigr)\;
          \QQ
          \Bigl(
          \sum_{j \in J} T_{j,x_j}^\xi \leq t
       \Bigr)
\\
& \leq &
 \max_{k=[n/2]+1, \ldots ,n-1} \Bigl\{ \frac{(n^\alpha\,t)^k}{k !} \Bigr\}
\mbox{ . }
\end{eqnarray*}
Thanks to the Stirling formula $k! \sim k^k e^{-k} \sqrt{ 2 \pi k}$ as $k \to \infty$,
the last expression can be bounded by
a constant times $(2\, e \,t)^{n/2} \, n^{-n(1-\alpha)/2}$ and is thus exponentially small.
We now turn to  $\EE_{\Pp_0} ( h_n)$, $n \geq 1$. Clearly,
\begin{equation*}
\tilde{\PP}_0^\xi \Bigl( \bigl( A^\xi_n \bigr)^c \Bigr)
\;\leq \;
\frac{2}{n} \,
 \EE_{\tilde{\PP}^\xi_0}
 \Bigl(
  \chi \bigl( \tilde{X}_0^\xi \notin B_n^\xi  \bigr) + \ldots +
     \chi \bigl( \tilde{X}_{n-1}^\xi \notin B_n^\xi  \bigr)
 \Bigr)
\;=\;
\frac{2}{n}\sum_{m=0}^{n-1}
 \EE_{\tilde{\PP}_\xi} \bigl( \lambda _0(\xi_m) > n^\a \bigr)
\,.
\end{equation*}
By Proposition  \ref{prop-ergodicity}  and
invoking Chebyshev's inequality, one obtains for any $l>0$
\begin{eqnarray*}
\EE_{\Pp_0}
 \bigl( h_n \bigr)
& \leq  &
\int {\Pp_0}( d \xi) \,
\tilde{\PP}_0^\xi\otimes \QQ
 \Bigl( \bigl\{ n^\xi_\ast (t ) \geq 1 \bigr\} \cap \bigl( A^\xi_n\bigr)^c  \Bigr)
\; \leq \; t
 \int {\Pp_0}( d \xi) \,
  \lambda_0 (\xi)
   \tilde{\PP}_0^\xi  \Bigl(  \bigl( A^\xi_n  \bigr)^c \Bigr)
\\
& \leq &
 \frac{2 t}{n} \,\sum_{m=0}^{n-1}
 \int {\Pp_0}( d \xi) \,
   \lambda_0(\xi) \,
    \EE_{\tilde{\PP}_\xi} \bigl( \lambda _0(\xi_m) > n^\a \bigr)
   \; = \; 2 t \, \EE_{\Pp_0} \bigl( \lambda_0 \,\chi ( \lambda_0 > n^\alpha ) \bigr)
\\
&  \leq &    \frac{2 t}{n^{\alpha l}} \, \EE_{\Pp_0} \bigl( \lambda_0^{l+1}
 \bigr)\;,
\end{eqnarray*}
where the second inequality follows from the same argument leading to
(\ref{eq-pietro}) and the equality follows from the stationarity of $\tilde\PP$.
This proves (\ref{eq-esmalda}).
We may now choose  $p= \alpha^{-1} >1$ arbitrarily close to $1$ so that
$\gamma q>1$ and such that one may
take for $l$ the smallest integer strictly greater
than $\gamma +1$. For such a choice the sum (\ref{eq-sofia}) converges.
We can now invoke Lemma~\ref{prop-finite_moment} to get the result.
\finpro

%%%%%%%%%%%%%%%%%%%%%%%%%%%%%%%%%%%%%%%%%%%

\end{document}